\documentclass[
  10pt,
  DIV=12,
  headings=standardclasses,
  headings=big
]{scrartcl}

\usepackage[T1]{fontenc}
\usepackage{amsmath,mathtools,stmaryrd}
\usepackage{newtxtext,newtxmath}
\usepackage{booktabs,array}
\usepackage{enumitem}
\usepackage{xcolor}
\usepackage{tikz}
\usetikzlibrary{arrows.meta,calc,decorations.pathreplacing,shapes.geometric}
\usepackage{pgfplots}
\pgfplotsset{compat=1.18}
\usepackage[most]{tcolorbox}
\usepackage{wrapfig}
\usepackage{authblk}

\renewcommand\Affilfont{\small}
\makeatletter
\renewcommand\AB@affilsepx{\protect\quad\protect\Affilfont}
\newcommand*\AB@breaks{}
\newcommand*\authornewline{%
  \begingroup
  \@tempcnta=\value{authors}%
  \advance\@tempcnta by 1\relax
  \edef\@tempa{\noexpand\@namedef{@sep\the\@tempcnta}{,\noexpand\\}}%
  \expandafter\g@addto@macro\expandafter\AB@breaks\expandafter{\@tempa}%
  \endgroup
}
\let\orig@AB@maketitle\maketitle
\renewcommand\maketitle{\AB@breaks\orig@AB@maketitle}
\makeatother

\usepackage[colorlinks=true,linkcolor=blue!65!black,citecolor=blue!65!black,urlcolor=blue!65!black]{hyperref}
\usepackage[backend=biber,sorting=none,sortcites,url=false,maxbibnames=5,backref=false,style=nature,autopunct=true]{biblatex}
\usepackage{changepage}
\DeclareAutoPunctuation{.,}
\let\cite\autocite
\DeclareCiteCommand{\citenum}
  {\usebibmacro{cite:init}\usebibmacro{prenote}}
  {\usebibmacro{citeindex}\usebibmacro{cite:comp}}
  {}
  {\usebibmacro{cite:dump}\usebibmacro{postnote}}
\DeclareCitePunctuationPosition{\citenum}{r}
\definecolor{frameblue}{HTML}{1E4E79}
\definecolor{panelbg}{HTML}{F7FAFC}
\definecolor{accentblue}{HTML}{2B6CB0}
\definecolor{accentgreen}{HTML}{2F855A}
\definecolor{accentamber}{HTML}{B7791F}
\definecolor{accentred}{HTML}{9B2C2C}

\setlist[itemize]{leftmargin=1.1em,itemsep=1.2pt,topsep=2.0pt,parsep=0pt}
\setcapindent{0pt}
\newcommand{\ket}[1]{|#1\rangle}
\newcommand{\bra}[1]{\langle#1|}
\usetikzlibrary{quantikz2}
\newcommand{\norm}[1]{\left\lVert #1 \right\rVert}
\usepackage{xcolor}
\usepackage{listings}

\definecolor{leankeyword}{HTML}{0033B3}
\definecolor{leansorry}{HTML}{C62828}
\definecolor{leancomment}{HTML}{6A737D}
\definecolor{leanstring}{HTML}{067D17}

\lstdefinelanguage{lean4}{
  morekeywords={def,theorem,lemma,example,abbrev,structure,class,instance,
    inductive,where,extends,deriving,let,have,show,from,fun,match,with,
    if,then,else,do,return,by,calc,namespace,section,end,open,variable,
    universe,noncomputable,private,protected,public,partial,import,module,
    notation,Type,Prop,Sort},
  morekeywords=[2]{sorry},
  sensitive=true,
  morecomment=[l]{--},
  morecomment=[s]{/-}{-/},
  morestring=[b]",
}

\begin{document}
\baselineskip14pt

\title{Discovering formally verified quantum programs with AI}

\author[1,\textasteriskcentered,\textdagger]{Sergey~Shirobokov}
\author[1,\textasteriskcentered,\textdagger]{Ben~Jaderberg}
\author[{}\textdaggerdbl,\textasteriskcentered]{Katerina~Hristova}
\author[1,\textasteriskcentered]{George~Tsoukalas}
\author[1,\textasteriskcentered]{Johannes~Bausch}\authornewline
\author[1]{Eric~Wieser}%
\author[1]{Francisco~J.H.~Heras} %
\author[2]{Nicholas C.~Rubin}
\author[2]{Guang~Hao~Low}
\author[1]{Sébastien~Racanière}\authornewline
\author[1]{Nenad~Tomašev}
\author[1]{Lei~M.~Zhang}
\author[2]{James~D.~Watson}
\author[2]{Tanuj~Khattar}
\author[{}\textdaggerdbl]{Aws~Albarghouthi}\authornewline
\author[{}\textdaggerdbl]{Stavan~Jain}
\author[1]{Harun~Khan}
\author[1]{Leopoldo~Sarra} %
\author[1]{Andrew~W.~Senior}
\author[2]{Michael~Broughton}\authornewline
\author[1]{Hussain~Masoom}
\author[2]{Sergio~Boixo}
\author[2]{Ryan~Babbush}
\author[1]{Swarat~Chaudhuri}
\author[1]{Pushmeet~Kohli}

\affil[1]{Google~DeepMind}
\affil[2]{Google~Quantum~AI}

\date{\today}
\maketitle
{\renewcommand{\thefootnote}{\fnsymbol{footnote}}%
\footnotetext[1]{Equal contributions.}%
\footnotetext[2]{Corresponding authors: \href{mailto:shirobokov@google.com}{\texttt{shirobokov@google.com}}, \href{mailto:benjaderberg@google.com}{\texttt{benjaderberg@google.com}}.}%
\footnotetext[3]{These authors performed this work as research consultants on assignment at Google.}}

\begin{refsegment}

\begin{abstract}
The timeline by which quantum computers can solve breakthrough problems in science depends on improved quantum algorithms as much as on better hardware~\cite{beverland2022assessing,preskill2025megaquop}. Such algorithmic advances are typically achieved through many years of expert human collaboration~\cite{reiher2017elucidating,lee2021thc}. In mathematics and computer science, large language models have accelerated progress through automated discovery loops~\cite{novikov2025alphaevolve,hubert2026alphaproof}. However, quantum algorithms can only be evaluated by classical computers on restricted or small instances~\cite{babbush2026kickmix,crichigno2026automating}. Here we demonstrate automated discovery of quantum programs beyond these regimes including order-of-magnitude improvements on scientific applications. To achieve this, we build a quantum program library in the Lean theorem prover in which Gemini~4 agents synthesize novel programs against formal problem specifications and verify their correctness and cost. Agent teams collaborate, sharing results and knowledge through wikis and discussing ideas at organized conferences. Our results improve on prior baselines for parallel search~\cite{grover1996fast} (up to $2.57\times$), quantum arithmetic~\cite{haner2018optimizing} ($290\times$), and simulations relevant to fusion, photochemistry and drug-metabolism by $12\times$, $8\times$ and $2.5\times$ respectively. Combining human and artificial intelligence at scale, under the rigor of mathematical proof, lets researchers design algorithms that today's classical and quantum computers cannot test, shortening the path to practical quantum computation.

\end{abstract}

\begin{figure}[!t]
\input{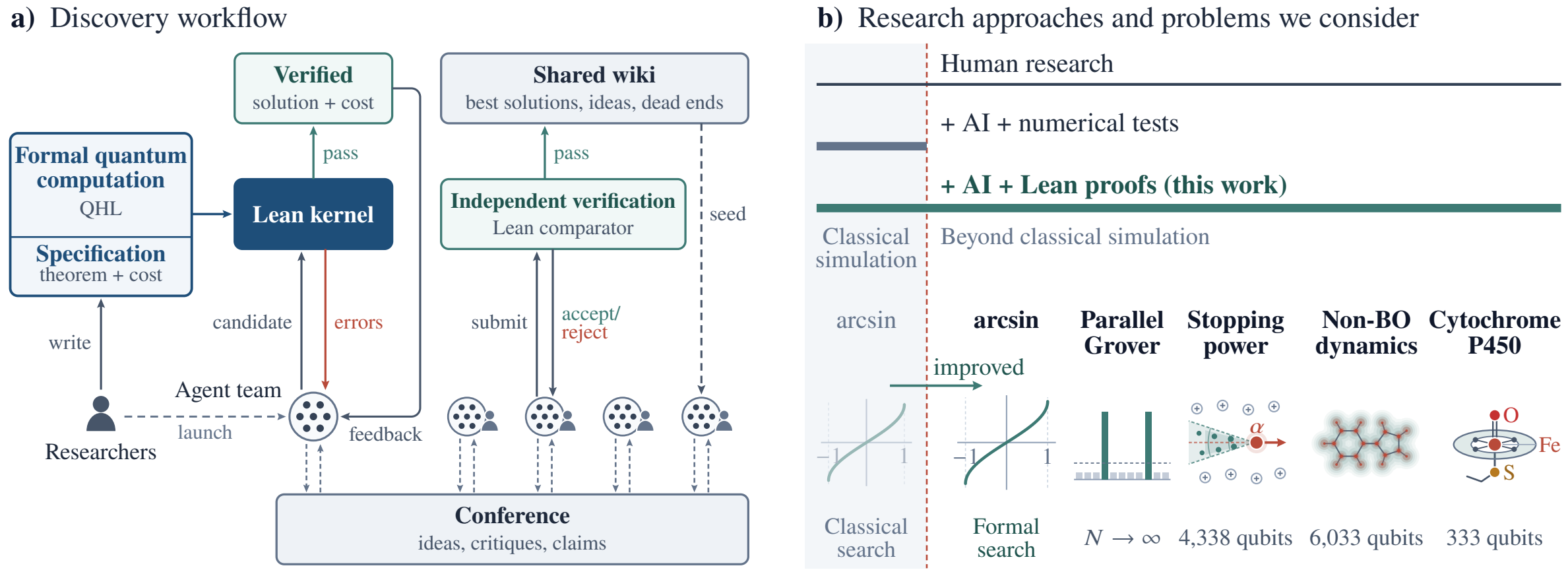}
\caption{\textbf{Formally verified multi-agent quantum program discovery.} \textbf{a)} Researchers write a Lean quantum Hoare logic specification of correctness and cost, then launch their agent team, which iterates on error feedback from the Lean kernel. Teams publish independently re-verified solutions, partial results and dead ends to a shared wiki, and exchange ideas and critiques in scheduled conferences.
\textbf{b)} Bars indicate which problem each approach can reach. AI search checked by numerical tests can only produce solutions that can be simulated classically (dashed line), which for $\arcsin$ means a restricted gate set~\cite{babbush2026kickmix}.}
\label{fig:overview}
\end{figure}

Quantum computers promise to tackle problems beyond classical reach across cryptanalysis~\cite{shor1997polynomial}, optimization~\cite{jordan2025dqi}, chemistry and drug discovery~\cite{aspuruguzik2005simulated,kassal2008polynomial,santagati2024drug}, materials science~\cite{bauer2020quantum,delgado2022battery}, and high-energy and plasma physics~\cite{jordan2012quantum,bauer2023quantum,joseph2023fusion}.
Many of the most promising applications are estimated to require a device with thousands of error-corrected (logical) qubits executing $10^{9}$ to $10^{15}$ logical operations~\cite{beverland2022assessing,dalzell2025survey}, well beyond the roughly one million operations projected for early fault-tolerant processors~\cite{preskill2025megaquop}.
Beyond improvements in hardware and quantum error correction to build better logical qubits from noisy physical qubits~\cite{acharya2025willow}, closing this gap relies on cumulative algorithmic resource reductions that lower qubit and gate counts~\cite{gidney2018halving,babbush2018linear,low2019qubitization,gilyen2019qsvt,su2021firstquant}.
Historically, such reductions take sustained scientific effort over many years~\cite{reiher2017elucidating,lee2021thc}. 

Artificial intelligence (AI) offers a promising route to accelerate these improvements. 
Traditional machine learning has already been successfully applied to e.g.~quantum hardware control~\cite{fosel2018reinforcement,sivak2023realtime,krenn2023artificial}, error-correction decoding~\cite{bausch2024learning}, and small-scale circuit synthesis~\cite{ruiz2025quantum}. While large language models (LLMs) are rapidly advancing mathematics and classical algorithms via execution-based search and automated reasoning~\cite{fawzi2022alphatensor,mankowitz2023faster,romeraparedes2024funsearch,novikov2025alphaevolve,lange2026shinkaevolve,alman2026trulysubquadratic3sumtruly,openai2026eth,anthropic2026eth} as well as formal theorem proving for fixed conjectures~\cite{hubert2026alphaproof,achim2025aristotle}, their application to quantum computing lags behind due to the challenge of \emph{verifying} general correctness without fault-tolerant hardware to test on. High-throughput validation of LLM-generated quantum programs has thus been restricted to circuit regimes that can be classically simulated~\cite{babbush2026kickmix,crichigno2026automating}, limiting the design space and scale of discovery (Fig.~\ref{fig:overview}b), whereas concurrent LLM-assisted formalization has focused on verifying known algorithms (Appendix~\ref{app:qhl}).
Bringing AI discovery to general quantum algorithms instead requires pairing formal models of quantum computation with mathematical proofs of correctness and cost, with access to tools such as numerical heuristics where they are effective. Formal reasoning is particularly critical when targeting asymptotic speedups and large-scale applications spanning thousands of qubits---just like scientists have reasoned about quantum algorithms over the past decades.

In this work, we introduce a Lean~\cite{demoura2021lean4} verification library based on quantum Hoare logic (QHL)~\cite{feng2021qhl}, engineer a multi-agent research workflow built around it, and prove new resource reductions across asymptotic algorithms and large-scale scientific simulations.
Our library pairs a language for quantum programs and classical control flow with formal proofs of correctness and resource cost, chosen per specification. For fault-tolerant circuits we use the spacetime cost $Q\cdot\mathcal{T}$, the number $Q$ of qubits times the number $\mathcal{T}$ of $T$ gates, which are costly in error-corrected quantum computers~\cite{litinski2019game}.
Researchers launch heterogeneous teams of LLM agents that iterate on algorithms, and exchange concepts, partial results, and dead ends through a shared wiki and scheduled cross-team virtual conferences, while an independent verifier audits every proof against a frozen specification written in our library (Fig.~\ref{fig:overview}a). Across five problems (Fig.~\ref{fig:overview}b)---computing $\arcsin$, parallel Grover search, cytochrome P450 energy levels, fusion alpha-particle stopping power and non-Born--Oppenheimer molecular dynamics---this workflow discovers new programs and circuits with machine-checked proofs of correctness and cost, 
reducing resource costs by $290\times$ for $\arcsin$ (combining classical and formal search), and by up to $12\times$ for our scientific applications compared to the strongest prior baselines (Fig.~\ref{fig:results}). Together, these advances shift researchers effort from manual circuit design, optimization, and resource tracking to problem specification and scientific interpretation.

\section{Formally Verified Multi-Agent Discovery}
\label{sec:framework}

\subsection{Formal Verification of Quantum Algorithms}
\label{sec:qhl}

We implement our verification and resource-counting library~\cite{googledeepmind2026fqc} in the Lean theorem prover~\cite{demoura2021lean4}, building on Mathlib~\cite{mathlib2020} and Physlib's Lean-QuantumInfo~\cite{physlib, meiburg2024quantuminfo}.
These formalize core objects of quantum information, but reasoning about quantum programs also requires a programming language to express them and a program logic to prove them correct.
Our library fills this gap by combining a classical-quantum programming language, a classical-quantum Hoare logic to reason about program correctness~\cite{feng2021qhl} and support for resource counting.
Besides classical operations such as assignment, conditional branching and unbounded loops, the formalized language supports core quantum operations: unitary transformations, mid-program qubit measurement, and state reset. We include further details in Appendix~\ref{app:qhl}. 

\begin{figure}[!t]
\input{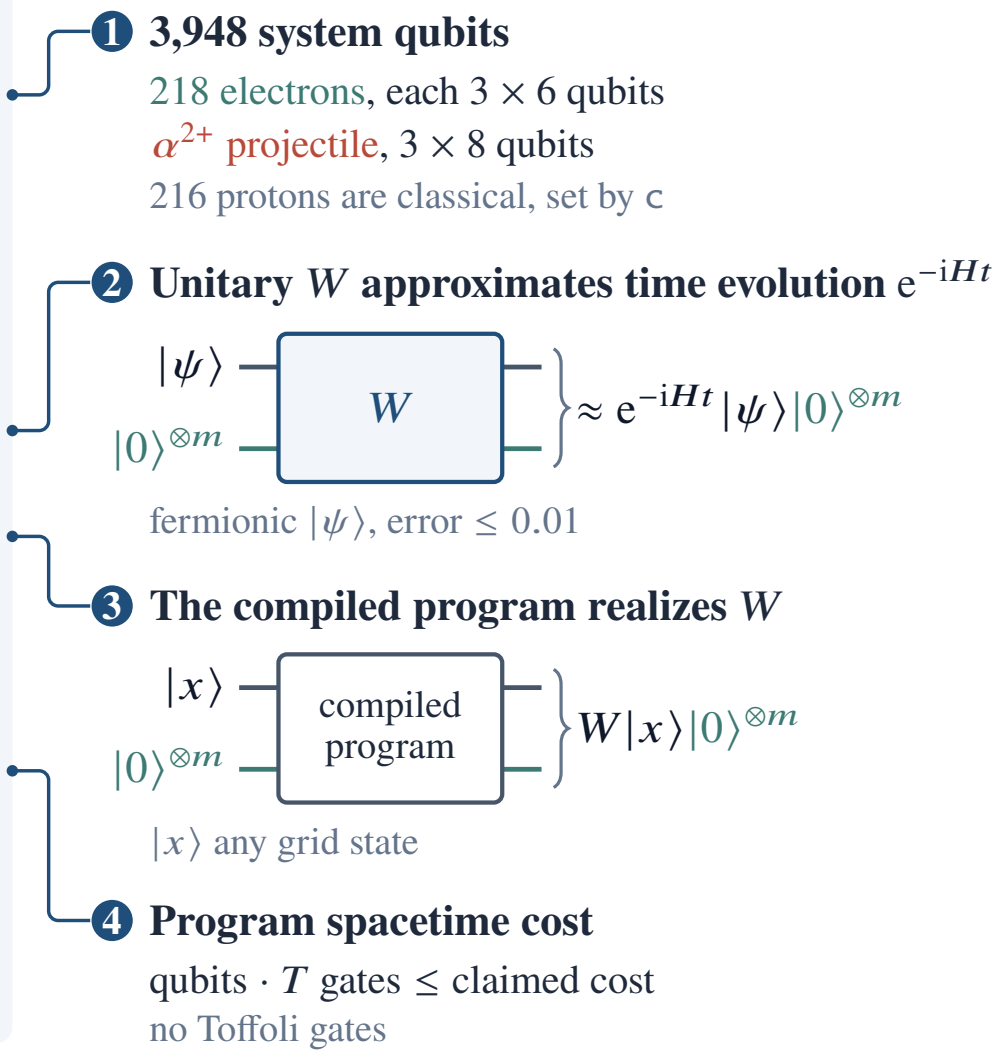}
\caption{\textbf{Lean specifications for discovery tasks.} An abridged specification for electronic stopping power in warm-dense hydrogen~\cite{rubin2024stopping}, where the goal is to construct a quantum program that simulates time evolution under the system Hamiltonian $H$. The agents fill in each \texttt{sorry} statement, which is Lean's placeholder for an unfilled definition or proof: the ancilla qubit count, the circuit and its claimed cost, and the proofs of the correctness and cost theorems. (\textbf{1})~The system qubits store the momentum of every quantum particle. (\textbf{2})~A unitary matrix $W$ on the system and ancilla qubits (extra qubits used to aid computation) must simulate time evolution $e^{-iH(c)t}$ for every physically allowed electron state, up to a specified error, and return the ancillas to $\ket{0}$. (\textbf{3})~The compiled program must act as $W$ on grid states whenever the ancillas start in $\ket{0}$. (\textbf{4})~The cost is a function of the logical spacetime volume $Q\cdot \mathcal{T}$ multiplied by the repeated runs of the stopping power routine: 50 repeats $\times$ 10 time steps $\times$  $Q\cdot \mathcal{T}$~(Sec.~\ref{sec:fusion}).}
\label{fig:specification}
\end{figure}

Correctness theorems are formulated as so-called \emph{Hoare triples} of (precondition, program, postcondition), 
which state that all executions of a program starting in a state satisfying a specified precondition end in a state satisfying a chosen postcondition. Triples are proved backwards from the postcondition by computing the weakest preconditions for each step of the program ~\cite{feng2021qhl}.
Alongside correctness, our library proves bounds on both expected and worst-case resource costs across branches and loops, including $T$-gate count and number of qubits required to execute the algorithm. To pose a program discovery task, we write a \emph{problem specification} as a Lean file (Fig.~\ref{fig:specification}) that states the target correctness and cost theorems while leaving placeholders (\texttt{sorry}) to be filled in with a candidate program, its cost bound, and their proofs.

\subsection{Multi-Agent Discovery Architecture}
\label{sec:agents}
To begin the search on a problem, researchers write the specification and launch independent teams of LLM agents with high-level optimization prompts rather than problem-specific ideas (Fig.~\ref{fig:overview}a). The agent teams then explore candidate algorithms and prove their correctness and cost against the independent verifier. Each researcher chooses their own team setup, from a single agent iterating on a candidate to hierarchical teams targeting individual subroutines, using tools such as analytical derivations, simulation, and numerical optimization (Appendix~\ref{app:swarm}). Unlike multi-agent science systems with a fixed architecture~\cite{swanson2025virtuallab,gottweis2026coscientist,lu2026aiscientist,ghareeb2026robin} or single search loop~\cite{novikov2025alphaevolve, lange2026shinkaevolve, yuksekgonul2026tttdiscover, achim2025aristotle, tsoukalas2026advancing}, we prescribe no internal team design and fix only the shared specification, verifier, and workflow. All large-scale experimental runs of autonomous discovery were performed using Gemini~4 Carbon models.

Agent teams can share progress through two channels. A wiki stores each problem's best verified solution, reusable Lean lemmas, and reviewed notes on promising ideas and dead ends~\cite{qu2026coral}. Newly launched teams can start from the wiki's best solution. Teams also meet in scheduled forum conferences to critique proposals and discuss unverified conjectures, which count as results only once verified.

Because agents can freely modify files in their workspace, they may exploit loopholes by weakening theorem statements, altering definitions, or adding axioms~\cite{zhong2025impossiblebench}. To prevent this, an external verifier checks every submission against the frozen specification using the Lean Comparator~\cite{leanfro2025comparator}, confirming that all target theorems, underlying definitions, and allowed axioms are unchanged (Appendix~\ref{app:verifier}). Researchers thus review only the specification before a run and can trust that any found and verified solution is a valid algorithm.

\section{Results}
\label{sec:results}

\begin{figure}[!t]
\input{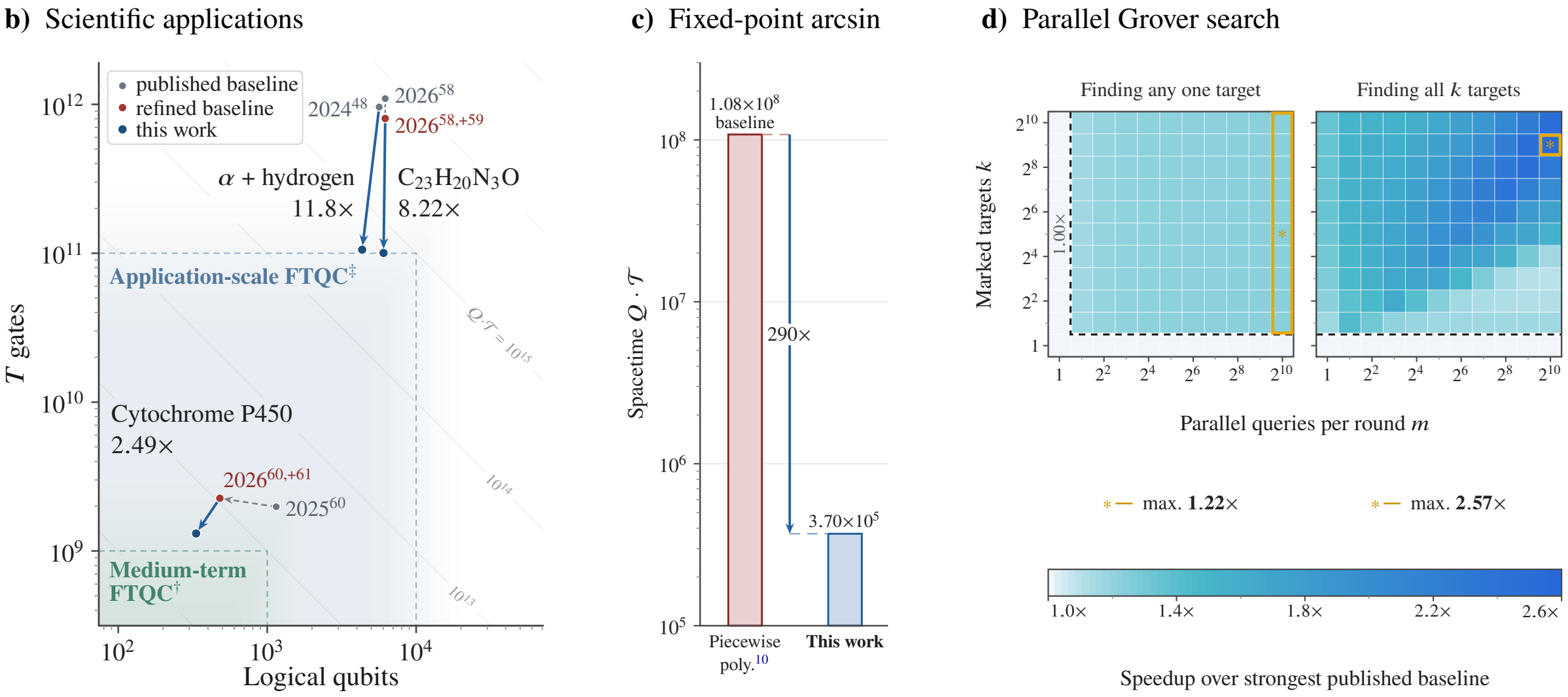}
\caption{\textbf{Summary of verified results.} \textbf{a)} Quantum algorithm cost for baselines and solutions from this work across application domains. 
\textbf{b)} Logical-qubit and $T$-gate improvements on the three scientific applications, with feathered hardware horizons (dashed lines at geometric midpoints) for medium-term fault-tolerant quantum computing (FTQC$^{\dagger}$; teraquop planar architectures~\cite{low2026denser,babbush2026kickmix}) and application-scale FTQC$^{\ddagger}$ (fault-tolerant application-scale quantum computers, FASQ~\cite{preskill2025megaquop,beverland2022assessing,low2026denser}). Grey points denote published baselines, red points denote refined baselines incorporating subsequent methodological improvements from the literature, and blue points denote this work. For cytochrome P450, the published ground-state estimate is extended to all eigenstates within $0.1$~Ha of $E_0$, raising its cost by $0.85\%$.
\textbf{c)} Cost for $25$-bit fixed-point $\arcsin$ ($\varepsilon \le 1.25\times 10^{-7}$) against piecewise-polynomial evaluation~\cite{haner2018optimizing}. \textbf{d)} Speedup in expected number of rounds for large $N$ of our verified parallel Grover circuits over the strongest published baseline~\cite{boyer1998tight,zalka1999grover,gingrich2000generalized,grover2004parallel,fang2026weakly} for powers of two $m, k \le 2^{10}$ when finding any one or all of $k$ marked targets with $m$ parallel queries per round (dashed contours: $1.00\times$ boundary; $*$ markers: peak speedups of $1.22\times$ and $2.57\times$).}
\label{fig:results}
\end{figure}

We demonstrate automated quantum program discovery on five applications (Fig.~\ref{fig:results}): parallel unstructured search (Sec.~\ref{sec:grover}), fault-tolerant arithmetic (Sec.~\ref{sec:arcsin}), and three physical and chemical simulations (Secs.~\ref{sec:fusion}--\ref{sec:p450}).

\subsection{Parallel Grover Search}
\label{sec:grover}
Unstructured search, the problem of finding one or all of $k$ marked items in a list of $N$ candidates, provided one of the earliest provable quantum advantages, with Grover's algorithm achieving a quadratic speedup over classical search~\cite{grover1996fast}.
Marked items are identified by an \emph{oracle} subroutine assumed to dominate runtime, so cost is measured solely in oracle queries. Allowing up to $m$ parallel queries per round on a shared quantum register (rather than $m$ independent, uncoupled devices)~\cite{grover1996fast,zalka1999grover}, we require algorithms to succeed with certainty and score them by their expected oracle rounds for large $N$, evaluated at powers of two up to $1{,}024$ for both $m$ and $k$ (Fig.~\ref{fig:results}d).
We compare against the strongest published protocol in each regime~\cite{boyer1998tight,zalka1999grover,gingrich2000generalized,grover2004parallel,fang2026weakly}, including Zalka's partition combined with weakly measured search at $k = 1$, evaluated under the same metric (Appendix~\ref{app:grover}).

Our Lean-verified circuits match this baseline on the $m = 1$ and $k = 1$ boundaries for finding a single element and outperform it at all multi-target parallel settings ($m, k \ge 2$), reducing expected rounds by up to $1.22\times$; for finding all elements, they outperform it at all multi-target settings ($k \ge 2$), by up to $2.57\times$. Part of this gain comes from combining published ideas via classical control, such as pairing domain partitioning or parallel restarts with weak measurements, which collapse coherence only partially at the price of only partial success information.
The largest find-all gains come from a new coherent mechanism: instead of splitting the list into $m$ blocks with unknown target counts, the circuit holds a superposition over all sets of $m$ candidates, and each parallel query imprints a sign set by the parity of how many remaining targets a set contains. Between queries, a fixed, target-independent walk on the Johnson graph of these sets steers amplitude toward sets containing every remaining target, which a final query records at once (Appendix~\ref{app:grover_results}).

\subsection{Fixed-Point arcsin}
\label{sec:arcsin}
Computing the function $\arcsin(y)$ on a quantum superposition of inputs is an important subroutine for loading numbers into a wavefunction, used in many algorithms including for solving linear systems of equations. We specifically consider the case where the input $y$ is represented by 25-classical bits, and the output must be accurate to an error $\le 1.25\times 10^{-7}$ across the domain $y \in [-1, 1]$. For this setting, the best-known published baseline implementation is a piecewise-polynomial construction~\cite{haner2018optimizing}, which requires a spacetime cost exceeding $1.08\times 10^8$ for an error of $\sim10^{-6}$. 

Unlike the other examples in this work, $\arcsin$ is known to have quantum implementations that can be efficiently simulated classically by using a restricted gate set called \emph{kickmix}~\cite{babbush2026kickmix}. Therefore, we run automated discovery in two steps. First, the agents only consider kickmix circuits, exploiting that solutions can be verified in a matter of seconds. This leads to fast improvements, overall reducing the spacetime cost to $4.15\times 10^{5}$ ($4{,}152$ $T$ gates on $100$ qubits).
Subsequently, we use our formal verification library to lift the kickmix restriction and further optimize the algorithm, using a new quantum gate set that makes the solution hard to verify outside Lean. From this, the agents discover a solution in which a certain component, the coefficient lookup, can be implemented using $33\%$ fewer gates, leading to a further overall $11\%$ lower cost of $3.70\times 10^5$ ($3{,}698$ $T$ gates on $100$ qubits), $290\times$ below the baseline~\cite{haner2018optimizing}.

\subsection{Stopping Power in Inertial-Confinement Fusion}
\label{sec:fusion}

\begin{figure}[!tp]
\input{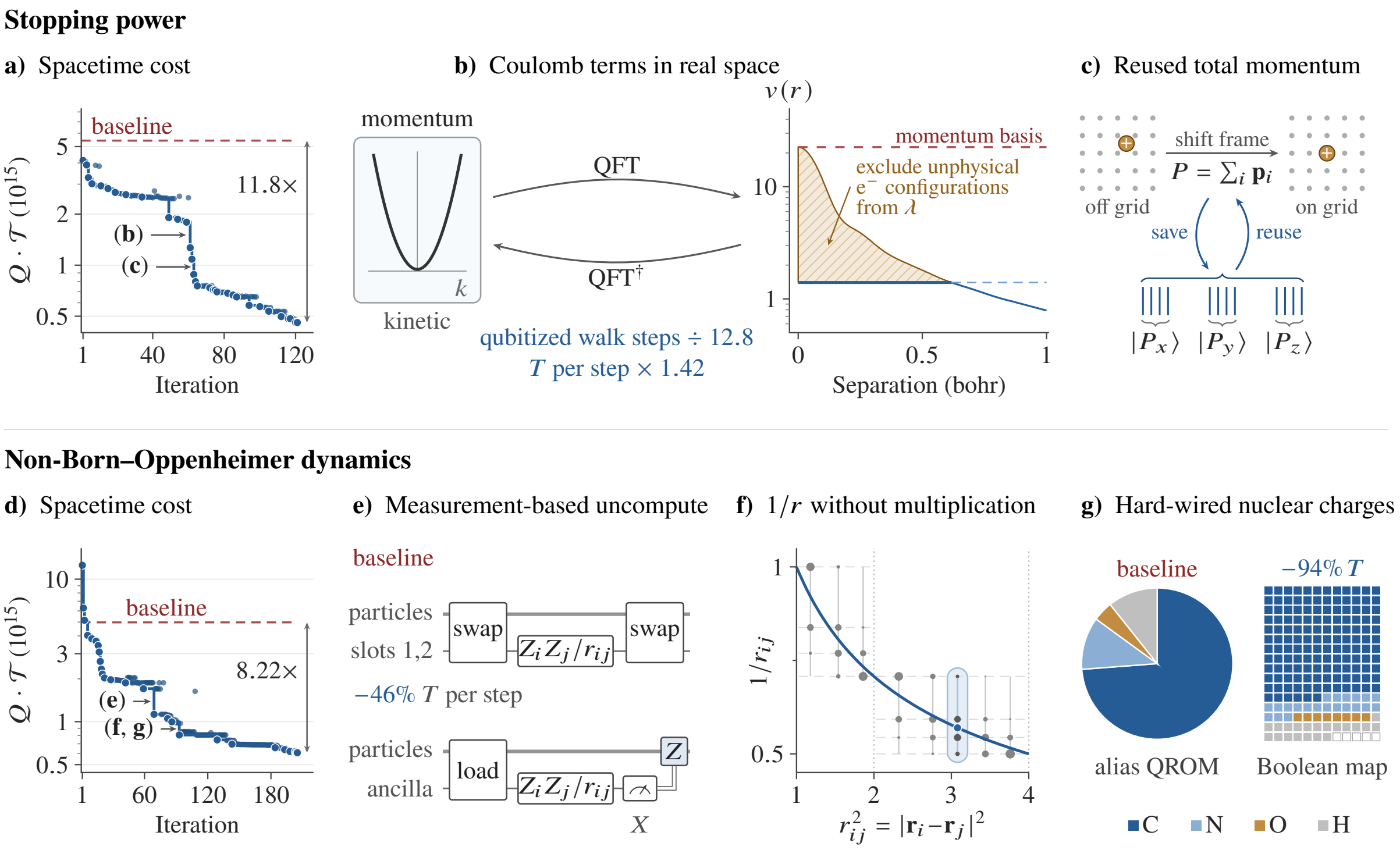}
\caption{\textbf{Where program improvements come from.} \textbf{a, d)} Cost of verified solutions. Arrows mark where the mechanisms in the other panels were introduced. Interestingly, the drop at iteration 49 in \textbf{a)} is not present in the final solution. \textbf{b)} A quantum Fourier transform (QFT) moves the electrons onto a real-space grid and back. On the grid, the pair potential $v(r)$ is split into long-range (blue) and short-range (hatched) parts. The long-range part costs about 30 times less per pair than in the momentum basis (red dashed). For the short-range part, a subroutine bounds the interaction strength of configurations that violate the Pauli exclusion principle, further reducing its cost. \textbf{c)} Protons sit off the grid, so to apply an electron--proton interaction the circuit shifts all electrons to put the proton on a grid point, a phase set by the sum of all $218$ electron momenta. Keeping this sum in a $42$-qubit register between walk steps halves the summations per step. \textbf{e)} To apply the electrostatic interaction $Z_i Z_j / r_{ij}$, rather than swapping the coordinates of the interacting pair into fixed slots and back, a copy is loaded into ancilla qubits. This can be cleared cheaply by measurement without destroying the originals. \textbf{f)} From the loaded coordinates, $r_{ij}^2$ is calculated cheaply and rescaled within each interval to $\tilde{r}_{ij}^2 \in [0,1)$. Through random sampling, one of four classically precomputed coefficients is chosen, with probabilities set by $\tilde{r}_{ij}^2$ (dot area). In expectation, this gives $1/r_{ij}$ as a Bernstein polynomial~\cite{goyal2012bernstein}, as shown for one value of $r_{ij}^2$ (shaded). \textbf{g)} Each nucleus has integer charge $Z_n \in \{1,6,7,8\}$. In the electron--nucleus attraction, rather than preparing these weights via alias sampling~\cite{babbush2018linear}, a uniform superposition is created and reversible logic assigns $Z_n$ of them to each nucleus. Applying a unit-charge interaction on every branch then gives each nucleus its full charge $Z_n$.}
\label{fig:applications}
\end{figure}

In inertial-confinement fusion, hydrogen fuel ignites only if the energetic $\alpha$ particles it produces deposit enough of their energy back into the fuel to sustain the reaction~\cite{hentschel2023improving,hurricane2023physics}. Modeling this requires knowing the fuel's stopping power, the rate at which it absorbs the kinetic energy of an $\alpha$ particle passing through it. No classical method for these dynamics is both efficient and systematically improvable~\cite{rubin2024stopping,grabowski2013molecular,correa2018calculating,ding2018ab,grabowski2020review}.

We consider an $\alpha$ particle and $218$ electrons moving among $216$ fixed protons. A candidate circuit must simulate the time evolution of the system $e^{-iHt}$ with error $\varepsilon = 10^{-2}$ on every state antisymmetric under exchange of same-spin electrons. The full stopping-power calculation repeats this evolution $500$ times.

Our baseline is constructed from Ref.~\citenum{rubin2024stopping}: we take the $5{,}650$ logical qubits and parameters in its Table~2 and evaluate its equation~(12), giving a spacetime cost of $5.42\times 10^{15}$ per run, averaged over the $500$ runs (Appendix~\ref{app:physics}). This builds $e^{-iHt}$ from a sequence of walk steps, each encoding $H/\lambda$, where $\lambda$ is the sum of the absolute weights when $H$ is written as a linear combination of unitaries (LCU). The number of walk steps~\cite{low2019qubitization} scales as $\lambda t$.

Running our discovery workflow, we find a solution with $4{,}338$ logical qubits and $11.8\times$ lower spacetime cost (Fig.~\ref{fig:applications}a). To understand the main improvement, we note that over $90\%$ of the baseline's $\lambda$ comes from electron--proton and electron--electron interactions, and that $\lambda$ must account for every state the qubits can be in, including unphysical ones. Our main innovation is to quantum Fourier transform all electrons to a real-space grid to apply these interactions (Fig.~\ref{fig:applications}b). Importantly, in real space the solution can exploit Pauli exclusion, which allows at most two opposite spin electrons on any grid point. For the short-range part of the potential, a subroutine kicks back a phase of $i$ in one LCU branch and $-i$ in the other for each electron on a nearby grid point. The branches interfere to apply the correct interaction, up to a constant, for two or fewer electrons, while keeping it no larger for three or more, so these grid points no longer set $\lambda$. In the final circuit, this cuts the walk steps by $12.8\times$, at $1.42\times$ the $T$ gates per step.

The largest other improvement comes from keeping the total electron momentum, needed to apply the electron--proton interaction, in a $42$-qubit register between walk steps. This halves how often it is recomputed and cuts the $T$ gates per step by $10\%$ (Fig.~\ref{fig:applications}c). More details are given in Appendix~\ref{app:physics}.

\subsection{\texorpdfstring{Non-Born--Oppenheimer Dynamics of $\mathrm{C_{23}H_{20}N_3O}$}{C23H20N3O}}
\label{sec:photochem}
Light-driven chemical processes such as solar energy
conversion~\cite{scholes2011lessons,lennox2017pcet},
photocatalysis~\cite{gentry2016pcet} and biological
photoreception~\cite{polli2010conical} proceed through excited-state
pathways where electronic and nuclear motion occur on comparable
timescales. This invalidates the Born--Oppenheimer approximation~\cite{worth2004beyond,mai2020photochem,hammesschiffer2010pcet}, 
meaning the nuclei must also be included in the quantum wavefunction. 
Methods that treat the nuclei classically
struggle to simulate quantum tunneling and superpositions of electronic states~\cite{subotnik2016surface,crespootero2018nonadiabatic}, while exact classical simulation 
scales exponentially with system size~\cite{worth2004beyond,kassal2008polynomial}.

Following Reaction~5 of Ref.~\citenum{pocrnic2026prebo}, we
simulate time evolution of $\mathrm{C_{23}H_{20}N_3O}$, modeled on
molecules proposed for photo-activated, proton-coupled electron
transfer~\cite{ren2018eddpt,lennox2017pcet}. Unlike the fixed protons of
Sec.~\ref{sec:fusion}, all $47$ nuclei are treated quantum mechanically
alongside the $187$ electrons ($5{,}616$ system qubits). Here a candidate circuit must implement time evolution $e^{-iHt}$
for $t = 1$~fs, to within error $\varepsilon = 10^{-2}$, for every initial state.

Improving the published estimate~\cite{pocrnic2026prebo} by the authors' subsequent work~\cite{pocrnic2026constants} gives a baseline cost of $4.98\times 10^{15}$ $Q\cdot \mathcal{T}$ (Appendix~\ref{app:nonbo}). 
In this work we find a solution that costs only $6.06\times 10^{14}$ $Q\cdot \mathcal{T}$, a reduction of $8.22\times$ (Fig.~\ref{fig:applications}d).

Within each walk step, three changes lead to the most significant
improvements. First, rather than
swapping interacting particles into fixed sets of qubits and
back~\cite{pocrnic2026prebo}, our solution loads their coordinates into ancilla qubits, which can be cleared by cheap
measurement~\cite{gidney2018halving,su2021firstquant}
(Fig.~\ref{fig:applications}e). Second, instead of evaluating $1/r$ with a multiplication step~\cite{pocrnic2026prebo}, we obtain $1/r$ in expectation from a piecewise 
Bernstein polynomial in $r^2$~\cite{goyal2012bernstein}, using only comparisons and a table lookup (Fig.~\ref{fig:applications}f). 
Finally, a nucleus of charge $Z_n$ attracts an electron just as $Z_n$ unit charges at the same point would. The solution exploits this by creating an equal superposition and assigning $Z_n$ of the branches to each nucleus via classical Boolean
logic~\cite{su2021firstquant}, rather than preparing
unequal weights by alias sampling~\cite{babbush2018linear,pocrnic2026prebo}
(Fig.~\ref{fig:applications}g). More details are given in Appendix~\ref{app:nonbo}.

\subsection{Cytochrome P450 Compound~I}
\label{sec:p450}
Cytochrome P450 enzymes break down most human medicines, so understanding how they react with a drug candidate helps predict how quickly the drug is cleared from the body and whether it produces toxic byproducts~\cite{denisov2005p450,guengerich2008toxicol,rittle2010compound1}.
These reactions are driven by a short-lived intermediate of the enzyme, Compound~I, whose reactivity depends on several closely-spaced spin states. Approximate classical methods such as density functional theory disagree on the energy ordering of these states~\cite{shaik2010p450}, making it a central target for quantum phase estimation, a quantum algorithm to obtain eigenstate energies~\cite{reiher2017elucidating,goings2022p450}.

We consider a model of Compound~I with 63 electrons in 58 orbitals (active space $X$~\cite{goings2022p450}) with the fixed double-factorized
tensor-hypercontraction (DFTHC) Hamiltonian~\cite{low2025spectrum}. For
every eigenstate within $\Delta = 0.1$~Ha of the ground state, the
circuit must estimate the energy with root-mean-square error at
most $\varepsilon = 1$~mHa, including all approximation errors.

For this problem, the best published spacetime cost is $2.26\times 10^{12}$~\cite{low2025spectrum}. To be conservative, we
further integrate recent compilation improvements~\cite{caesura2025blissthc,low2026denser}, which roughly halve the cost. Finally, since the published estimate is for the ground-state energy only, we adjust it for our task of predicting all energies within $\Delta$ of the ground state, giving a cumulative baseline of
$1.09\times 10^{12}$ on $481$ logical qubits (Appendix~\ref{app:p450}).

For this problem, we find a solution that achieves a spacetime cost of $4.37\times10^{11}$, a reduction of
$2.49\times$. To understand the main improvement, consider the baseline. To apply a Hamiltonian term, each walk step performs $57$ rotations in orbital
space, to define new coordinates in which the term acts on one
orbital~\cite{lee2021thc}. The chosen term is stored as a binary number, which is used to read the rotation angles needed into a $247$-qubit register from a
quantum read-only memory~\cite{babbush2018linear}. Our circuit instead gives
each term its own qubit, and sets only the qubit of the chosen term to $1$ (a
one-hot register). Each angle can then be read from these qubits with only
cheap (Clifford) gates once the register is built, with no table lookup. 
Prior work~\cite{caesura2025blissthc} found this narrowly worse than the
baseline because it needs more qubits, and suggested mixing one-hot and binary
encodings. Our solution uses this both to look up the term weights and to supply the
rotation angles, accounting for over $95\%$ of the $T$-gate saving per step and
most of the qubit saving. More details are given in
Appendix~\ref{app:p450}.

\subsection{Collaboration Across Agent Teams}
\label{sec:collaboration}

\begin{figure}[!t]
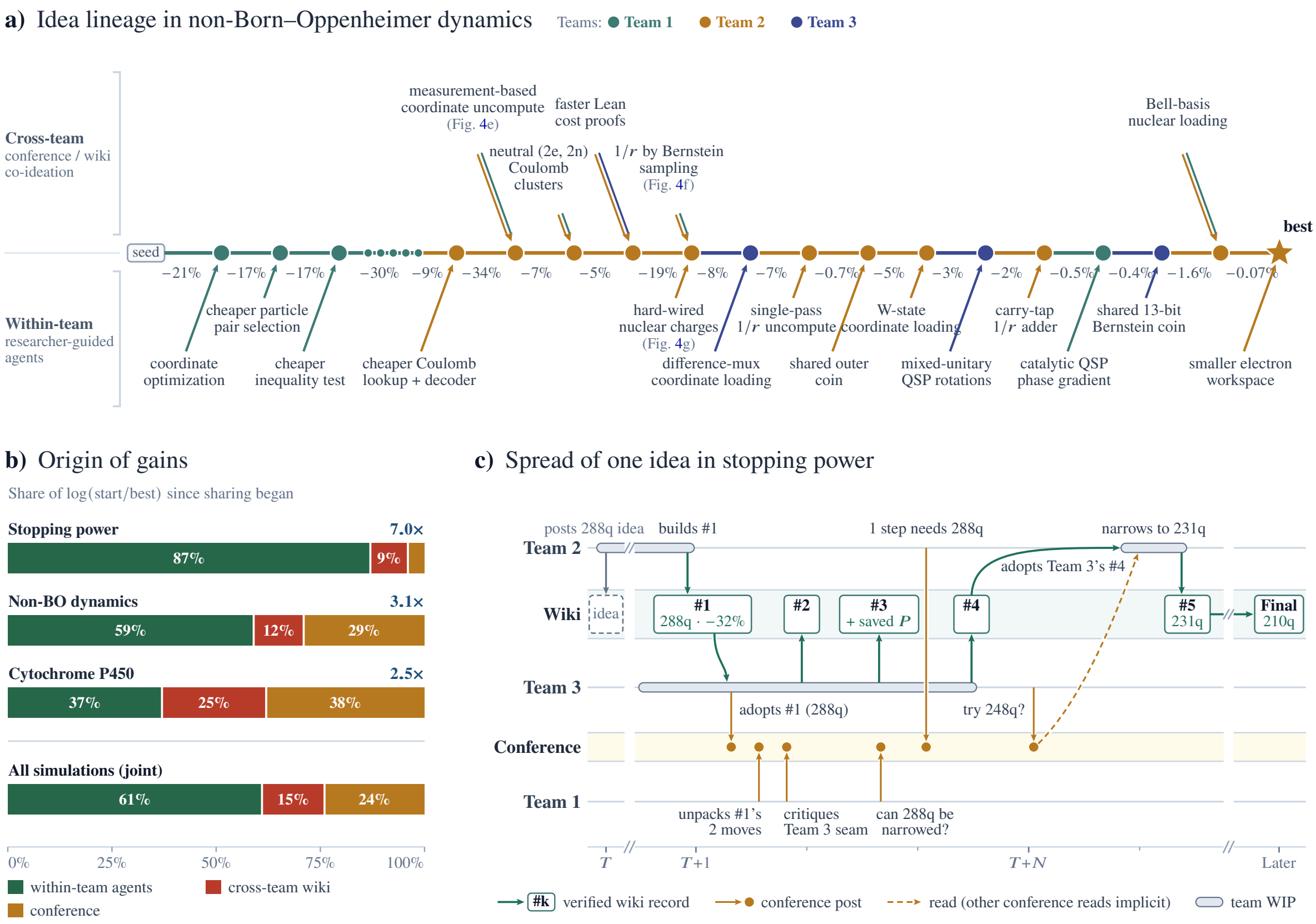

\input{figures/fig4a_lineage}
\par\vspace{4pt}
\input{figures/fig4bc_spread}
\caption{\textbf{Cross-team collaboration and idea lineage.} \textbf{a)} Lineage of verified circuits for non-Born--Oppenheimer dynamics across Teams~1--3 (of seven participating teams) from the baseline (seed) to the final circuit (star), with the $Q\cdot \mathcal{T}$ reduction of each step. Within-team discoveries enter from below; cross-team conference and wiki ideas enter from above.
\textbf{b)} Share of the logarithmic $Q\cdot \mathcal{T}$ reduction from within-team ideas, cross-team wiki entries, and conference exchanges once both channels were active on each scientific application, with reduction factors over this window shown at right and the unweighted average at bottom. \textbf{c)} Cross-team evolution of one stopping-power idea across wiki records \#1--\#5 and conference critiques, narrowing the shared ancilla register from $288$ to $231$ qubits; later records narrowed it further to $210$ qubits.}
\label{fig:collaboration}
\label{fig:lineage}
\label{fig:spread}
\end{figure}

We now discuss the dynamics of the agentic runs to better understand the role of multiple agent teams in search. Our best solutions emerged from collaboration across teams rather than any single team working alone. In non-Born--Oppenheimer dynamics, for example, the leading record changed teams eight times after Team~1's initial run, and the final circuit combines techniques from three agent teams, superseding intermediate improvements from other teams (Figs.~\ref{fig:applications}e--g, \ref{fig:collaboration}a). Whereas the wiki transferred verified code, conferences exchanged unverified proposals and critique: the largest single drop ($-34\%$) began as a Team~1 conference proposal that Team~2 refined and proved (Fig.~\ref{fig:applications}e), and a second key improvement grew from a similar exchange (Fig.~\ref{fig:applications}f).

Such cross-team contributions were sustained across all three scientific applications. To estimate each channel's share once both were active, we classify every verified gain by where its new idea originated (Fig.~\ref{fig:collaboration}b). Averaged across the three scientific applications, $61\%$ of the total cost reduction came from within-team ideas, $15\%$ from cross-team wiki entries or Lean files, and $24\%$ from conferences. The cross-team share ranges from $13\%$ in stopping power to $63\%$ in cytochrome P450. Credit is often hard to separate cleanly, as one proof often combines ideas from a team's own search, the wiki, and conferences.

Figure~\ref{fig:collaboration}c traces one such cross-team exchange in stopping power, where Teams~2 and~3 refined a shared ancilla register across five successive wiki records with conference input from Team~1, narrowing it from $288$ to $231$ qubits; later records narrowed it further to $210$ qubits. Beyond complete proofs, dead-end notes and partial proofs also transferred across teams: a note proving why a proposed shortcut fails stopped two proposals from other teams, while helper Lean modules proved by one team were imported by a second and reused in a third team's stopping-power circuit.

Machine-checked verification is also what makes open collaboration safe at this scale, letting teams build on one another's Lean proofs without manual code review. Flawed candidates and claims fell into two categories (Appendix~\ref{app:verifier}): broken proofs or inaccurate script-based estimates (such as candidates that bypassed local Lean checks), which our external verifier and conference critique filtered out, and loopholes in the specifications themselves (such as exploiting how Toffoli gates were counted without improving the circuit), which researchers and teams closed by revising the specification.

\section{Discussion and Outlook}
\label{sec:discussion}
In this work, we show that combining formal verification with LLM agents can push the frontier of quantum algorithm development. We demonstrate this across fundamental algorithms, fault-tolerant subroutines and full-scale scientific applications, achieving up to a $2.57\times$ reduction in expected oracle rounds for parallel Grover search, a $290\times$ spacetime reduction for fixed-point $\arcsin$, and $11.8\times$, $8.22\times$ and $2.49\times$ reductions on simulations relevant to fusion, photochemistry and drug metabolism. Where classical simulation can check candidate circuits, as with $\arcsin$, numerical testing with formal proving on top is effective. For asymptotic bounds and general algorithms on hundreds or thousands of qubits, formal proof is essential.

Beyond these results, our experiments point to a new research workflow. Rather than replacing human researchers, formal AI discovery shifts where human effort is applied. Researchers and domain experts choose the problems and carefully prepare faithful formal specifications. External verification then lets independent agent teams build on one another's work without manual review, and carry out the search and formal proofs that drive down costs. Finally, researchers interpret the verified solutions to guide future research directions.

Our end-to-end simulations still demand far more resources than projected for the first fault-tolerant quantum computers~\cite{preskill2025megaquop} (Fig.~\ref{fig:results}b; Appendix~\ref{app:horizons}), and reaching early hardware will require further orders-of-magnitude improvements. Part of the challenge is for us to choose specifications well, as agents cannot optimize beyond them. For example, to match the baseline, our non-Born--Oppenheimer dynamics specification requires correctness on \emph{all} input states, even unphysical ones that break exchange symmetry. Similarly, fixing the factorization in cytochrome P450 excludes the representation advances behind recent gains~\cite{lee2021thc,goings2022p450,low2025spectrum}.

Bringing algorithms onto real hardware also means extending formal specifications to error-correcting codes, qubit routing and physical layout~\cite{litinski2019game,low2026denser}, so human--agent teams can optimize algorithms for a given code or co-design both.
Furthermore, qubits are expected to be a more limited resource than gates for early machines, suggesting optimizing each factor unequally~\cite{gidney2025rsa}. Other hardware designs, where $T$ gates can run in parallel, may instead favor minimizing qubits times $T$ depth~\cite{litinski2019game}. Formalizing these architectural constraints allows us to optimize algorithms and fault-tolerant layouts together before large-scale quantum processors exist. Over time, collaborative proof search can thus turn what were once isolated, multi-year derivations into cumulative resource reductions, closing the gap to useful quantum computation from above while physical hardware scales from below.

\section*{Code availability}
Our Lean library, together with the problem specifications and verified solutions discussed in this manuscript, is available at \url{https://github.com/google-deepmind/formal-quantum-computation}~\cite{googledeepmind2026fqc}.

\printbibliography[segment=\therefsegment, title={References}]
\end{refsegment}

\clearpage
\appendix
\setcounter{figure}{0}
\renewcommand{\thefigure}{S\arabic{figure}}
\renewcommand{\theHfigure}{S\arabic{figure}}
\setcounter{table}{0}
\renewcommand{\thetable}{S\arabic{table}}
\renewcommand{\theHtable}{S\arabic{table}}

\begin{refsegment}

\section{Formal Quantum Computation Library}
\label{app:qhl}
We formalize our verification and resource-counting framework for quantum algorithms in Lean~\cite{demoura2021lean4}, an interactive theorem proving language based on dependent type theory.
We chose Lean for its already established rich library ecosystem for broad mathematics, physics, and quantum information theory ~\cite{mathlib2020, physlib, meiburg2024quantuminfo}.
Moreover, Lean has seen significant traction in recent months for verification of AI-generated proofs. In particular, formalizing quantum algorithms in Lean allows us to model a broader semantics of quantum programming, complementing fast fuzz-testing evaluators like kickmix~\cite{babbush2026kickmix}. Prior formal frameworks for quantum program verification have been developed across several interactive theorem provers. In Rocq (formerly Coq), SQIR~\cite{hietala2021sqir} verifies unitary quantum circuits, while CoqQ~\cite{zhou2023coqq} formalizes quantum Hoare logic~\cite{ying2011floyd} for pure quantum \texttt{while} programs. In Isabelle/HOL, QHLProver~\cite{liu2019qhlprover} provides an earlier mechanization of quantum Hoare logic for imperative quantum algorithms. Concurrent work has also explored quantum verification in Lean~\cite{zhang2026leanquantumalg}, most notably QEDIA~\cite{ren2026qedia}, which formalizes classical-quantum Hoare logic~\cite{practicalQHLpart1} for partial and total correctness. Ref.~\citenum{endo2026trust} also separates human-reviewed specifications from machine-checked proofs to reduce the human ``trust cost'' of AI-generated quantum circuits, a division our workflow mirrors. Beyond correctness verification, our framework includes formal support for measuring the worst-case and expected resource costs of algorithms, which is critical for quantum algorithm discovery.

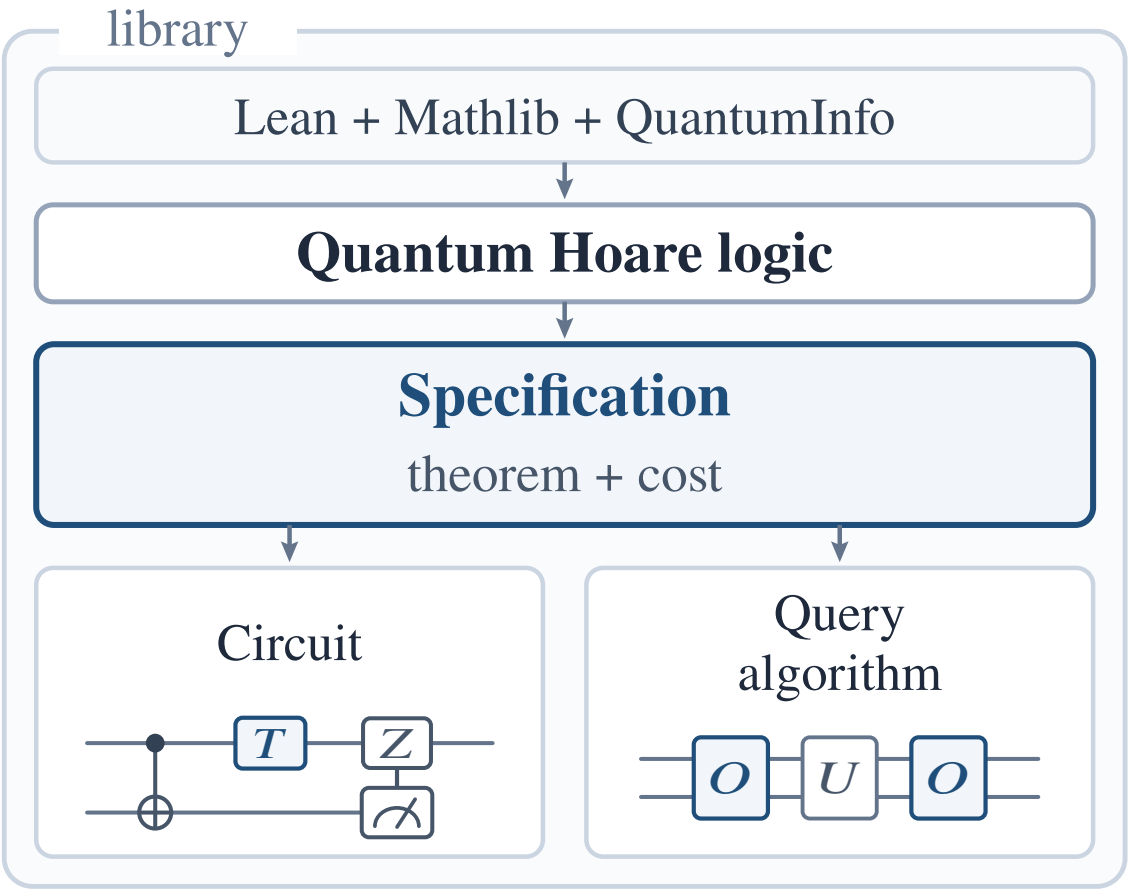
\begin{figure}[t]
\providecolor{frameblue}{HTML}{1E4E79}%
\providecolor{specbg}{HTML}{F2F6FA}%
\providecolor{libbg}{HTML}{FAFBFC}%
\providecolor{basebg}{HTML}{F8FAFC}%
\providecolor{inkbody}{HTML}{1E293B}%
\providecolor{inklabel}{HTML}{334155}%
\providecolor{inkbond}{HTML}{475569}%
\providecolor{inksub}{HTML}{64748B}%
\providecolor{inkmuted}{HTML}{94A3B8}%
\providecolor{gridline}{HTML}{CBD5E1}%
\def\fqMid{\fontsize{7.2}{8.4}\selectfont}%
\def\fqSpec{\fontsize{7.6}{8.8}\selectfont}%
\def\fqSub{\fontsize{6.6}{7.7}\selectfont}%
\def\fqBox{\fontsize{6.5}{7.6}\selectfont}%
\def\fqGateQ{\fontsize{5.8}{6.6}\selectfont}%
\def\fqGateC{\fontsize{5.5}{6.3}\selectfont}%
\centering
\resizebox{0.50\textwidth}{!}{%
\begin{tikzpicture}[x=1cm, y=1cm,
  fqArr/.style={-{Stealth[length=2.55pt, width=2.05pt, inset=0.38pt]}, line width=0.56pt}]
  \path[use as bounding box] (0,0) rectangle (5.2000,4.1000);
  \fill[white] (0,0) rectangle (5.2000,4.1000);

  \filldraw[fill=libbg, draw=gridline, line width=0.60pt, rounded corners=3.5pt] (0.1530,0.1300) rectangle (5.0483,3.8641);
  \fill[white] (0.3925,3.7630) rectangle (1.4293,3.9747);
  \node[anchor=base, inner sep=0pt, font=\fqSub, text=inksub] at (0.9109,3.8015) {library};

  \filldraw[fill=basebg, draw=gridline, line width=0.55pt, rounded corners=2.1pt] (0.2928,3.2915) rectangle (4.9086,3.7005);
  \node[anchor=base, inner sep=0pt, font=\fqBox, text=inklabel] at (2.6007,3.4146) {Lean + Mathlib + QuantumInfo};
  \draw[fqArr, inksub] (2.6007,3.2915) -- (2.6007,3.1294);

  \filldraw[fill=white, draw=inkmuted, line width=0.62pt, rounded corners=2.1pt] (0.2928,2.6827) rectangle (4.9086,3.1061);
  \node[anchor=base, inner sep=0pt, font=\fqMid, text=inkbody] at (2.6007,2.8145) {\textbf{Quantum Hoare logic}};
  \draw[fqArr, inksub] (2.6007,2.6827) -- (2.6007,2.5206);

  \filldraw[fill=specbg, draw=frameblue, line width=0.75pt, rounded corners=2.2pt] (0.2928,1.7083) rectangle (4.9086,2.4975);
  \node[anchor=base, inner sep=0pt, font=\fqSpec, text=frameblue] at (2.6007,2.1870) {\textbf{Specification}};
  \node[anchor=base, inner sep=0pt, font=\fqSub, text=inkbond] at (2.6007,1.8560) {theorem + cost};
  \draw[fqArr, inksub] (1.3992,1.7083) -- (1.3992,1.5447);
  \draw[fqArr, inksub] (3.8021,1.7083) -- (3.8021,1.5447);

  \filldraw[fill=white, draw=gridline, line width=0.56pt, rounded corners=2.1pt] (0.2928,0.2600) rectangle (2.5057,1.5207);
  \node[anchor=base, inner sep=0pt, font=\fqBox, text=inkbody] at (1.3992,1.1182) {Circuit};
  \draw[inksub, line width=0.50pt, line cap=round] (0.5149,0.7556) -- (2.2861,0.7556);
  \draw[inksub, line width=0.50pt, line cap=round] (0.5149,0.4524) -- (1.7204,0.4524);
  \filldraw[fill=white, draw=inkbond, line width=0.50pt] (0.8125,0.4524) circle [radius=0.0700];
  \draw[inkbond, line width=0.50pt] (0.8125,0.7556) -- (0.8125,0.3824);
  \draw[inkbond, line width=0.50pt] (0.7425,0.4524) -- (0.8825,0.4524);
  \fill[inklabel] (0.8125,0.7556) circle [radius=0.0410];
  \filldraw[fill=specbg, draw=frameblue, line width=0.52pt, rounded corners=0.9pt] (1.1629,0.6445) rectangle (1.4682,0.8664);
  \node[anchor=center, inner sep=0pt, font=\fqGateC, text=frameblue] at (1.3155,0.7556) {$T$};
  \draw[inkbond, line width=0.52pt] (1.8685,0.5602) -- (1.8685,0.6470);
  \filldraw[fill=white, draw=inkbond, line width=0.50pt, rounded corners=0.9pt] (1.7202,0.6470) rectangle (2.0168,0.8640);
  \node[anchor=center, inner sep=0pt, font=\fqGateC, text=inkbond] at (1.8685,0.7556) {$Z$};
  \filldraw[fill=white, draw=inkbond, line width=0.50pt, rounded corners=0.9pt] (1.7144,0.3444) rectangle (2.0226,0.5602);
  \draw[inkbond, line width=0.46pt, line cap=round] (1.7720,0.3928) arc[start angle=180, end angle=0, x radius=0.0960, y radius=0.0800];
  \draw[inkbond, line width=0.46pt, line cap=round] (1.8685,0.3890) -- (1.9477,0.5106);

  \filldraw[fill=white, draw=gridline, line width=0.56pt, rounded corners=2.1pt] (2.6956,0.2600) rectangle (4.9086,1.5207);
  \node[anchor=base, inner sep=0pt, font=\fqBox, text=inkbody] at (3.8021,1.2484) {Query};
  \node[anchor=base, inner sep=0pt, font=\fqBox, text=inkbody] at (3.8021,0.9880) {algorithm};
  \draw[inksub, line width=0.50pt, line cap=round] (2.9341,0.6866) -- (4.6693,0.6866);
  \draw[inksub, line width=0.50pt, line cap=round] (2.9341,0.5205) -- (4.6693,0.5205);
  \filldraw[fill=specbg, draw=frameblue, line width=0.52pt, rounded corners=0.9pt] (3.1645,0.4262) rectangle (3.4885,0.7809);
  \node[anchor=center, inner sep=0pt, font=\fqGateQ, text=frameblue] at (3.3265,0.6036) {$O$};
  \filldraw[fill=white, draw=inksub, line width=0.48pt, rounded corners=0.9pt] (3.6397,0.4262) rectangle (3.9637,0.7809);
  \node[anchor=center, inner sep=0pt, font=\fqGateQ, text=inkbond] at (3.8017,0.6036) {$U$};
  \filldraw[fill=specbg, draw=frameblue, line width=0.52pt, rounded corners=0.9pt] (4.1149,0.4262) rectangle (4.4389,0.7809);
  \node[anchor=center, inner sep=0pt, font=\fqGateQ, text=frameblue] at (4.2769,0.6036) {$O$};
\end{tikzpicture}%
}%
\caption{\textbf{Verification library and specification stack.} Built on \texttt{Mathlib}~\cite{mathlib2020}, \texttt{Physlib}'s \texttt{Lean-QuantumInfo}~\cite{meiburg2024quantuminfo, physlib} in Lean, our quantum Hoare logic~\cite{feng2021qhl,practicalQHLpart1} formalizes classical--quantum states, weakest preconditions and resource costs, supporting problem specifications (correctness theorem and cost bound) across dynamic Clifford+$T$ circuits and black-box query algorithms.}
\label{fig:app_qhl}
\end{figure}

The core of our Lean library is formed by a formalization of a classical-quantum imperative \texttt{while} language with an associated classical-quantum Hoare logic~\cite{feng2021qhl}. Supporting classical variables and classical control flow is crucial in practice: realistic quantum algorithms routinely interleave quantum operations with classical arithmetic, loops, and measurement-dependent branching. Handling classical states natively avoids the state-space overhead of encoding classical data into quantum registers. 

Correctness in this work is expressed through total Hoare triples $\{\!\!\{\Theta\}\!\!\}\;P\;\{\!\!\{\Psi\}\!\!\}$.
These assert that for every initial (classical-quantum) state $\Delta$ satisfying the \emph{pre}condition $\Theta$, all executions of the program $P$ terminate in a (classical-quantum) state satisfying the \emph{post}condition $\Psi$.
In the classical case, pre-- and postconditions are predicates on the state.
In the quantum case, a state no longer simply satisfies or violates an assertion, the condition is quantitative and takes value in $[0,1]$. We call this the \emph{expectation} of satisfaction.
A triple then asserts that, for every initial state $\Delta$, the expectation of $\Delta$ satisfying an assertion $\Theta$ is bounded above by the expectation of the resulting state $\Delta'$ (obtained after running the program $P$ on $\Delta$) to satisfy $\Psi$.

Verification is performed via reasoning about \emph{weakest preconditions}. Starting from a given postcondition $\Psi$, we compute the weakest precondition for this to hold by inductively reasoning about the program backwards.
The full semantics of weakest preconditions is given in Figure \ref{fig:semantics}. We follow the source closely but not fully faithfully: we have made certain modifications both to fit our verification purposes and to address ambiguities in the paper. We next outline the main ingredients of the logic and where our formalization differs from Ref.~\citenum{feng2021qhl}.

\paragraph{Variable types.} Programs operate on two kinds of variables: classical and quantum. In Ref.~\citenum{feng2021qhl} classical variables can have two types, boolean and integer.
We focus on integer-valued classical variables.
This is without loss of generality, since boolean values can be encoded as integers.
More precisely, given a type $\mathsf{Var}$ of classical variables, the type of classical states (stores) is:
\begin{lstlisting}
abbrev Store (Var : Type) : Type := Var → ℤ
\end{lstlisting}

For quantum variables, Feng and Ying~\cite{feng2021qhl} consider a countably infinite set $\mathsf{qVar}$ of quantum variables $q_1, q_2, \ldots$ each associated with a Hilbert space of dimension $d_{q_{1}}, d_{q_{2}}, \ldots$ respectively.
We restrict our set-up to a single quantum variable $q$ with associated Hilbert space $\mathcal{H}(d)$ of dimension $d$, where $d$ is a positive natural number.
In practice, $\mathcal{H}(d)$ can be identified with $\mathbb{C}^d$.
This loses no expressiveness if we limit ourselves to finitely many quantum variables: if these are of dimensions $d_1, \dots, d_n$ can be combined into one register of dimension $d = \prod_{i=1}^{n} d_i$.
Although, this requires lifting operations to the full space explicitly.
Supporting multiple named quantum variables is left for future work.
    
\paragraph{Classical-quantum states.} Let $\mathcal{L}(d)$ be the set of Hermitian operators (matrices) on $\mathcal{H}(d)$ and $\mathcal{D}(d) = \{\rho \in \mathcal{L}(d) \mid 0 \preceq \rho,\; \mathrm{Tr}(\rho) \le 1\}$ the set of partial density operators, ordered by the Löwner order.
A \emph{classical-quantum state} $\Delta$ is a function $\Delta: \texttt{Store Var} \to \mathcal{D}(d)$ with countable support whose total trace
$\sum_{\sigma} \mathrm{tr}(\Delta(\sigma))$ is at most $1$.
In particular, any deficit $1 - \|\Delta\|$ represents nontermination.
In other words, for a classical store $\sigma$, $\mathrm{tr}(\Delta(\sigma))$ is the probability of being in the classical branch $\sigma$ and $\Delta(\sigma)$, suitably normalized, is the quantum state in that branch. We represent this slightly more generally as:
\begin{lstlisting}
def CQState (State : Type*) (d : ℕ) : Type :=
  {Δ : State → mcalD d // ∃ c, HasSum (fun σ ↦ (Δ σ).trace) c ∧ c ≤ 1}
\end{lstlisting}
and our formalization shows that \texttt{CQState (Store Var) d} is equivalent to Definition~3.1 of Ref.~\citenum{feng2021qhl}; countable support need not be separately imposed, since a suitable family of non-negative reals has countable support (covered by \texttt{HasSum}).
The formalization and proofs of this can be found in \texttt{CQState.lean}~\cite{googledeepmind2026fqc}. 

\paragraph{Classical-quantum assertions.} We define the set of unit-interval observables~\cite{feng2021qhl} (quantum predicates) as $\mathcal{O}(d) = \{M \in \mathcal{L}(d) \mid 0 \preceq M \preceq I\}$, where $0$ and $I$ are the zero and identity matrix respectively.
A \emph{classical-quantum assertion} is a map $\Theta: \texttt{Store Var} \to \mathcal{O}(d)$.
This approach is inspired by classical probabilistic program logics, mirroring the use of by real-valued expectations there~\cite{morgan1996probabilistic}. We formalize this (again slightly more generally) as follows, where \texttt{Observable} $d$ is our formalization of $\mathcal{O} (d)$:
\begin{lstlisting}
abbrev CQAssertion (State : Type*) (d : ℕ) : Type :=
  State → Observable d
\end{lstlisting}
As an example, Figure~\ref{fig:clean-ancillas-property} shows how \texttt{CQAssertion} is used to express the clean-ancilla property for subroutines that initialize workspace qubits in $|0\rangle$ and uncompute them upon completion.

\begin{figure}[t]
\centering
\begin{minipage}{0.6\textwidth}
\begin{tcolorbox}[colback=panelbg,colframe=frameblue!25,boxrule=0.5pt,arc=2pt,left=5pt,right=5pt,top=3pt,bottom=3pt]
\begin{lstlisting}[basicstyle=\ttfamily\scriptsize]
def RealizesCleanSubroutine 
    {Var : Type} [DecidableEq Var] {n k m : ℕ}
    (rTarget : QRegister k n) (rAnc : QRegister m n)
    (_hdisj : rTarget.DisjointWires rAnc)
    (U : Unitary (2 ^ k))
    (P : QProg Var (2 ^ n)) : Prop :=
  ∀ (M : Observable (2 ^ n)),
    let post : CQAssertion (Store Var) (2 ^ n) :=
      .restrictToZeroAncilla (fun _ ↦ M) rAnc
    ⦃ post.qAssign (rTarget.embedUnitary U)
        |>.restrictToZeroAncilla rAnc ⦄
    P
    ⦃ post ⦄
\end{lstlisting}
\end{tcolorbox}
\end{minipage}
\caption{A formal definition of the clean-ancilla property. Here, a program \texttt{P} implements the unitary $U$ on register \texttt{rTarget} and returns the disjoint ancilla register \texttt{rAnc} to $|0\cdots0\rangle$: for every predicate $M$, from any input with clean ancillas, \texttt{P} must end with clean ancillas and satisfy $M$ at least as often as applying $U$ would. Since $M$ is arbitrary, \texttt{P} acts exactly as $U$ on all such inputs.}
\label{fig:clean-ancillas-property}
\end{figure}
Similarly to \texttt{CQState}, \texttt{CQAssertion} has a partial order: we can lift the Löwner order on observables by declaring $\Theta_1 \preceq \Theta_2$, if $\Theta_1 (\sigma) \preceq \Theta_2(\sigma)$ for all classical stores $\sigma$.
Note that since we work with a single quantum variable, in contrast to Ref. ~\citenum{feng2021qhl}, we need not impose further conditions on the quantum variables used when defining the order. 
\texttt{CQAssertion} is a pointed $\omega$-CPO with respect to the Löwner order: it has a least element, and every increasing chain $\Theta_0 \preceq \Theta_1 \preceq \cdots$ has a supremum.
This structure is crucial as it allows us to define the weakest precondition on a \texttt{while} loop as the supremum of the weakest preconditions of its finite unrollings and to prove the soundness of the \texttt{while} loop rule.
In most logics, assertions form a complete lattice, a stronger structure in which every set of assertions has a supremum and an infimum.
This matters in practice: existing verification generator frameworks such as Lean's \texttt{vcgen} tactic~\cite{lean2026vcgen} and Loom~\cite{gladshtein2026loom, loomrepo} require this.
The observables on a Hilbert space of dimension at least $2$ \cite{gudder1996counterexamples}, do not even form a lattice under the Löwner order, so these frameworks are not presently applicable.
Therefore, we implemented the quantum Hoare logic and weakest precondition calculus from scratch.
    
At first glance, the formalization of \texttt{CQAssertion (Store Var)} above differs from Definition~3.5 of Ref.~\citenum{feng2021qhl}, which includes the two conditions on $\Theta : \texttt{Store Var} \to \mathcal{O}(d)$:
\begin{enumerate}
    \item The image set $\Theta(\texttt{Store Var})$ is countable;
    \item For each $M \in\Theta(\texttt{Store Var})$, the preimage $\Theta^{-1}(M)$ is definable by a classical assertion.
\end{enumerate}
We do not need to impose condition (1) separately: when the set of classical variables \texttt{Var} is finite,
$\texttt{Store Var} = \texttt{Var} \to \mathbb{Z}$ is countable, so the image of
$\Theta : \texttt{Store Var} \to \mathcal{O}(d)$ is automatically countable.
We prove this in \texttt{CQAssertion.equivCountable}.
Condition (2) holds automatically since we represent classical assertions as predicates $\texttt{Store}\ \texttt{Var} \to \texttt{Prop}$.
However, if the classical assertions are first-order logic formulas over \texttt{Var} as is stated in the beginning of Section~3.2 of Ref.~\citenum{feng2021qhl}, cq-assertions are no longer closed under suprema of increasing chains in the Löwner order and therefore, not an $\omega$-CPO.
We encounter the same roadblock to the $\omega$-CPO structure if we explicitly impose condition (1) on the cq-assertions while allowing countably infinitely many classical variables, as opposed to obtaining it naturally by restricting to finitely many classical variables. The full formalization can be found in \texttt{CQAssertion.lean}~\cite{googledeepmind2026fqc}.

\paragraph{Expectation.} To reason about program correctness, we need a quantifiable notion of \emph{how much} a cq-state satisfies a cq-assertion. In the quantum case, this is no longer a boolean predicate; instead, we consider the expectation of a cq-state $\Delta$ satisfying a cq-assertion $\Theta$:
\begin{equation}
  \mathbb{E}[\Delta \models \Theta] \;\overset{\triangle}{=}\; \sum_{\sigma} \mathrm{tr}\!\left(\Theta(\sigma)\,\Delta(\sigma)\right) \;\in\; [0, 1].
  \label{eq:expectationdef}
\end{equation}
This can be thought of as the expected satisfaction of $\Theta$ in state $\Delta$, averaging over all possible classical branches $\sigma$. Note that since we have one quantum variable, our formula is a simplified version of Definition ~3.7 in Ref.~\citenum{feng2021qhl}.

We formalize Equation ~\ref{eq:expectationdef} as follows:
\begin{lstlisting}
def expect (Δ : CQState State d) (θ : CQAssertion State d) : ℝ :=
  ∑' σ, ⟪(θ σ).val, Δ σ⟫_ℝ
notation "bbE[" Δ " ⊨ " θ "]" => expect Δ θ
\end{lstlisting}
This definition and Equation~\ref{eq:expectationdef} are equivalent: the inner product of the Hermitian matrices $\Theta(\sigma)$ and $\Delta(\sigma)$ is the real value of the trace of their product as defined in Equation~\ref{eq:expectationdef}.
The prime around the sum stands for Mathlib's \texttt{tsum}, or topological sum, the limit of finite partial sums of possibly infinite index type.
Finally, $\mathtt{expect}\ \Delta\ \Theta$ takes values in the unit-interval, even though its return type is the real numbers.
This follows from the trace of positive semi-definite matrices being non-negative combined with lemmas we prove in our library establishing the term inside the sum is bounded above by $1$ (see theorems \texttt{expect\_mono}, \texttt{expect\_top}, \texttt{CQState.trace\_le\_one} in \texttt{Expectation.lean}~\cite{googledeepmind2026fqc}).

\paragraph{Syntax.} Programs are expressed in the following variant of the classical-quantum imperative \texttt{while}-language of Feng and Ying~\cite{feng2021qhl}:
\begin{equation}
\begin{aligned}
  P \;\Coloneqq\;& \texttt{skip} \;\mid\; \texttt{abort} \;\mid\; x \coloneqq e \;\mid\; \texttt{qset} \;\mid\; \texttt{apply}\;U \\
  \mid\;& x \coloneqq \texttt{measure}\;\mathcal{M} \;\mid\; P_1 \;;\; P_2 \\
  \mid\;& \texttt{qif}\;b\;\texttt{then}\;P_1\;\texttt{else}\;P_2 \;\mid\; \texttt{qwhile}\;b\;\texttt{loop}\;P 
\end{aligned}
\label{eq:dsl}
\end{equation}

Our formalization diverges from Ref.~\citenum{feng2021qhl} in the following ways: we omit random assignment $x \mathrel{:=_{\$}} g$, we work with a single quantum system rather than several quantum variables, and we represent expressions $e$ and guards $b$ directly as Lean functions on the classical store. We embed this language in Lean as an inductive datatype \texttt{QProg}, with the  commands as constructors.

A \texttt{QProg} program is parameterized by a finite type \texttt{Var} of integer-valued classical variables $x$ and on a single $d$-dimensional quantum system ($d$ an arbitrary positive natural number, with $d = 2^n$ for $n$ qubits).
Classical expressions $e : \texttt{Store}\ \texttt{Var} \to \mathbb{Z}$ and decidable guards $b : \texttt{Store}\ \texttt{Var} \to \texttt{Prop}$ are shallowly embedded as functions of the classical store $\sigma$, avoiding the need to encode boolean guards as nonzero integers and aligning branch conditions directly with Lean's propositional tactics.
The command $\texttt{qset}$ resets the quantum system to $|0\rangle$, $\texttt{apply}\;U$ applies a unitary operator $U$, and $x \coloneqq \texttt{measure}\;\mathcal{M}$ performs an $i$-outcome quantum measurement with Kraus operators $\mathcal{M} = \{M_o\}_{o < i}$, $\sum_o M_o^\dagger M_o \preceq I$, storing the outcome $o$ in $x$.
The commands $\texttt{qif}$ and $\texttt{qwhile}$ respectively branch and loop on classical guards.
As an example, simple applications of $\texttt{qwhile}$ provide support for repeat-until-success loops.
Denotational semantics for \texttt{QProg} are given in Figure~\ref{fig:semantics}.  

\paragraph{Weakest Preconditions.}
Program verification is performed backwards using the weakest precondition (wp) transformer (Figure~\ref{fig:semantics}), which takes a program and a target postcondition and computes the weakest assertion that must hold before execution:
\begin{lstlisting}
def wp : QProg Var d → CQAssertion (Store Var) d → CQAssertion (Store Var) d
\end{lstlisting}
This formalizes the predicate transformer of Table~3 in Ref.~\citenum{feng2021qhl}. Since we are interested in terminating classical-quantum programs, we focus our attention on total correctness. Therefore, we formalize the weakest precondition calculus from Ref. \citenum{feng2021qhl} and omit the weakest liberal preconditions, which are only needed for partial correctness. 

Particularly, given a classical-quantum program $P$ and a postcondition $\Psi$, $\texttt{wp}\ P\ \Psi$ computes the ``weakest'' cq-assertion whose expectation in any initial state $\Delta$ equals the expectation of $\Psi$ in the state reached after executing $P$. Since cq-assertions form an $\omega$-CPO under the Löwner order, the \emph{weakest} precondition is the largest element with respect to the Löwner order for which the desired postcondition still holds.

The wp rules on basic commands are described in Figure~\ref{fig:semantics}. They are all faithfully formalized to the source, apart from the syntax and variable-type choices in \texttt{QProg}, and can be found in \texttt{WP.lean}~\cite{googledeepmind2026fqc}.

We use the weakest precondition calculus to prove total-correctness Hoare triples $\{\!\!\{\Theta\}\!\!\}\;P\;\{\!\!\{\Psi\}\!\!\}$, formalized as propositions:
\begin{lstlisting}
def Triple (Θ : CQAssertion (Store Var) d) (c : QProg Var d)
    (Ψ : CQAssertion (Store Var) d) : Prop :=
  ∀ Δ : CQState (Store Var) d, bbE[Δ ⊨ Θ] ≤ bbE[(interp c).toStateFun Δ ⊨ Ψ]
\end{lstlisting}
Here $(\texttt{interp}\ c).\texttt{toStateFun}\ \Delta$ formalizes the output state $\llbracket c \rrbracket(\Delta)$ of Figure~\ref{fig:semantics}.
The central property connecting the forward denotational semantics and the backward wp transformer is exact expectation duality (Lemma~4.14 in Ref.~\citenum{feng2021qhl}):
\begin{lstlisting}
theorem expect_wp (c : QProg Var d) (Δ : CQState (Store Var) d)
    (Ψ : CQAssertion (Store Var) d) :
    bbE[Δ ⊨ wp c Ψ] = bbE[(interp c).toStateFun Δ ⊨ Ψ]
\end{lstlisting}
Exact expectation duality (\texttt{expect\_wp}) immediately implies \emph{soundness}: that $\texttt{wp}\ c\ \Psi$ is itself a valid precondition, and that it is indeed the weakest such precondition. These are formalized in \texttt{wp\_soundness} and \texttt{wp\_weakest} in \texttt{WP.lean}. 

\paragraph{Resource Counting and Cost Models.}
Alongside correctness, our library supports two ways of calculating the resource cost of a program: syntactic gate counting over structured circuits, and semantic expected-cost accounting over program executions with branching and unbounded loops. At the circuit level, straight-line circuits are represented as the inductive type \texttt{Circuit} parametrized by the number of wires $n$. This is built sequentially from nine primitive unitary gates \texttt{PrimitiveGate} (Clifford+$T$, SWAP and TOFF gates). The count of each gate type is obtained simply by summing its occurrences along the sequence. Because each gate type is tracked separately, the cost assigned to each gate is tunable in a specification (for instance, counting only $T$ gates while treating Cliffords as free). Circuits can also include mid-circuit single-qubit measurements and classical branching, packaged in the type \texttt{DynamicCircuit} which appears in the problems described in this work. For these, we set the resource counting to have measurements add $0$ to gate counts, while the cost of a classical branch is the maximum of its two branch counts, giving a worst-case bound. Through the typing of the output circuit, we also keep track of the number of qubits used to include in resource counts. All of this can be found in the folder \texttt{Circuits}, with basic definitions in \texttt{Basic.lean}, and the resource counting framework in \texttt{ResourceCounting.lean}.

To calculate exact expected resource usage directly on \texttt{QProg}, a user-supplied cost model (\texttt{CostModel}) assigns nonnegative integer weights $w$ to each primitive command (\texttt{assignmentCost}, \texttt{resetCost}, \texttt{measurementCost}, and \texttt{unitaryCost}). Unitary and measurement weights may depend on the operator applied. For instance, one can charge $1$ for a designated oracle unitary and $0$ for all other operations.

Given $w$ and an initial cq-state $\Delta$ with total trace $\|\Delta\| = \sum_\sigma \mathrm{tr}(\Delta(\sigma))$, the branch-weighted expected cost $\mathrm{cost}_w(P, \Delta)$ (\texttt{expectedCost}) is defined by structural recursion
on $P$ alongside the denotational semantics $\llbracket P \rrbracket(\Delta)$:
\begin{equation}
\begin{aligned}
  \mathrm{cost}_w(\texttt{skip}, \Delta) &= \mathrm{cost}_w(\texttt{abort}, \Delta) = 0, \qquad
  \mathrm{cost}_w(c, \Delta) = \|\Delta\| \cdot w(c), \\
  \mathrm{cost}_w(P_1 \;;\; P_2, \Delta) &= \mathrm{cost}_w(P_1, \Delta) + \mathrm{cost}_w(P_2, \llbracket P_1 \rrbracket(\Delta)), \\
  \mathrm{cost}_w(\texttt{qif}\;b\;\texttt{then}\;P_1\;\texttt{else}\;P_2, \Delta) &= \mathrm{cost}_w(P_1, \Delta|_b) + \mathrm{cost}_w(P_2, \Delta|_{\neg b}), \\
  \mathrm{cost}_w(\texttt{qwhile}\;b\;\texttt{loop}\;P, \Delta) &= \bigsqcup_{n \ge 0} C_n(\Delta), \quad
  C_0(\Delta) = 0,\; C_{n+1}(\Delta) = \mathrm{cost}_w(P, \Delta|_b) + C_n(\llbracket P \rrbracket(\Delta|_b)),
\end{aligned}
\label{eq:expectedCost}
\end{equation}
for primitive commands $c$, taking values in $[0, \infty]$ so that loop suprema always exist. The total trace, which could also be seen as the probability mass of entering the current branch of the program, is formalized as \texttt{stateMass}. Our library provides compositional rules to upper-bound $\mathrm{cost}_w(P, \Delta)$ across sequences, branches, and loops, bounding a \texttt{qwhile} loop by its per-iteration cost times the expected number of iterations, which can itself be bounded via \texttt{wp}. These definitions and compositional rules on the costs (formalized as \texttt{CostBound}) are mechanized in \texttt{QuantumHoareLogic/ResourceCounting}.

Each problem specification specializes this framework along two axes. First, the counted resource is chosen to match the computational model, such as gate counts \& spacetime, or black-box oracle calls (charging weight $1$ per oracle application and $0$ for all other operations, as in parallel Grover search; Appendix~\ref{app:grover}). Second, costs can be measured either as concrete numerical bounds at a fixed problem size (like the \texttt{arcsin} problem) or as asymptotic bounds. In many cases, these asymptotic bounds are enforced to be of the form $c \cdot N^a$ as the input size $N$ grows, ordered lexicographically on $(a, c)$, which gives a straightforward ordering on discovered solutions.

\begin{figure}[t]
\centering
{\small
\renewcommand{\arraystretch}{1.32}
\begin{tabular*}{\textwidth}{@{\extracolsep{\fill}}lll@{}}
\toprule
Program $P$ & Denotational semantics $\llbracket P \rrbracket(\Delta)$ & Weakest precondition $\mathrm{wp}(P, \Psi)$ \\
\midrule
$\texttt{skip}$ & $\Delta$ & $\Psi$ \\
$\texttt{abort}$ & $0$ & $0$ \\
$x \coloneqq e$ & $\sigma' \mapsto \sum_{\sigma:\,\sigma[x\mapsto e(\sigma)]=\sigma'} \Delta(\sigma)$ & $\sigma \mapsto \Psi(\sigma[x \mapsto e(\sigma)])$ \\
$\texttt{qset}$ & $\sigma \mapsto \mathrm{Tr}(\Delta(\sigma))\,|0\rangle\langle 0|$ & $\sigma \mapsto \langle 0|\Psi(\sigma)|0\rangle\, I$ \\
$\texttt{apply}\;U$ & $\sigma \mapsto U\,\Delta(\sigma)\,U^\dagger$ & $\sigma \mapsto U^\dagger\,\Psi(\sigma)\,U$ \\
$x \coloneqq \texttt{measure}\;\mathcal{M}$ & $\sigma' \mapsto \sum_{o}M_o\,\llbracket x \coloneqq o \rrbracket(\Delta)\,M_o^\dagger$ & $\sigma \mapsto \sum_{o} M_o^\dagger\,\Psi(\sigma[x \mapsto o])\,M_o$ \\
$P_1 \;;\; P_2$ & $\llbracket P_2 \rrbracket(\llbracket P_1 \rrbracket(\Delta))$ & $\mathrm{wp}(P_1, \mathrm{wp}(P_2, \Psi))$ \\
$\texttt{qif}\;b\;\texttt{then}\;P_1\;\texttt{else}\;P_2$ & $\llbracket P_1 \rrbracket(\Delta|_b) + \llbracket P_2 \rrbracket(\Delta|_{\neg b})$ & $\mathrm{wp}(P_1, \Psi)|_b + \mathrm{wp}(P_2, \Psi)|_{\neg b}$ \\
$\texttt{qwhile}\;b\;\texttt{loop}\;P$ & $\bigsqcup_{n \ge 0} W_n(\Delta)$ & $\bigsqcup_{n \ge 0} \Psi_n$ \\
\bottomrule
\end{tabular*}}
\vspace{2pt}
{\footnotesize
\[
  \mathrm{where} \hspace{0.16cm} W_{n+1}(\Delta) = \Delta|_{\neg b} + W_n(\llbracket P \rrbracket(\Delta|_b)), \qquad
  \Psi_{n+1} = \mathrm{wp}(P, \Psi_n)|_b + \Psi|_{\neg b}.
\]
}
\vspace{-8pt}
\caption{Inductive definitions of the denotational semantics $\llbracket P \rrbracket(\Delta)$ and weakest precondition $\mathrm{wp}(P, \Psi)$. For a guard $b$, $\Delta|_b$ retains $\Delta(\sigma)$ where $b(\sigma) \neq 0$ and sets all other branches to $0$ (similarly for $\Psi|_b$ and $\neg b$). Loop unrollings start from $W_0(\Delta) = 0$ and $\Psi_0 = 0$.}
\label{fig:semantics}
\end{figure}

\section{Verification Pipeline}
\label{app:verifier}
Each problem is defined by a specification file pinned to a frozen repository commit, declaring the target correctness theorem and resource-cost function with typed \texttt{sorry} placeholders for the candidate construction, workspace parameters, and proofs (Sec.~\ref{sec:qhl}). Because agents may game local checks without discovering or improving the algorithm, for instance by weakening theorem statements, modifying helper definitions, adding axioms, disabling local kernel checks, or misreporting costs~\cite{baker2025monitoring,zhong2025impossiblebench}, candidate grading is isolated from the agent environment.

Every submission is replayed in a clean sandbox by an external verifier using the Lean Comparator~\cite{leanfro2025comparator} against the frozen specification. A submission passes only if: (i) all target theorem statements and every transitive definition they depend on match the specification; (ii) the compiled proof terms depend solely on standard Lean axioms (\texttt{propext}, \texttt{Quot.sound}, and \texttt{Classical.choice}), with no \texttt{sorry} or custom axioms; and (iii) the proved resource theorem reduces to the claimed closed-form cost. The same check runs automatically before any solution is merged into the shared wiki. Numerical estimates and agent-reported figures are used only internally by teams to prioritize search directions, whereas researchers need only audit the physical model and cost conventions of the specification itself, revising the pinned specification if an unintended modeling loophole is found.

\section{Multi-Agent Orchestration}
\label{app:swarm}
\paragraph{Agent teams.} Each researcher configures and operates an independent team of LLM agents, choosing the agent framework, models, prompts, and division of labor. Setups ranged from a single agent iterating on a candidate construction to multi-agent hierarchical organizations that split subroutine design, error analysis, and Lean proof engineering across specialized workers. In this study, researchers put their domain knowledge, discussion, and design into the problem specification itself rather than seeding domain knowledge or steering the research while underway, though the workflow naturally supports active human seeding and steering as well. During exploration, agents may invoke any external tools, including natural-language derivations, classical simulation, numerical optimization, and remote compute clusters.

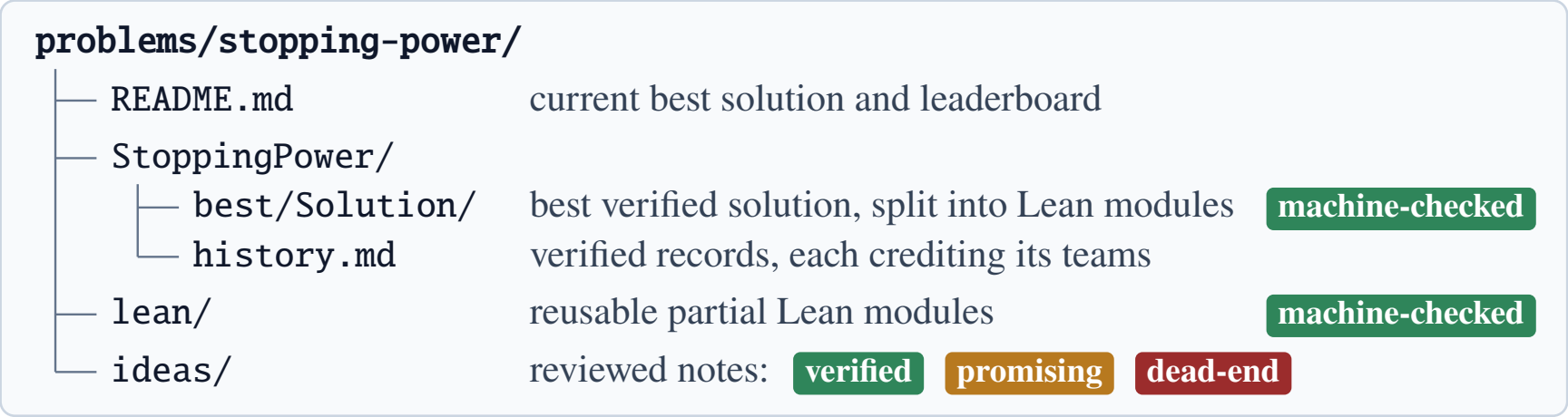
\begin{figure}[!htbp]
\centering
\providecolor{inkdark}{HTML}{0F172A}%
\providecolor{inklabel}{HTML}{334155}%
\providecolor{inksub}{HTML}{64748B}%
\providecolor{gridline}{HTML}{CBD5E1}%
\providecolor{basebg}{HTML}{F8FAFC}%
\providecolor{accentgreen}{HTML}{2F855A}%
\providecolor{accentamber}{HTML}{B7791F}%
\providecolor{accentred}{HTML}{9B2C2C}%
\begin{tikzpicture}[
    x=1cm, y=1cm,
    fname/.style={anchor=base west, inner sep=0pt, font=\ttfamily\footnotesize, text=inkdark},
    fdesc/.style={anchor=base west, inner sep=0pt, font=\footnotesize, text=inklabel},
    tree/.style={draw=inksub, line width=0.45pt},
    tag/.style={anchor=base west, inner xsep=2.4pt, inner ysep=1.3pt, rounded corners=1.5pt,
                font=\scriptsize\bfseries, text=white,
                execute at begin node=\vphantom{bdfgp}},
  ]
  \def\dy{0.07}%
  \filldraw[fill=basebg, draw=gridline, line width=0.5pt, rounded corners=3pt] (-0.25,-2.65) rectangle (11.17,0.38);
  \node[fname, font=\ttfamily\footnotesize\bfseries] at (0,0) {problems/stopping-power/};
  \draw[tree] (0.15,-0.12) -- (0.15,-2.40+\dy);
  \foreach \y in {-0.42,-0.84,-1.98,-2.40} {\draw[tree] (0.15,\y+\dy) -- (0.45,\y+\dy);}
  \draw[tree] (0.75,-0.96) -- (0.75,-1.56+\dy);
  \foreach \y in {-1.20,-1.56} {\draw[tree] (0.75,\y+\dy) -- (1.05,\y+\dy);}

  \node[fname] at (0.55,-0.42) {README.md};
  \node[fdesc] at (3.60,-0.42) {current best solution and leaderboard};

  \node[fname] at (0.55,-0.84) {StoppingPower/};
  \node[fname] at (1.15,-1.20) {best/Solution/};
  \node[fdesc] at (3.60,-1.20) {best verified solution, split into Lean modules};
  \node[tag, fill=accentgreen] at (8.97,-1.20) {machine-checked};
  \node[fname] at (1.15,-1.56) {history.md};
  \node[fdesc] at (3.60,-1.56) {verified records, each crediting its teams};

  \node[fname] at (0.55,-1.98) {lean/};
  \node[fdesc] at (3.60,-1.98) {reusable partial Lean modules};
  \node[tag, fill=accentgreen] at (8.97,-1.98) {machine-checked};

  \node[fname] at (0.55,-2.40) {ideas/};
  \node[fdesc] (rnotes) at (3.60,-2.40) {reviewed notes:};
  \node[tag, fill=accentgreen] (tver) at ([xshift=4.5pt]rnotes.base east) {verified};
  \node[tag, fill=accentamber] (tpro) at ([xshift=4.0pt]tver.base east) {promising};
  \node[tag, fill=accentred]   (tdea) at ([xshift=4.0pt]tpro.base east) {dead-end};
\end{tikzpicture}
\caption{\textbf{Layout of one problem directory in the shared wiki (stopping power).} The best solution and the reusable partial Lean modules are checked by the external verifier against the frozen specification before merging, while idea notes are reviewed summaries labeled verified, promising, or dead-end. Each verified record in \texttt{history.md} lists the teams credited for it.}
\label{fig:app_wiki}
\end{figure}

\paragraph{Shared wiki.} Teams exchange persistent artifacts through a version-controlled repository with one directory per problem~\cite{schmidgall2025agentrxiv,qu2026coral}. Each entry contains: (i) the current best verified solution, which any new run can adopt as its starting point; (ii) reusable partial Lean files verified to compile against the shared library; and (iii) reviewed notes classifying explored idea families as verified, promising, or dead ends, alongside short technical lessons. Whereas solutions and partial Lean files are machine-checked, idea notes are reviewed summaries rather than formal proofs (see Fig.~\ref{fig:app_wiki}). A candidate solution is merged only if it passes automatic re-verification against the frozen specification and clears a minimum relative cost-improvement threshold ($5\%$ by default) so that teams do not spend effort adapting to frequent minor updates. Because verified Lean solutions grow large across successive iterations, problems with established solutions also accept updates that substantially shorten the Lean code (by at least $20\%$) while strictly reducing cost, keeping files within agent context limits and reducing Lean compilation time. Recording negative results with concrete obstructions (such as numerical counterexamples or Lean lemmas) prevents later runs across all teams from repeating known dead ends.

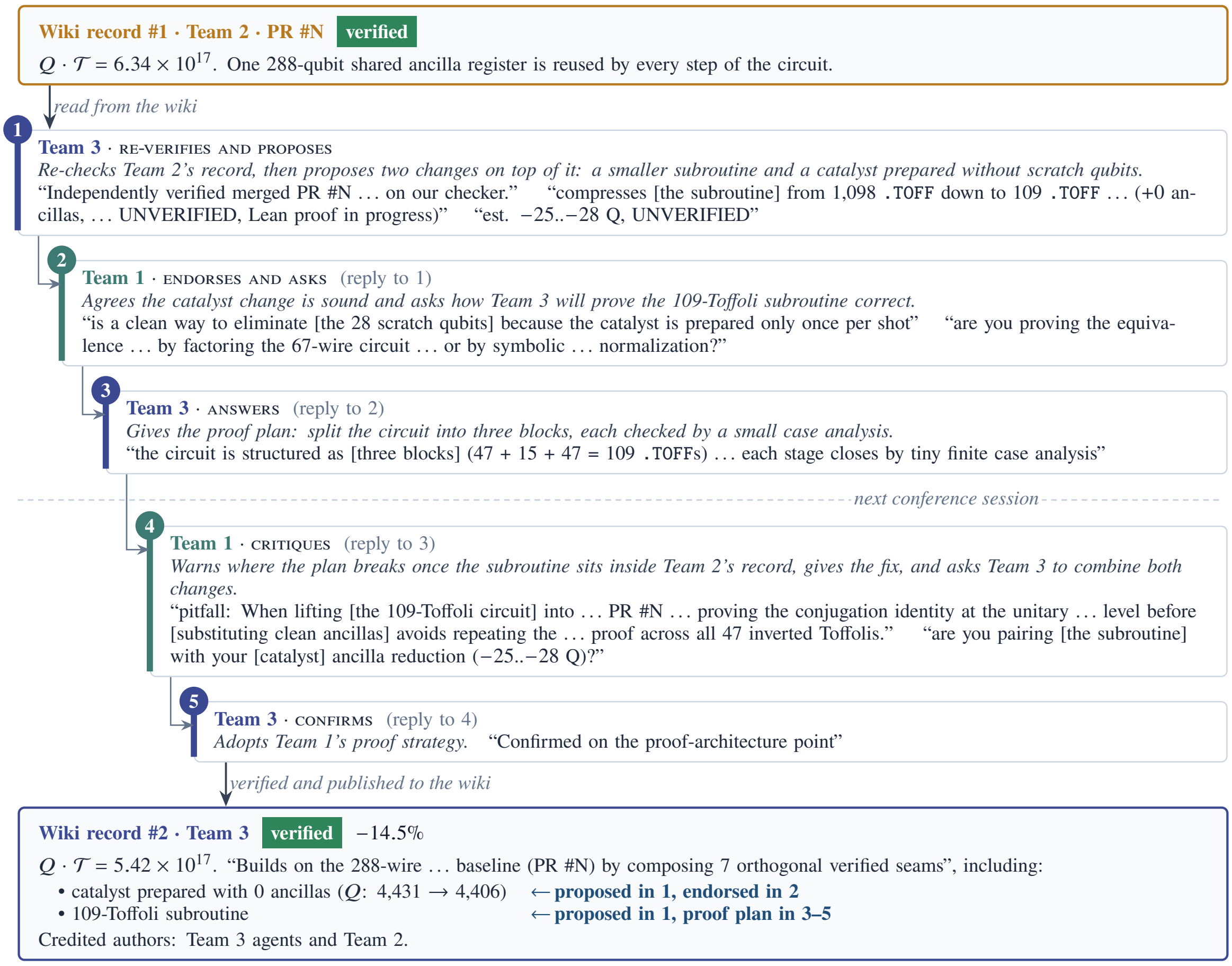
\begin{figure}[!htbp]
\centering
\providecolor{teamR1}{HTML}{3D7A74}%
\providecolor{teamR2}{HTML}{B7791F}%
\providecolor{teamR3}{HTML}{3B4992}%
\providecolor{inkdark}{HTML}{0F172A}%
\providecolor{inkbody}{HTML}{1E293B}%
\providecolor{inklabel}{HTML}{334155}%
\providecolor{inksub}{HTML}{64748B}%
\providecolor{gridline}{HTML}{CBD5E1}%
\providecolor{basebg}{HTML}{F8FAFC}%
\providecolor{accentgreen}{HTML}{2F855A}%
\providecolor{ideaBlue}{HTML}{1E4E79}%
\begin{tikzpicture}[
    x=1cm, y=1cm,
    post/.style={draw=gridline, fill=white, line width=0.45pt, rounded corners=2pt,
                 inner xsep=7pt, inner ysep=3.5pt, align=left, anchor=north west,
                 font=\scriptsize, text=inkbody},
    record/.style={post, fill=basebg, line width=0.8pt, text width=14.6cm},
    badge/.style={circle, inner sep=0pt, minimum size=3.6mm, font=\scriptsize\bfseries,
                  text=white},
    reply/.style={draw=inksub, line width=0.5pt, -{Stealth[length=1.6mm,width=1.3mm]}},
    flow/.style={draw=inklabel, line width=0.7pt, -{Stealth[length=2mm,width=1.6mm]}},
    note/.style={font=\scriptsize\itshape, text=inksub, inner sep=1pt},
  ]
  \providecommand{\acfVtag}{\colorbox{accentgreen}{\textcolor{white}{\bfseries verified}}}
  \providecommand{\acfRole}[1]{\textsc{#1}}
  \providecommand{\acfGloss}[1]{{\itshape\color{inklabel}#1}}
  \providecommand{\acfFrom}[1]{{\color{ideaBlue}$\leftarrow$\,\bfseries #1}}
  \providecommand{\acfCut}{\mbox{.\kern0.16em.\kern0.16em.}}
  \def\acfInd{0.55}%

  \node[record, draw=teamR2] (r1) at (0,0) {%
    {\bfseries\color{teamR2}Wiki record \#1 \textperiodcentered\ Team 2 \textperiodcentered\ PR \#N}\ \ \acfVtag\\[1pt]
    $Q\cdot \mathcal{T} = 6.34\times 10^{17}$. One $288$-qubit shared ancilla register is reused by every step of the circuit.};

  \node[post, text width=14.6cm] (p1) at ([yshift=-5.5mm]r1.south west) {%
    {\bfseries\color{teamR3}Team 3} \textperiodcentered\ \acfRole{re-verifies and proposes}\\
    \acfGloss{Re-checks Team 2's record, then proposes two changes on top of it: a smaller subroutine and a catalyst prepared without scratch qubits.}\\
    ``Independently verified merged PR \#N \acfCut{} on our checker.'' \quad ``compresses [the subroutine] from 1,098 \texttt{.TOFF} down to 109 \texttt{.TOFF} \acfCut{} ($+0$ ancillas, \acfCut{} UNVERIFIED, Lean proof in progress)'' \quad ``est. $-25$..$-28$ Q, UNVERIFIED''};

  \node[post, text width=14.6cm-\acfInd cm] (p2) at ([xshift=\acfInd cm, yshift=-3mm]p1.south west) {%
    {\bfseries\color{teamR1}Team 1} \textperiodcentered\ \acfRole{endorses and asks}\ \ {\color{inksub}(reply to 1)}\\
    \acfGloss{Agrees the catalyst change is sound and asks how Team 3 will prove the 109-Toffoli subroutine correct.}\\
    ``is a clean way to eliminate [the 28 scratch qubits] because the catalyst is prepared only once per shot'' \quad ``are you proving the equivalence \acfCut{} by factoring the 67-wire circuit \acfCut{} or by symbolic \acfCut{} normalization?''};

  \node[post, text width=14.6cm-2*\acfInd cm] (p3) at ([xshift=\acfInd cm, yshift=-3mm]p2.south west) {%
    {\bfseries\color{teamR3}Team 3} \textperiodcentered\ \acfRole{answers}\ \ {\color{inksub}(reply to 2)}\\
    \acfGloss{Gives the proof plan: split the circuit into three blocks, each checked by a small case analysis.}\\
    ``the circuit is structured as [three blocks] (47 + 15 + 47 = 109 \texttt{.TOFF}s) \acfCut{} each stage closes by tiny finite case analysis''};

  \coordinate (sd) at ([yshift=-3.2mm]p3.south west);
  \draw[gridline, line width=0.6pt, dash pattern=on 2pt off 1.6pt] (0,0 |- sd) -- (15.0,0 |- sd);
  \node[note, fill=white, anchor=center] at (11.6,0 |- sd) {next conference session};

  \node[post, text width=14.6cm-3*\acfInd cm] (p4) at ([xshift=\acfInd cm, yshift=-6.4mm]p3.south west) {%
    {\bfseries\color{teamR1}Team 1} \textperiodcentered\ \acfRole{critiques}\ \ {\color{inksub}(reply to 3)}\\
    \acfGloss{Warns where the plan breaks once the subroutine sits inside Team 2's record, gives the fix, and asks Team 3 to combine both changes.}\\
    ``pitfall: When lifting [the 109-Toffoli circuit] into \acfCut{} PR \#N \acfCut{} proving the conjugation identity at the unitary \acfCut{} level before [substituting clean ancillas] avoids repeating the \acfCut{} proof across all 47 inverted Toffolis.'' \quad ``are you pairing [the subroutine] with your [catalyst] ancilla reduction ($-25$..$-28$ Q)?''};

  \node[post, text width=14.6cm-4*\acfInd cm] (p5) at ([xshift=\acfInd cm, yshift=-3mm]p4.south west) {%
    {\bfseries\color{teamR3}Team 3} \textperiodcentered\ \acfRole{confirms}\ \ {\color{inksub}(reply to 4)}\\
    \acfGloss{Adopts Team 1's proof strategy.}\quad ``Confirmed on the proof-architecture point''};

  \node[record, draw=teamR3] (r2) at ([yshift=-5.5mm]p5.south west -| r1.west) {%
    {\bfseries\color{teamR3}Wiki record \#2 \textperiodcentered\ Team 3}\ \ \acfVtag\ \ {\bfseries\color{inkdark}$-14.5\%$}\\[1pt]
    $Q\cdot \mathcal{T} = 5.42\times 10^{17}$. ``Builds on the 288-wire \acfCut{} baseline (PR \#N) by composing 7 orthogonal verified seams'', including:\\[1pt]
    \begin{tabular}{@{\hspace*{1em}}l@{\hspace{1.2em}}l@{}}
    \textbullet\ catalyst prepared with 0 ancillas ($Q$: 4,431 $\to$ 4,406) & \acfFrom{proposed in 1, endorsed in 2}\\
    \textbullet\ 109-Toffoli subroutine & \acfFrom{proposed in 1, proof plan in 3--5}
    \end{tabular}\\[1pt]
    Credited authors: Team 3 agents and Team 2.};

  \foreach \p/\c/\n in {p1/teamR3/1, p2/teamR1/2, p3/teamR3/3, p4/teamR1/4, p5/teamR3/5} {
    \draw[\c, line width=2.2pt] ([yshift=-2pt]\p.north west) -- ([yshift=2pt]\p.south west);
    \node[badge, fill=\c] at (\p.north west) {\n};
  }

  \draw[flow] ([xshift=4mm]r1.south west) -- ([xshift=4mm]p1.north west)
     node[midway, right, note] {read from the wiki};
  \foreach \a/\b in {p1/p2, p2/p3, p3/p4, p4/p5} {
    \draw[reply] ([xshift=2.6mm]\a.south west) |- ([yshift=-3.0mm]\b.north west);
  }
  \draw[flow] ([xshift=4mm]p5.south west) -- ([xshift=4mm]p5.south west |- r2.north)
     node[midway, right, note] {verified and published to the wiki};
\end{tikzpicture}
\caption{\textbf{A cross-team conference exchange between two verified records in the stopping power problem (records \#1 and \#2 of Fig.~\ref{fig:collaboration}c).} Each numbered post replies directly to the one above it, across two conference sessions. Colored bars mark the team, the italic line is an analysis summary, and quoted text is verbatim from the conference; ``\dots\unkern'' marks omitted text, and words in square brackets, such as [the subroutine], replace Lean code snippets used in the original posts.}

\label{fig:app_conference}
\end{figure}

\paragraph{Conferences.} Because many useful exchanges discuss unverified conjectures, teams convene periodically every four hours on a shared forum open across participating researchers and agent frameworks. During each session, agents pose questions, critique claims, and propose candidate constructions for other teams to test (see Fig.~\ref{fig:app_conference}). Discussion rules encourage direct replies between teams as opposed to standalone status reports, preventing the one-way broadcasting typical of unstructured agent boards~\cite{jiang2026moltbook,chen2026colmad}. To uphold the correctness of intermediate work, statements and cost projections posted in a conference carry no weight until independently verified and published to the wiki.

\section{Fault-Tolerant Cost Metric and Hardware Horizons}
\label{app:horizons}
\subsection{Logical spacetime volume and architectural assumptions}
\label{app:cost_metric}

Throughout the fault-tolerant benchmarks in this work, candidate circuits and literature baselines are compared at matched system encodings and error budgets by their logical spacetime volume $Q \cdot \mathcal{T}$, defined as total peak logical qubits $Q$ (including clean workspace ancillas) multiplied by total $T$-gate count $\mathcal{T}$, pricing clean-ancilla Toffoli gates at $4\,\mathcal{T}$ via measurement-based temporary-$\mathrm{AND}$ uncomputation~\cite{jones2013toffoli,gidney2018halving,khattar2025rise}. This metric models compilation onto two-dimensional planar surface-code architectures using lattice surgery~\cite{fowler2012surface,horsman2012surface,litinski2019game}. Because Clifford gates can be tracked in the classical Pauli frame or absorbed into multi-qubit Pauli product measurements, fault-tolerant overhead is dominated by preparing non-Clifford magic states through distillation or cultivation~\cite{bravyi2005universal,gidney2019efficient,gidney2024magic,litinski2019game}.
Under Pauli-based or serial-factory execution---where magic states are consumed at a fixed logical cycle rate---execution time tracks $\mathcal{T}$ while memory footprint tracks $Q$, making $Q \cdot \mathcal{T}$ proportional to $Q_{\mathrm{logical}} \cdot D_{\mathrm{cycles}}$, the topological spacetime volume~\cite{litinski2019game,gidney2021honeycomb,beverland2022assessing,kim2022fault,caesura2025blissthc,preskill2025megaquop}. Optimizing $Q \cdot \mathcal{T}$ also regularizes space--time tradeoffs against degenerate extremes---neither permitting unbounded clean-ancilla QROAM replication to suppress $T$-gate count~\cite{babbush2018linear,berry2019qubitization,lee2021thc} nor forcing deep ancilla-starved recomputation to save a handful of qubits~\cite{haner2018optimizing,gu2026discovering}---and aligns directly with the $(Q, \text{Toffoli}/\mathcal{T})$ metrics reported across standard resource-estimation benchmarks~\cite{reiher2017elucidating,vonburg2021catalysis,su2021firstquant,goings2022p450,rubin2023materials,rubin2024stopping,low2025spectrum,pocrnic2026prebo}.

Physical device cost departs from $Q \cdot \mathcal{T}$ in two principal architectural regimes. First, provisioning parallel magic-state factories compresses execution depth from $\mathcal{T}$ toward $T$-depth at the expense of dedicated factory footprint and routing overhead~\cite{litinski2019game,beverland2022assessing,kim2022fault}.
Second, architectures incorporating non-local connectivity, high-rate quantum low-density parity-check (qLDPC) memory, or Active Volume compilation decouple idle memory storage from active reaction layers, lowering the spatial cost of inactive data qubits relative to active workspace~\cite{litinski2022active,bravyi2024high,caesura2025blissthc,gidney2025rsa,low2026denser}. Because every verified scientific simulation in Fig.~\ref{fig:results}b reduces both $Q$ and $\mathcal{T}$ relative to its baseline, our improvements hold under any monotone weighting of logical space and non-Clifford-gate count.

\subsection{Fault-tolerant capability regimes and hardware horizons}
\label{app:horizon_regimes}

\begin{table}[!t]
\addtokomafont{caption}{\footnotesize}
\caption{\textbf{Fault-tolerant capability regimes and hardware horizons.} Summary of the capability regimes referenced in Fig.~\ref{fig:results}b, listing plotted dashed-line midpoints $(Q, \mathcal{T})$, nominal upper-bound ceilings $(Q_{\max}, \mathcal{T}_{\max})$ (rendered in Fig.~\ref{fig:results}b with 14-step normal-quantile feathering spanning $Q \approx 5.9\times 10^{2}\text{--}1.7\times 10^{3}$, $\mathcal{T} \approx 3.5\times 10^{8}\text{--}2.9\times 10^{9}$ for teraquop planar and $Q \approx 5.0\times 10^{3}\text{--}2.0\times 10^{4}$, $\mathcal{T} \approx 3.5\times 10^{10}\text{--}2.9\times 10^{11}$ for FASQ), nominal spacetime volume $Q\cdot \mathcal{T}$, and representative benchmarks and references.}
\label{tab:horizons}
\vspace{2pt}
{\fontsize{7.5}{9.2}\selectfont
\setlength{\tabcolsep}{3.0pt}%
\renewcommand{\arraystretch}{1.22}%
\begin{tabular*}{\textwidth}{@{\extracolsep{\fill}}lllll>{\raggedright\arraybackslash}p{4.3cm}@{}}
\toprule
\textbf{Capability regime} & \textbf{Midpoint $(Q, \mathcal{T})$} & \textbf{$Q$ ceiling} & \textbf{$\mathcal{T}$ ceiling} & \textbf{$Q{\cdot}\mathcal{T}$} & \textbf{Benchmarks \& references} \\
\midrule
Medium-term FTQC (teraquop) & $(10^{3},\, 10^{9})$ & $5{\times}10^{2}\text{--}2{\times}10^{3}$ & $3{\times}10^{8}\text{--}3{\times}10^{9}$ & $10^{12}$ & Cytochrome P450; compact catalysis \& ECC~\cite{gidney2021honeycomb,vonburg2021catalysis,goings2022p450,gidney2024magic,caesura2025blissthc,low2025spectrum,low2026denser,babbush2026kickmix} \\
Application-scale FTQC (FASQ) & $(10^{4},\, 10^{11})$ & $5{\times}10^{3}\text{--}2{\times}10^{4}$ & $3{\times}10^{10}\text{--}3{\times}10^{11}$ & $10^{15}$ & Stopping power, $\mathrm{C_{23}H_{20}N_3O}$ dynamics; catalysis, materials \& RSA~\cite{reiher2017elucidating,gidney2021rsa,su2021firstquant,beverland2022assessing,kim2022fault,litinski2022active,rubin2023materials,rubin2024stopping,preskill2025megaquop,pocrnic2026prebo} \\
\bottomrule
\end{tabular*}}
\end{table}

To place end-to-end logical resource counts in hardware context, Fig.~\ref{fig:results}b overlays two fault-tolerant capability regimes (Table~\ref{tab:horizons}) as upper-bound resource envelopes $(Q \le Q_{\max},\, \mathcal{T} \le \mathcal{T}_{\max})$, with dashed lines at reference midpoints and feathered boundaries reflecting spread across estimates.

\paragraph{Medium-term FTQC (teraquop planar).}
Gidney \emph{et al.}~\cite{gidney2021honeycomb} (Sec.~2.1) define the teraquop scale at $10^{12}$ reliable logical operations ($\sim 10^{3}$ logical qubits over $\sim 10^{9}$ steps at per-step error $\sim 10^{-12}$), and Low \emph{et al.}~\cite{low2026denser} quantify how denser planar surface-code layouts and magic-state cultivation set the corresponding qubit and $T$-gate ceilings across planar code distances. This regime sits at midpoint $Q = 10^{3}$, $\mathcal{T} = 10^{9}$, and $Q\cdot \mathcal{T} = 10^{12}$, with feathered ceiling $Q \in [5\times 10^{2},\, 2\times 10^{3}]$ and $\mathcal{T} \in [3\times 10^{8},\, 3\times 10^{9}]$.

\paragraph{Application-scale FTQC (FASQ / petaquop).}
Preskill~\cite{preskill2025megaquop} defines the fault-tolerant application-scale quantum (FASQ) regime for full-scale scientific and cryptographic workloads, and Beverland \emph{et al.}~\cite{beverland2022assessing} (Table~I) establish the logical-qubit and $T$-gate budgets required across materials, chemistry, dynamics, and factoring. This regime sits at midpoint $Q = 10^{4}$, $\mathcal{T} = 10^{11}$, and $Q\cdot \mathcal{T} = 10^{15}$, with feathered ceiling $Q \in [5\times 10^{3},\, 2\times 10^{4}]$ and $\mathcal{T} \in [3\times 10^{10},\, 3\times 10^{11}]$.

\section{Parallel Grover Search}
\label{app:grover}
\subsection{Problem specification and oracle model}
\label{app:grover_spec}

\begin{figure}[!tp]
\addtokomafont{caption}{\footnotesize}
\providecolor{frameblue}{HTML}{1E4E79}
\providecolor{accentblue}{HTML}{2B6CB0}
\providecolor{inkdark}{HTML}{0F172A}
\providecolor{inkbody}{HTML}{1E293B}
\providecolor{inklabel}{HTML}{334155}
\providecolor{inksub}{HTML}{64748B}
\protected\def\fgHdr{\fontsize{8.4}{10.0}\selectfont}
\protected\def\fgSix{\fontsize{6.2}{7.3}\selectfont}
\protected\def\fgTiny{\fontsize{5.1}{6.0}\selectfont}
\protected\def\fgMicro{\fontsize{4.7}{5.5}\selectfont}

\tikzset{
  fgphead/.style={font=\fgHdr, text=black, anchor=base west, inner sep=0pt},
  fgqstd/.style={fill=white, draw=black!80, text=black, line width=0.45pt, rounded corners=1.0pt, inner sep=2.3pt},
  fgqora/.style={fill=accentblue!16, draw=frameblue, text=frameblue, line width=0.58pt, rounded corners=1.2pt, inner sep=2.3pt},
  fgqstdb/.style={fill=white, draw=black!80, text=black, line width=0.45pt, rounded corners=1.0pt, inner xsep=0.8pt, inner ysep=0.4pt},
  fgqorab/.style={fill=accentblue!16, draw=frameblue, text=frameblue, line width=0.58pt, rounded corners=1.2pt, inner xsep=0.8pt, inner ysep=0.4pt},
  fgqmet/.style={fill=white, draw=black!80, line width=0.45pt},
  fgqgrp/.style={fill=none, draw=black!50, densely dashed, line width=0.45pt, rounded corners=2.0pt, inner xsep=3.5pt, inner ysep=2.6pt},
  fgqgrplbl/.style={label position=above, anchor=south, yshift=-0.03cm, font=\fgSix, text=black!85},
}

\centering
\begin{tikzpicture}[x=1cm, y=1cm]
  \path[use as bounding box] (0,0) rectangle (8.02, 3.16);

  \begin{scope}[xshift=-7.50cm, yshift=-10.20cm]
  \def\spBadge{\rmfamily\fontsize{5.50}{6.20}\selectfont\bfseries}%
  \def\spNumbadge#1#2#3{%
    \fill[frameblue] (#1,#2) circle [radius=0.102cm];
    \node[anchor=center, inner sep=0pt, font=\spBadge, text=white] at (#1,#2) {#3};
  }%

  \node[fgphead] at (7.50, 13.18) {\color{inkbody}Fundamental primitives of the adaptive query model};

  \spNumbadge{7.60}{12.80}{1}
  \node[anchor=west, inner sep=0pt, font=\fgTiny, text=inkdark] at (7.75, 12.80)
    {\texttt{qset}\;{\color{inksub}($+0$)}};

  \spNumbadge{9.42}{12.80}{2}
  \node[anchor=west, inner sep=0pt, font=\fgTiny, text=inkdark] at (9.57, 12.80)
    {\texttt{step}\;{\color{inksub}($+0$)}};

  \spNumbadge{10.62}{12.80}{3}
  \node[anchor=west, inner sep=0pt, font=\fgTiny, text=inkdark] at (10.77, 12.80)
    {\texttt{oracle}\;{\color{frameblue}\textbf{($+1$)}}};

  \spNumbadge{11.90}{12.80}{4}
  \node[anchor=west, inner sep=0pt, font=\fgTiny, text=inkdark] at (12.05, 12.80)
    {\texttt{measure}\;{\color{inksub}($+0$)}};

  \spNumbadge{13.38}{12.80}{5}
  \node[anchor=west, inner sep=0pt, font=\fgTiny, text=inkdark] at (13.53, 12.80)
    {\texttt{qif}\,/\,\texttt{qwhile}};

  \node[anchor=west, inner sep=0pt, font=\fgSix, text=inkdark] at (7.50, 11.73) {$\ket{0}^{\otimes Q}$};
  \draw[draw=inksub, line width=0.45pt, decorate, decoration={brace, amplitude=2.4pt}]
    (8.20, 11.08) -- (8.20, 12.38);

  \node[anchor=west, inner sep=0pt, font=\fgTiny, text=inklabel] at (8.26, 12.30) {slot $1$};
  \foreach \dy in {-0.07, 0.00, 0.07} {
    \fill[inksub] (8.50, 11.97+\dy) circle [radius=0.015cm];
  }
  \node[anchor=west, inner sep=0pt, font=\fgTiny, text=inklabel] at (8.26, 11.64) {slot $m$};
  \node[anchor=west, inner sep=0pt, font=\fgTiny, text=inklabel] at (8.26, 11.16) {work};
  \node[anchor=east, inner sep=0pt, font=\fgSix, text=inkdark]   at (8.75, 10.72) {$\sigma_0 \in \Sigma$};

  \filldraw[fill=accentblue!6, draw=frameblue!48, densely dotted, line width=0.42pt, rounded corners=1.5pt]
    (10.62, 11.42) rectangle (11.50, 12.51);

  \draw[draw=frameblue!68, densely dashed, line width=0.48pt, rounded corners=2.4pt]
    (9.25, 10.56) rectangle (14.12, 12.55);

  \draw[draw=black!80, line width=0.40pt] (12.238, 11.47) -- (12.238, 10.72);
  \draw[draw=black!80, line width=0.40pt] (12.282, 11.47) -- (12.282, 10.72);
  \fill[white] (12.19, 11.11) rectangle (12.33, 11.21);
  \fill[black!85] (12.26, 10.72) circle [radius=0.036cm];
  \node[anchor=west, inner sep=0pt, font=\fgTiny, text=inkdark] at (12.34, 10.94) {$y$};

  \draw[draw=black!80, line width=0.40pt] (13.518, 10.72) -- (13.518, 11.00);
  \draw[draw=black!80, line width=0.40pt] (13.562, 10.72) -- (13.562, 11.00);
  \fill[black!85] (13.54, 10.72) circle [radius=0.036cm];

  \draw[draw=black!80, line width=0.45pt] (8.80, 12.30) -- (14.48, 12.30);
  \draw[draw=black!80, line width=0.45pt] (8.80, 11.64) -- (14.48, 11.64);
  \draw[draw=black!80, line width=0.45pt] (8.80, 11.16) -- (14.48, 11.16);
  \draw[draw=black!80, line width=0.40pt] (8.80, 10.742) -- (14.48, 10.742);
  \draw[draw=black!80, line width=0.40pt] (8.80, 10.698) -- (14.48, 10.698);

  \draw[draw=black!80, line width=0.45pt] (8.97, 12.23) -- (9.05, 12.37);
  \node[anchor=south, inner sep=0pt, font=\fgMicro, text=inklabel] at (9.01, 12.40) {$n{+}1$};
  \draw[draw=black!80, line width=0.45pt] (8.97, 11.57) -- (9.05, 11.71);
  \node[anchor=south, inner sep=0pt, font=\fgMicro, text=inklabel] at (9.01, 11.74) {$n{+}1$};
  \draw[draw=black!80, line width=0.45pt] (8.97, 11.09) -- (9.05, 11.23);
  \node[anchor=south, inner sep=0pt, font=\fgMicro, text=inklabel] at (9.01, 11.26) {$w$};

  \filldraw[fgqstd] (9.42, 11.00) rectangle (10.22, 12.47);
  \node[anchor=center, align=center, inner sep=0pt, font=\fgSix, text=inkdark] at (9.82, 11.735)
    {$U$\\[-0.6pt]{\fgMicro\color{inksub}$\in \mathrm{U}(2^Q)$}};

  \filldraw[fgqora] (10.70, 12.13) rectangle (11.42, 12.47);
  \node[anchor=center, inner sep=0pt, font=\fgSix, text=frameblue] at (11.06, 12.30) {$O_S$};
  \foreach \dy in {-0.07, 0.00, 0.07} {
    \fill[frameblue!85] (11.06, 11.97+\dy) circle [radius=0.016cm];
  }
  \filldraw[fgqora] (10.70, 11.47) rectangle (11.42, 11.81);
  \node[anchor=center, inner sep=0pt, font=\fgSix, text=frameblue] at (11.06, 11.64) {$O_S$};

  \foreach \xpos in {10.42, 11.72, 12.83, 14.30} {
    \foreach \dy in {-0.065, 0.00, 0.065} {
      \fill[inksub!75] (\xpos, 11.97+\dy) circle [radius=0.013cm];
    }
  }

  \filldraw[fill=white, draw=black!80, line width=0.45pt, rounded corners=1.0pt]
    (11.98, 11.47) rectangle (12.54, 11.81);
  \draw[draw=black!82, line width=0.42pt]
    (12.101, 11.535) arc [start angle=152, end angle=28, radius=0.180cm];
  \draw[-{Stealth[length=1.6pt, width=1.3pt]}, draw=black!85, line width=0.42pt]
    (12.26, 11.515) -- (12.375, 11.735);

  \filldraw[fgqstd] (13.14, 11.00) rectangle (13.94, 12.47);
  \node[anchor=center, align=center, inner sep=0pt, font=\fgSix, text=inkdark] at (13.54, 11.735)
    {\texttt{qif}\\[-0.6pt]{\fgMicro\color{inksub}$b'(\sigma)$}};

  \draw[-{Stealth[length=2.1pt, width=1.7pt]}, draw=frameblue!85, line width=0.46pt, rounded corners=1.8pt]
    (13.94, 10.56) -- (13.94, 10.32) -- (9.43, 10.32) -- (9.43, 10.55);
  \node[fill=white, inner xsep=2.2pt, inner ysep=0.5pt, font=\fgTiny, text=frameblue] at (11.685, 10.32)
    {\texttt{qwhile}\ $b(\sigma)$\;{\color{inksub}(trace-weighted)}};

  \draw[draw=inksub, line width=0.45pt, decorate, decoration={brace, amplitude=2.4pt}]
    (14.53, 12.38) -- (14.53, 10.64);
  \node[anchor=west, align=left, inner sep=0pt, font=\fgTiny, text=inkdark] at (14.66, 11.51)
    {$\neg b(\sigma)\to\mathrm{out}$\\[-0.3pt]{\fgMicro\color{inksub}($+0$)}};
  \end{scope}
\end{tikzpicture}
\caption{\textbf{Parallel Grover query model.} Workspace allocation (\texttt{qset}, $+0$ query cost) initializes $m$ query registers of $n+1$ qubits and $w$ ancilla qubits in $\ket{0}^{\otimes Q}$ together with a classical state $\sigma \in \Sigma$. Unitary evolution (\texttt{step}, $+0$) applies any $S$-independent unitary $U \in \mathrm{U}(2^Q)$ across all $Q$ qubits. Parallel oracle querying (\texttt{oracle}, $+1$) applies $O_S^{\otimes m}$ (Eq.~\eqref{eq:parallel_oracle}) simultaneously across $m$ disjoint $(n+1)$-qubit subregisters. Projective measurement (\texttt{measure}, $+0$) measures a subregister into classical variable $y$ in $\sigma$ while leaving unmeasured wires coherent. Classical control (\texttt{qif}/\texttt{qwhile}; Appendix~\ref{app:qhl}) branches and loops on predicates of $\sigma$, weighting each oracle round by the active-branch probability $\sum_{\sigma:\,b(\sigma)}\mathrm{Tr}(\Delta_t(\sigma))$ until $\neg b(\sigma)$ outputs the result.}
\label{fig:app_grover}
\end{figure}
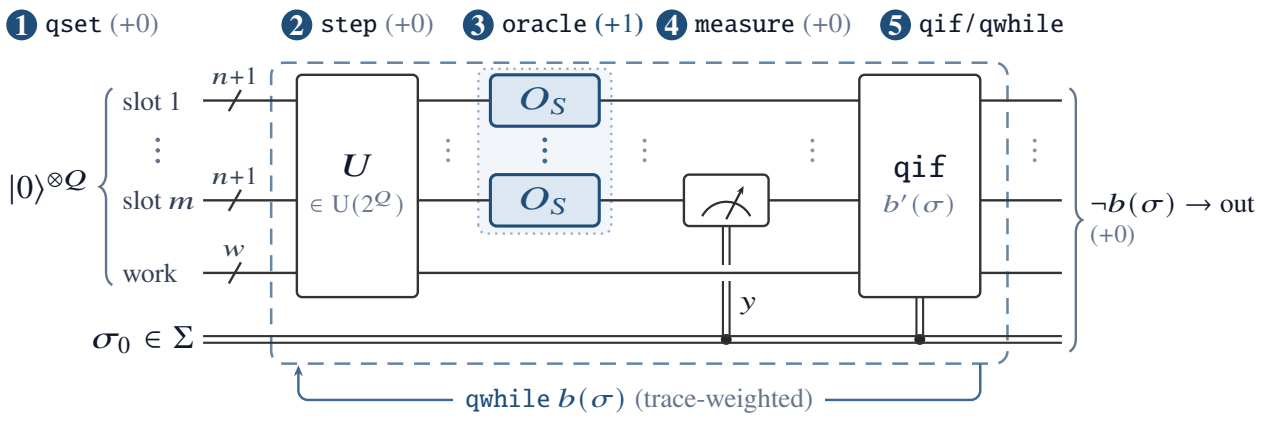

Given a search space $\{0,\dots,N-1\}$ of size $N = 2^n$ ($n \ge 1$) and an unknown marked subset $S \subseteq \{0,\dots,N-1\}$ of known size $|S| = k \le N$, we study two exact (zero-error) search problems across $(m, k) \in \{1, 2, 4, \dots, 1024\}^2$ ($121$ settings): \emph{Find-Any}, which must terminate with probability~$1$ and output an index $x \in S$, and \emph{Find-All}, which must terminate with probability~$1$ and output all $k$ elements of $S$ (so algorithms cannot trade error probability for fewer queries). We allow up to $m$ parallel queries per round~\cite{zalka1999grover,Jeffery2016} (Fig.~\ref{fig:app_grover}). Access to $S$ is given by the $(n+1)$-qubit reversible marking oracle
\begin{equation}
  O_S\ket{b}\ket{x} = \ket{b \oplus [x \in S]}\ket{x}, \qquad b \in \{0, 1\},\; x \in \{0,\dots,N-1\},
  \label{eq:marking_oracle}
\end{equation}
where $b$ is a single-qubit target bit and $x$ is the $n$-qubit index. Starting from $\ket{0}^{\otimes Q}$ on an unconstrained $Q$-qubit workspace, an algorithm interleaves zero-cost $S$-independent unitaries, projective measurements, and classical branching or loops with parallel $m$-query rounds
\begin{equation}
  O_S^{\otimes m} = \bigotimes_{j=1}^{m} O_S^{(j)}
  \label{eq:parallel_oracle}
\end{equation}
on any $m$ disjoint $(n+1)$-qubit registers, weighted in expectation by the probability of being in an active loop branch. To query fewer than $m$ registers in a round, the remaining target qubits are prepared in $\ket{+}$, on which $O_S$ acts as the identity. Our cost metric is the expected number of parallel query rounds (query depth).

Depending on how the target qubit $b$ is prepared (and optionally measured), a single call to $O_S$ implements three operations. Preparing $\ket{b} = \ket{-}$ applies the phase flip $(I - 2\Pi_S)$ for $\Pi_S = \sum_{x\in S}\ket{x}\!\bra{x}$, while preparing $\ket{b} = \ket{0}$ and measuring $b$ tests whether $x \in S$ (and once a subset $F \subseteq S$ has been found, unmarking $F$ via $\ket{b}\ket{x} \mapsto \ket{b \oplus [x \in F]}\ket{x}$ is independent of $S$ and costs zero queries). More generally, preparing $\ket{\chi_\gamma} = \cos\gamma\,\ket{0} - \sin\gamma\,\ket{1}$ for $\gamma \in (0, \pi/4)$ leaves $\ket{\chi_\gamma}$ unchanged for $x \notin S$ while mapping $\ket{\chi_\gamma} \mapsto -\sin(2\gamma)\ket{\chi_\gamma} + \cos(2\gamma)\ket{\chi_\gamma^\perp}$ for $x \in S$, where $\ket{\chi_\gamma^\perp} = \sin\gamma\,\ket{0} + \cos\gamma\,\ket{1}$. Measuring $b$ in $\{\ket{\chi_\gamma}, \ket{\chi_\gamma^\perp}\}$ therefore applies, with $\kappa = \cos^2(2\gamma) \in (0, 1)$,
\begin{equation}
  M_1 = \sqrt{\kappa}\,\Pi_S \;(\text{outcome }\ket{\chi_\gamma^\perp}), \qquad
  M_0 = (I - \Pi_S) - \sqrt{1 - \kappa}\,\Pi_S \;(\text{outcome }\ket{\chi_\gamma})\text{:}
  \label{eq:weak_kraus}
\end{equation}
a click ($\ket{\chi_\gamma^\perp}$) projects onto $S$ with no false positives, while a no-click ($\ket{\chi_\gamma}$) only slightly damps the marked amplitude and still flips its sign by $-1$ for the next Grover diffusion step.

To compare algorithms independent of the domain size $N = 2^n$, let $\mathbb{E}[D_n(S)]$ be a program's expected query depth on marked set $S$, and let $\overline{D}_{n,m,k} = \binom{N}{k}^{-1}\sum_{|S|=k}\mathbb{E}[D_n(S)]$ be its average over all $k$-element subsets (which equals the worst-case expected depth over $|S| = k$, since indices can be randomly permuted at zero query cost). We normalize $\overline{D}_{n,m,k}$ by the fixed reference scales
\begin{equation}
  D_{\mathrm{std}}^{\mathrm{any}}(N, m, k) = \frac{\pi}{4}\sqrt{\frac{N}{mk}}, \quad
  D_{\mathrm{std}}^{\mathrm{all}}(N, m, k) = \sum_{r=1}^{k} \frac{\pi}{4}\sqrt{\frac{N}{mr}} = \frac{\pi}{4}\,H_k^{(1/2)}\sqrt{\frac{N}{m}}, \quad
  H_k^{(1/2)} = \sum_{r=1}^{k}\frac{1}{\sqrt{r}},
  \label{eq:vanilla_scales}
\end{equation}
and certify in Lean a rational constant $R(m, k)$ satisfying $\overline{D}_{n,m,k} \le R(m, k)\,D_{\mathrm{std}}(2^n, m, k)$ for all sufficiently large $n$. Because $D_{\mathrm{std}}$ is only a fixed reference scale rather than a lower bound, all speedups are computed as ratios against published baselines in the same metric, so $D_{\mathrm{std}}$ cancels out.

\subsection{Published prior baselines}
\label{app:grover_baselines}

Every quantum search protocol obeys $\Theta(\sqrt{N})$ scaling in domain size~\cite{bennett1997strengths,boyer1998tight,zalka1999grover}, with parallel round lower bounds $\Omega(\sqrt{N/(mk)})$ for Find-Any (Zalka's bound~\cite{zalka1999grover} for $k = 1$; the general case follows from the parallel adversary bound~\cite{Jeffery2016}) and $\Omega(\sqrt{Nk/(m\min(m, k))})$ for Find-All~\cite{grover2004parallel}, so protocols at fixed $(m, k)$ compete on their asymptotic leading constant $r(m, k) = \lim_{N\to\infty}\overline{D}_{n,m,k}/D_{\mathrm{std}}(N, m, k)$. Prior work spans exact and fixed-point amplitude amplification~\cite{grover1996fast,hoyer2000arbitrary,long2001grover,brassard2002quantum,yoder2014fixed}; early-stopped projective restarts with domain partitioning, independent parallel runs, or sequential strike-out~\cite{boyer1998tight,zalka1999grover,gingrich2000generalized,aldenbro2024comparing}; and damped or weakly measured loops~\cite{mizel2009critically,andresmartinez2022weakly,fang2026weakly}. We compare against the strongest published protocol in each $(m, k)$ regime, together with the direct combination of Zalka's partition~\cite{zalka1999grover} and weakly measured search~\cite{fang2026weakly} on the $k = 1$ boundary (Fig.~\ref{fig:app_grover_baselines}). Parallel quantum-walk and learning-graph algorithms~\cite{Jeffery2016} target multi-element certificates and do not tighten these baselines.

\begin{figure}[!tp]
\addtokomafont{caption}{\footnotesize}
\input{figures/figs_parallel_grover_baselines}
\caption{\textbf{Published parallel Grover baselines and verified bounds across $m, k \in \{1, 2, 4, \dots, 1024\}$.} \textbf{a} Each row evaluates a prior baseline protocol in our zero-error asymptotic metric (${}^\star$ marks the winning baseline in each regime); the bottom rows summarize our Lean-verified circuits and speedups over the baseline envelope. Values report asymptotic expected parallel rounds as $N = 2^n \to \infty$ relative to $D_{\mathrm{std}}$ (Eq.~\eqref{eq:vanilla_scales}). \textbf{b, c} Winning baseline envelopes for finding any one (\textbf{b}) or all $k$ (\textbf{c}) targets, on a logarithmic colour scale, with dashed contours marking regime boundaries.}
\label{fig:app_grover_baselines}
\end{figure}

\paragraph{Boundary regime ($m = 1$ or $k = 1$; \textbf{BBHT}, \textbf{ZK}, \textbf{WG}).}
On a single register ($m = 1$) with $k$ marked items, each Grover iteration rotates the state toward $S$ by $2\theta = 2\arcsin\sqrt{k/N} \approx 2\sqrt{k/N}$, so $j$ iterations reach angle $\phi = (2j+1)\theta \approx 2j\sqrt{k/N}$ and find a marked item upon measurement with probability $\sin^2\phi$. Standard Grover runs to $\phi = \pi/2$ ($r = 1$), where $\sin^2\phi$ flattens out and the final queries buy little success probability. Boyer, Brassard, H{\o}yer and Tapp~\cite{boyer1998tight} (\textbf{BBHT}) instead stop early at $\phi < \pi/2$ and restart from scratch on failure: each run costs $j \approx \frac{\phi}{2}\sqrt{N/k}$ queries and succeeds after $1/\sin^2\phi$ expected runs, minimizing expected queries $\frac{\phi}{2\sin^2\phi}\sqrt{N/k}$ at
\begin{equation}
  r_{\mathrm{BBHT}} = \frac{2\phi^*}{\pi\sin^2\phi^*} \approx 0.878567, \qquad \tan\phi^* = 2\phi^* \;\;(\phi^* \approx 1.1656,\; \sin^2\phi^* \approx 0.845).
  \label{eq:bbht_zalka}
\end{equation}
Across the $m = 1$ and $k = 1$ boundaries of Find-Any and Find-All (which coincide when $k = 1$), $r_{\mathrm{BBHT}}$ sets the projective-restart baseline. For $k = 1$ and $m \ge 2$ in both tasks, Zalka~\cite{zalka1999grover} (\textbf{ZK}) splits $\{0,\dots,N-1\}$ into $m$ disjoint blocks of size $N/m$ so the unique block containing the target runs at $\phi^*$ ($r_{\mathrm{ZK}} = r_{\mathrm{BBHT}}$). For $m = 1$ and $k \ge 2$ in Find-All, repeating early-stopped search $k$ times while unmarking each found item at zero query cost~\cite{boyer1998tight,aldenbro2024comparing} achieves $r_{\mathrm{BBHT}}$ at every stage.

On the single-register boundary ($m = 1, k \ge 1$ for Find-Any, and $(1, 1)$ for Find-All), early-stopped projective search is superseded by Fang and Heunen's weakly measured loop~\cite{fang2026weakly} (\textbf{WG}), which avoids discarding the amplified state when a check misses. Instead, \textbf{WG} first runs standard Grover halfway to $\alpha^* = \pi/4$ ($50\%$ marked weight, costing $\frac{1}{2}\,D_{\mathrm{std}}^{\mathrm{any}}$), and then alternates calibrated weak checks (Eq.~\eqref{eq:weak_kraus}, $\kappa \approx 8\sqrt{k/N}$) with single zero-query diffusion steps. Whenever a check does not click, its slight damping and $-1$ phase flip are offset by the diffusion step to return the state back to $\pi/4$ without losing superposition; waiting for a click adds $\frac{1}{\pi}\,D_{\mathrm{std}}^{\mathrm{any}}$ expected queries, giving
\begin{equation}
  r_{\mathrm{WG}} = \frac{1}{2} + \frac{1}{\pi} \approx 0.818310.
  \label{eq:wg}
\end{equation}
For $k = 1$ and $m \ge 2$ in both tasks, running \textbf{WG} inside each of Zalka's $m$ disjoint blocks of size $N/m$ (\textbf{ZK}+\textbf{WG}) is a direct substitution---the $m - 1$ empty blocks never click while the unique target block evolves at scale $\sqrt{N/m}$---upgrading the $k = 1$ boundary from $r_{\mathrm{ZK}} \approx 0.879$ to $r_{\mathrm{WG}} \approx 0.818$.

\paragraph{Interior Find-Any ($m, k \ge 2$; \textbf{GW}).}
When both $m \ge 2$ and $k \ge 2$, naive domain splitting breaks down because each of the $m$ blocks receives a random, unknown number of targets (and most blocks are empty when $k < m$), causing multi-target blocks to rotate faster and overshoot $\pi/2$ under any fixed schedule. Gingrich, Williams and Cerf~\cite{gingrich2000generalized} (\textbf{GW}) avoid this unknown-load problem by running all $m$ registers independently over the full domain $\{0,\dots,N-1\}$. Stopping all $m$ registers synchronously after $j \approx \frac{\phi}{2}\sqrt{N/k}$ rounds finds a target unless all $m$ independent measurements miss (probability $\cos^{2m}\phi$); minimizing expected rounds $\frac{\phi}{2(1 - \cos^{2m}\phi)}\sqrt{N/k}$ selects $\phi^*(m) < \phi^*$ with
\begin{equation}
  r_{\mathrm{GW}}(m) = \frac{2\sqrt{m}\,\phi^*(m)}{\pi\bigl(1 - \cos^{2m}\phi^*(m)\bigr)}, \qquad
  1 - \cos^{2m}\phi^*(m)\bigl(1 + 2m\phi^*(m)\tan\phi^*(m)\bigr) = 0.
  \label{eq:gw}
\end{equation}
Because $m$ unentangled copies combine success probabilities classically rather than coherently, $r_{\mathrm{GW}}(m)$ is independent of $k$ and degrades from $0.942$ at $m = 2$ to $0.997$ at $m = 1024$, setting the published baseline on all $100$ interior cells (Fig.~\ref{fig:app_grover_baselines}b).

\paragraph{Interior Find-All ($m, k \ge 2$; \textbf{GR0}).}
To let $m$ registers find distinct targets in parallel at $\sqrt{N/m}$ block scale, Grover and Radhakrishnan~\cite{grover2004parallel} randomly partition $\{0,\dots,N-1\}$ into $m$ blocks of size $N/m$. Each block receives an unknown target count $c$, which they clear via a \emph{descending staircase} of exact amplitude-amplification stages~\cite{brassard2002quantum} at assumed counts $s = t, t-1, \dots, 1$ (total depth $H_t^{(1/2)}\frac{\pi}{4}\sqrt{N/m}$), testing and unmarking hits after each stage. Counting \emph{down} guarantees emptying every block with $c \le t$: even if stages $s > c$ miss, once $s$ reaches a block's remaining count $c$, the assumed and actual counts match, finding a target with certainty and stepping $c \to c - 1$ in lockstep with $s$ (while mismatched stages $s \neq c$ still hit with probability $\sin^2(\frac{\pi}{2}\sqrt{c/s})$). Because Theorem~3 of ref.~\citenum{grover2004parallel} leaves tail cleanup at $\Theta(\cdot)$ order, we evaluate a concrete two-pass schedule (\textbf{GR0}): Pass~1 runs a staircase from the optimal integer cap $1 \le t_1^*(m, k) < k$ down to $1$, and Pass~2 clears all $R_1 = k - |\text{found}|$ remaining targets via a second staircase from $R_1$ down to $1$ (Fig.~\ref{fig:app_grover_baselines}c).

\subsection{Algorithmic mechanisms in verified circuits}
\label{app:grover_results}

Beyond the $k = 1$ boundary (\textbf{ZK}+\textbf{WG}), our interior baselines are single-paper protocols; Fig.~\ref{fig:app_grover_results}a separates \emph{automatic recombinations}, which join unchanged published subroutines by new classical control, from \emph{new coherent mechanisms}, which alter the quantum dynamics or measurements within a search stage, listing a few of the many mechanisms found per class. Since mechanisms interact, Fig.~\ref{fig:app_grover_baselines}a reports only joint speedups, while Fig.~\ref{fig:app_grover_results}b,c details the verified relative round costs and winning circuit regimes across $(m, k)$.

Unlike distributed baselines such as \textbf{GW} and \textbf{GR0} that run $m$ uncoupled registers, our shared-workspace model allows inter-slot entanglement: coupling all $m$ slots via odd-parity fanout in Find-Any extends the boundary constant $r_{\mathrm{WG}}$ to every interior $(m, k \ge 2)$, while Find-All entangles slot and storage registers.
In our strongest Find-All mechanism ($k \lesssim m$; Fig.~\ref{fig:app_grover_johnson}), three ideas avoid the bin-load variance of partitioned search: (i)~preparing a uniform superposition over $m$-element subsets $\ket{E}$ so one $\ket{-}$-marked call to $O_S^{\otimes m}$ (unmarking archived hits $F$) supplies all target information via the hit-parity phase $(-1)^{|E \cap (S\setminus F)|}$; (ii)~alternating that query with a fixed, $S$-independent reflection on the Johnson-scheme eigenspaces of $m$-subsets~\cite{Jeffery2016}---analogous to Grover's uniform diffusion step and compiled via quantum singular value transformation (QSVT)~\cite{gilyen2019qsvt}, plus a terminal phase kick, driving the hit count $w = |E \cap (S\setminus F)|$ coherently from $0$ to $r = |S\setminus F|$ in under $\sqrt{N/m}$ rounds; and (iii)~batch readout, where one final query writes all $w \in [w_{\mathrm{lo}}, r]$ hits into $\ket{F}$ at once, retrying or clearing any remainder via exact loops.
Although $S$-independent unitaries carry zero query cost in the black-box specification, our verified circuits use $O((m+k)n)$ qubits and each unitary step compiles into $\mathrm{poly}(n, m, k)$ elementary gates (within error $2^{-\Omega(n)}$ for the QSVT-compiled diffusion steps).

\begin{figure}[!tp]
\addtokomafont{caption}{\footnotesize}
\input{figures/figs_parallel_grover_results}
\caption{\textbf{Verified parallel Grover mechanisms and relative round costs across $m, k \in \{1, 2, 4, \dots, 1024\}$.} \textbf{a} Selected algorithmic mechanisms in our Lean-verified circuits (not exhaustive). \textbf{b, c} Verified asymptotic expected parallel rounds relative to $D_{\mathrm{std}}$ (Eq.~\eqref{eq:vanilla_scales}) for finding any one (\textbf{b}) or all $k$ (\textbf{c}) targets, on a logarithmic colour scale, with dashed contours and labels marking the winning mechanism in each region (cf.~panel \textbf{a}; \textbf{WG} and \textbf{ZK+WG} denote ties with the baseline).}
\label{fig:app_grover_results}
\end{figure}

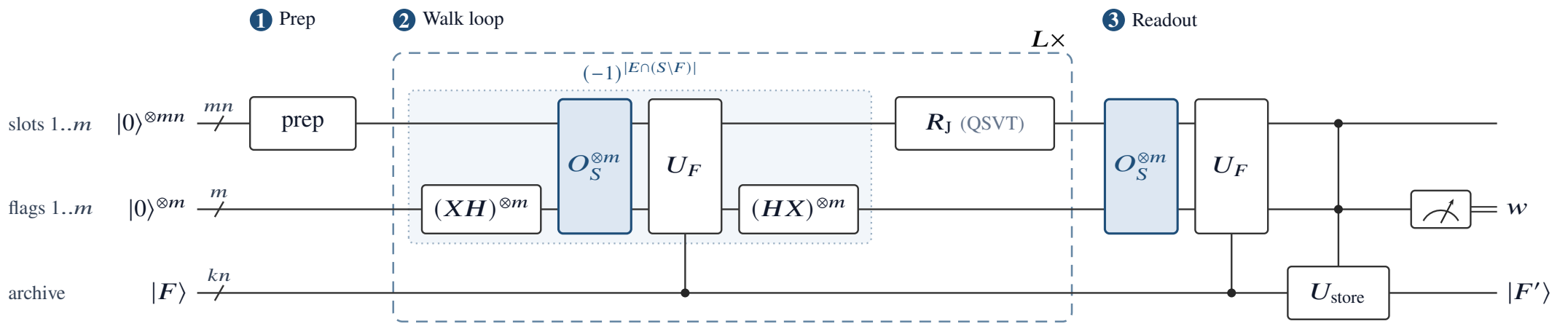
\begin{figure}[!tp]
\addtokomafont{caption}{\footnotesize}
\providecolor{frameblue}{HTML}{1E4E79}
\providecolor{accentblue}{HTML}{2B6CB0}
\providecolor{inkdark}{HTML}{0F172A}
\providecolor{inkbody}{HTML}{1E293B}
\providecolor{inklabel}{HTML}{334155}
\providecolor{inksub}{HTML}{64748B}
\protected\def\fgHdr{\fontsize{8.4}{10.0}\selectfont}
\protected\def\fgSix{\fontsize{6.2}{7.3}\selectfont}
\protected\def\fgTiny{\fontsize{5.1}{6.0}\selectfont}
\protected\def\fgMicro{\fontsize{4.7}{5.5}\selectfont}

\tikzset{
  fgphead/.style={font=\fgHdr, text=black, anchor=base west, inner sep=0pt},
  fgqstd/.style={fill=white, draw=black!80, text=black, line width=0.45pt, rounded corners=1.0pt, inner sep=2.3pt},
  fgqora/.style={fill=accentblue!16, draw=frameblue, text=frameblue, line width=0.58pt, rounded corners=1.2pt, inner sep=2.3pt},
  fgqmet/.style={fill=white, draw=black!80, line width=0.45pt, rounded corners=1.0pt},
}

\centering
\begin{tikzpicture}[x=1cm, y=1cm]
  \path[use as bounding box] (0.00, 0.32) rectangle (14.04, 3.32);

  \def\spBadge{\rmfamily\fontsize{5.50}{6.20}\selectfont\bfseries}%
  \def\spNumbadge#1#2#3{%
    \fill[frameblue] (#1,#2) circle [radius=0.102cm];
    \node[anchor=center, inner sep=0pt, font=\spBadge, text=white] at (#1,#2) {#3};
  }%

  \spNumbadge{2.32}{3.16}{1}
  \node[anchor=west, inner sep=0pt, font=\fgTiny, text=inkdark] at (2.48, 3.16)
    {Prep};

  \spNumbadge{3.62}{3.16}{2}
  \node[anchor=west, inner sep=0pt, font=\fgTiny, text=inkdark] at (3.78, 3.16)
    {Walk loop};

  \spNumbadge{10.06}{3.16}{3}
  \node[anchor=west, inner sep=0pt, font=\fgTiny, text=inkdark] at (10.22, 3.16)
    {Readout};

  \filldraw[fill=accentblue!6, draw=frameblue!45, densely dotted, line width=0.42pt, rounded corners=1.5pt]
    (3.66, 1.13) rectangle (7.86, 2.52);
  \node[anchor=center, inner sep=0pt, font=\fgTiny, text=frameblue] at (5.76, 2.68)
    {$(-1)^{|E\cap(S\setminus F)|}$};

  \draw[draw=frameblue!68, densely dashed, line width=0.48pt, rounded corners=2.4pt]
    (3.52, 0.42) rectangle (9.68, 2.86);
  \node[anchor=south east, inner sep=0pt, font=\fgSix, text=black] at (9.64, 2.91)
    {$L\times$};

  \node[anchor=west, inner sep=0pt, font=\fgTiny, text=inklabel] at (0.02, 2.22) {slots $1..m$};
  \node[anchor=east, inner sep=0pt, font=\fgSix,  text=inkdark]  at (1.66, 2.22) {$\ket{0}^{\otimes mn}$};

  \node[anchor=west, inner sep=0pt, font=\fgTiny, text=inklabel] at (0.02, 1.44) {flags $1..m$};
  \node[anchor=east, inner sep=0pt, font=\fgSix,  text=inkdark]  at (1.66, 1.44) {$\ket{0}^{\otimes m}$};

  \node[anchor=west, inner sep=0pt, font=\fgTiny, text=inklabel] at (0.02, 0.68) {archive};
  \node[anchor=east, inner sep=0pt, font=\fgSix,  text=inkdark]  at (1.66, 0.68) {$\ket{F}$};

  \draw[draw=black!80, line width=0.45pt] (1.74, 2.22) -- (13.54, 2.22);
  \draw[draw=black!80, line width=0.45pt] (1.74, 1.44) -- (12.76, 1.44);
  \draw[draw=black!80, line width=0.45pt] (1.74, 0.68) -- (13.54, 0.68);

  \draw[draw=black!80, line width=0.45pt] (1.90, 2.15) -- (1.98, 2.29);
  \node[anchor=south, inner sep=0pt, font=\fgMicro, text=inklabel] at (1.94, 2.32) {$mn$};

  \draw[draw=black!80, line width=0.45pt] (1.90, 1.37) -- (1.98, 1.51);
  \node[anchor=south, inner sep=0pt, font=\fgMicro, text=inklabel] at (1.94, 1.54) {$m$};

  \draw[draw=black!80, line width=0.45pt] (1.90, 0.61) -- (1.98, 0.75);
  \node[anchor=south, inner sep=0pt, font=\fgMicro, text=inklabel] at (1.94, 0.78) {$kn$};

  \filldraw[fgqstd] (2.22, 1.98) rectangle (3.18, 2.46);
  \node[anchor=center, inner sep=0pt, font=\fgSix, text=inkdark] at (2.70, 2.22) {prep};

  \filldraw[fgqstd] (3.78, 1.22) rectangle (4.86, 1.66);
  \node[anchor=center, inner sep=0pt, font=\fgSix, text=inkdark] at (4.32, 1.44) {$(XH)^{\otimes m}$};

  \filldraw[fgqora] (5.02, 1.22) rectangle (5.68, 2.44);
  \node[anchor=center, inner sep=0pt, font=\fgSix, text=frameblue] at (5.35, 1.83) {$O_S^{\otimes m}$};

  \draw[draw=black!80, line width=0.45pt] (6.17, 1.22) -- (6.17, 0.68);
  \fill[black!85] (6.17, 0.68) circle [radius=0.040cm];
  \filldraw[fgqstd] (5.84, 1.22) rectangle (6.50, 2.44);
  \node[anchor=center, inner sep=0pt, font=\fgSix, text=inkdark] at (6.17, 1.83) {$U_F$};

  \filldraw[fgqstd] (6.66, 1.22) rectangle (7.74, 1.66);
  \node[anchor=center, inner sep=0pt, font=\fgSix, text=inkdark] at (7.20, 1.44) {$(HX)^{\otimes m}$};

  \filldraw[fgqstd] (8.08, 1.98) rectangle (9.52, 2.46);
  \node[anchor=center, inner sep=0pt, font=\fgSix, text=inkdark] at (8.80, 2.22)
    {$R_{\mathrm{J}}$\;{\fgTiny\color{inksub}(QSVT)}};

  \filldraw[fgqora] (9.98, 1.22) rectangle (10.64, 2.44);
  \node[anchor=center, inner sep=0pt, font=\fgSix, text=frameblue] at (10.31, 1.83) {$O_S^{\otimes m}$};

  \draw[draw=black!80, line width=0.45pt] (11.13, 1.22) -- (11.13, 0.68);
  \fill[black!85] (11.13, 0.68) circle [radius=0.040cm];
  \filldraw[fgqstd] (10.80, 1.22) rectangle (11.46, 2.44);
  \node[anchor=center, inner sep=0pt, font=\fgSix, text=inkdark] at (11.13, 1.83) {$U_F$};

  \draw[draw=black!80, line width=0.45pt] (12.10, 0.92) -- (12.10, 2.22);
  \fill[black!85] (12.10, 2.22) circle [radius=0.040cm];
  \fill[black!85] (12.10, 1.44) circle [radius=0.040cm];
  \filldraw[fgqstd] (11.64, 0.44) rectangle (12.56, 0.92);
  \node[anchor=center, inner sep=0pt, font=\fgSix, text=inkdark] at (12.10, 0.68) {$U_{\mathrm{store}}$};

  \filldraw[fgqmet] (12.76, 1.27) rectangle (13.30, 1.61);
  \draw[draw=black!82, line width=0.42pt]
    (12.871, 1.335) arc [start angle=152, end angle=28, radius=0.180cm];
  \draw[-{Stealth[length=1.6pt, width=1.3pt]}, draw=black!85, line width=0.42pt]
    (13.03, 1.315) -- (13.145, 1.535);
  \draw[draw=black!80, line width=0.40pt] (13.30, 1.462) -- (13.54, 1.462);
  \draw[draw=black!80, line width=0.40pt] (13.30, 1.418) -- (13.54, 1.418);

  \node[anchor=west, inner sep=0pt, font=\fgSix, text=inkdark] at (13.62, 1.44) {$w$};
  \node[anchor=west, inner sep=0pt, font=\fgSix, text=inkdark] at (13.62, 0.68) {$\ket{F'}$};
\end{tikzpicture}
\caption{\textbf{High-level circuit for the collective Johnson-graph walk.} Registers hold $m$ address slots ($\ket{0}^{\otimes mn}$), $m$ flag qubits ($\ket{0}^{\otimes m}$), and a $kn$-qubit archive $\ket{F}$ of found targets. (1)~\textrm{prep} prepares the uniform superposition $\binom{N}{m}^{-1/2}\sum_{|E|=m}\ket{E}$ over $m$-subsets $E \subseteq \{0,\dots,N-1\}$. (2)~For $L < \sqrt{N/m}$ rounds ($L\times$), $(XH)^{\otimes m}$ ($X$ then $H$) and $(HX)^{\otimes m}$ switch the flags between $\ket{0}^{\otimes m}$ and $\ket{-}^{\otimes m}$; $O_S^{\otimes m}$ is the parallel marking oracle; $U_F$ unmarks slots already in $F$, so the dotted block applies $(-1)^{|E\cap(S\setminus F)|}$; and $R_{\mathrm{J}}$ is an $S$-independent Johnson-graph reflection compiled via QSVT. (3)~After one call to $O_S^{\otimes m}$ and $U_F$, $U_{\mathrm{store}}$ records new hits into $\ket{F'}$ and measuring the flags outputs the hit count $w = |E\cap(S\setminus F)|$.}
\label{fig:app_grover_johnson}
\end{figure}

\section{Fixed-Point \texorpdfstring{$\arcsin$}{arcsin}}
\label{app:arcsin}
\subsection{Problem and specification}
Many fault-tolerant quantum algorithms require converting amplitudes held in a quantum register into single-qubit rotation angles, including probability-distribution loading and state preparation~\cite{grover2002creating}, eigenvalue inversion in quantum linear-system solvers~\cite{harrow2009quantum}, and quantum rejection and Metropolis sampling~\cite{temme2011quantum,ozols2013rejection}. Implementing these rotations via phase kickback or controlled Pauli gates first evaluates $\arcsin(y)$ reversibly on a fixed-point register, a subroutine that often dominates the non-Clifford gate count~\cite{haner2018optimizing,sanders2019blackbox}. 

We consider reversible evaluation $\ket{y}\ket{b}\ket{0}^{\otimes m} \mapsto \ket{y}\ket{(b + z(y)) \bmod 2^{k}}\ket{0}^{\otimes m}$ on $k = 25$ signed two's-complement registers, where $y$ is the input register, $b$ is the output register, and $z(y)$ approximates $\arcsin(y)$. 
Because $\frac{\mathrm{d}}{\mathrm{d}y}\arcsin(y) = (1-y^2)^{-1/2}$ diverges as $|y|\to 1$, uniform polynomial approximations require high degree near the endpoints, while iterative shift-and-add methods accumulate deep chains of reversible multiplications~\cite{haner2018optimizing,burge2024cordic}.
Our signed two's complement representation ($1$ sign, $1$ integer and $23$ fractional bits) matches the \texttt{float32} mantissa size. With $23$ fractional bits, the smallest representable value is $2^{-23}\simeq 1.19\times 10^{-7}$. Allowing for some small rounding error, we require a maximum absolute error $|z(y) - \arcsin(y)| \le 1.25\times 10^{-7}$ across all $2^{24}+1 = 16{,}777{,}217$ inputs $y \in [-1, 1]$. Using $m$ clean ancillas, a candidate program in Clifford+$T$ gates and mid-circuit measurements must apply this exact action across all $2^{50}$ basis states $(y, b)$ and return all ancillas to $\ket{0}^{\otimes m}$.

\begin{figure}[t]
    \centering
    \input{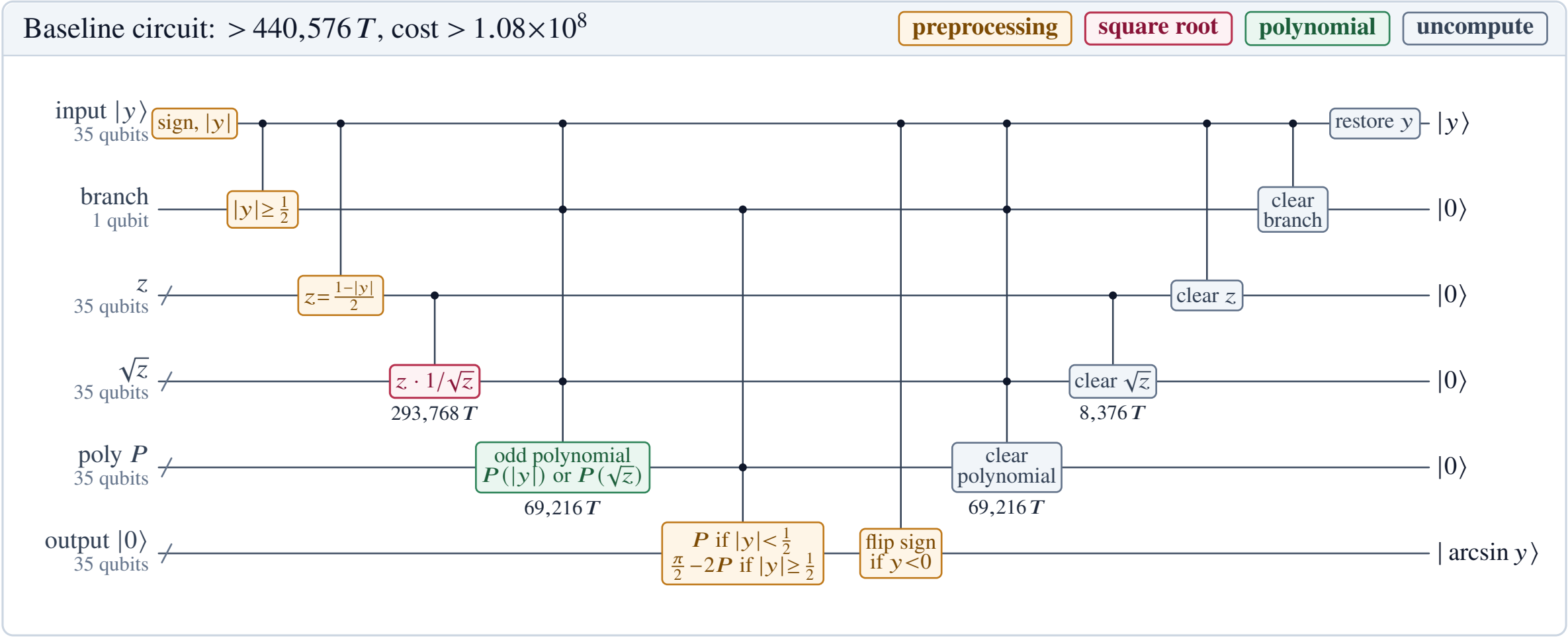}
    \caption{Baseline circuit to implement the $\arcsin$ function~\cite{haner2018optimizing}, using $n=35$, $m=3$, $d=3$. 
    After restricting to the positive range, the approximation splits the calculation of $\arcsin$ between a degree-$3$ polynomial in the interval $[0,0.5]$ and a Newton iteration for the square root function in $[0.5,1]$. Number of $T$ gates and number of qubits calculated as in their Appendix, for reversible computation.}
    \label{fig:app_arcsin_baseline}
\end{figure}

\subsection{Baseline}
We consider as baseline the algorithm by H\"aner et al.~\cite{haner2018optimizing}, which uses a piecewise-polynomial approach.\footnote{This approximation is stronger than the quantum CORDIC paper~\cite{burge2024cordic}. While their paper does not report results for our input type, we ran the code from their github repository and got a maximum error of $9\times 10^{-3}$.}
We compare against their $(n, m, d) = (35, 3, 3)$ fixed-point two's-complement configuration, whose maximum error $\sim 10^{-6}$ is driven by the error in the square root computation (Fig.~6 of ref.~\citenum{haner2018optimizing}), and
larger than our target bound of $1.25\times 10^{-7}$, thus providing a lower bound on the spacetime cost. 
As shown in Fig.~\ref{fig:app_arcsin_baseline}, this baseline splits the computation of $\arcsin$ on the interval $\left[-1, 1\right]$ between a degree-$3$ polynomial evaluation on $\left[-0.5, 0.5\right]$ and a Newton iteration for $\sqrt{z}=z \cdot 1/\sqrt{z}$.%
\footnote{Our baseline requires $\sqrt{z} = z / \sqrt{z}$. A recent circuit~\cite{khattar2026leveragingverifiablestructurequantum} computes $1/\sqrt{x}$ to $2^{-24}$ relative error in $7{,}735$ Toffoli gates and $260$ qubits. Using this here would slightly increase the total number of qubits but significantly reduce the number of gates. The number of gates for the square-root could be cut by a factor of up to $10$, and the baseline~\cite{haner2018optimizing} cost by up to $3\times$.}
As their Appendix~D describes, the former can be implemented using $140$ qubits and requires $17{,}304$ Toffoli gates and the same amount for reversing the computation, for a total of $34{,}608$ Toffoli gates ($138{,}432$ $T$ gates).
The latter can be implemented using $245$ qubits and requires $73{,}442$ Toffoli gates to implement it as the product of $z$ with $1/\sqrt{z}$, as described in their Appendix~C.2, and an additional $2{,}094$ Toffoli gates to also reverse the product operation, being $1/\sqrt{z}$ already a reversible operation, for a total of $75{,}536$ Toffolis ($302{,}144$ $T$ gates).

We calculate a lower bound of the full circuit spacetime cost of $1.08\times 10^{8}$ by using the maximum number of required qubits in the two subcircuits, $245$, and the sum of the $T$ gates required for them, $440{,}576$ $T$ gates. We thus ignore persistent registers and the smaller branch-selection and combination blocks, which would further increase the cost. 

\subsection{Program Discovery}
While our library lets us represent arbitrary gates and formally verify the cost of algorithms, the \emph{kickmix} model~\cite{babbush2026kickmix} restricts circuits to classical reversible gates ($X$, $\mathrm{CNOT}$, Toffoli and $\mathrm{CSWAP}$), Pauli-$X$ measurement-based uncomputation~\cite{gidney2018halving}, and diagonal $Z$, $CZ$ and $CCZ$ phase corrections so that every basis input stays a single signed basis state $(-1)^{\phi}\ket{x}$ along each measurement branch, enabling fast classical fuzz-testing without statevector simulation. The fact that agentic systems can produce compact arithmetic quantum circuits by using the feedback from kickmix testing is already known~\cite{gu2026discovering, khattar2026leveragingverifiablestructurequantum}, and has also been applied to a related $\arcsin(c/X)$ subroutine, but on a restricted input range rather than $[-1, 1]$. We exploited this speed by first running our multi-agent system to search for circuits using kickmix. The result was a circuit of spacetime cost $498{,}168$ ($102$ qubits, $4{,}884$ $T$ gates).

\begin{figure}
    \centering
    \input{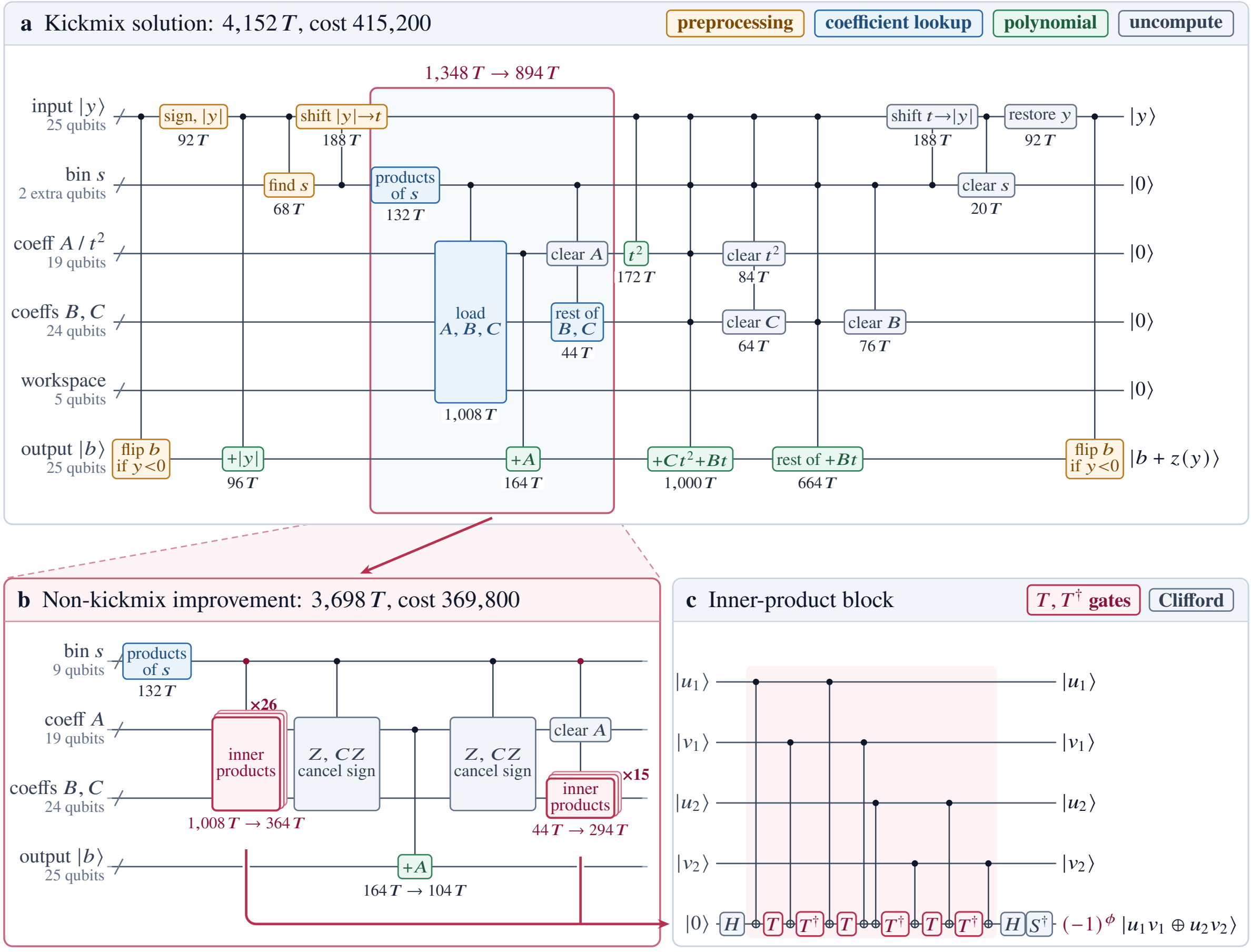}
    \caption{Sketch of the algorithm to implement the fixed-point $\arcsin$ module. 
    (a) Full kickmix circuit. It uses $50$ data qubits ($|y\rangle$ and $|b\rangle$, $25$ each) and $50$ ancilla qubits reused across stages, and a total of $4{,}152$ $T$ gates. 
    The bin index $s$ has $9$ bits but needs only $2$ extra ancillas. After "shift $|y|\to t$", $6$ bits of $s$ are simply the top $6$ bits of $t$, which identify the bin within its group. A $7$th bit replaces the $\tfrac12$ bit of $|y|$, which can be recomputed from $s$ and is erased. The remaining $2$ bits use one ancilla each: one is stored directly, and the other replaces the top bit of $|y\rangle$, whose value is moved into an ancilla. 
    $t^2$ is computed into the ancillas freed by clearing $A$. At peak, the ancillas hold $A$ ($19$), $B$ and $C$ ($24$), $5$ lookup work qubits, $1$ bit of $s$, and the moved bit of $|y\rangle$.
    (b) Improved block in the non-kickmix circuit. 
    Algorithm verification with our formal quantum computation library allows further optimization of the circuit.
    Specifically, the coefficient lookup block can be replaced with a more efficient block, saving overall $454$ $T$ gates. Each coefficient bit is computed by one inner-product block taking a variable number $r \in \{2, \dots, 7\}$ of product pairs and using $2r + 2$ $T$ gates. We have $26$ blocks totaling $364$ $T$ gates before $+A$, and $15$ blocks totaling $230$ $T$ gates on the freed qubits after clearing $A$. Each inner-product block leaves an extra sign phase that is then removed before returning the result. The whole circuit requires now $3{,}698$ $T$ gates instead of $4{,}152$ $T$ gates.
    (c) Example of the optimized inner-product block for two product pairs, using $6$ $T$ gates, which could not be expressed in the kickmix framework.}
    \label{fig:app_arcsin_circuit}
\end{figure}

However, the kickmix gate set is limited, and we expect that the full Clifford+$T$ gate set could allow for a more efficient implementation.
To obtain more general algorithms that are not limited by the kickmix restrictions, we need a different way to verify the circuit, and this is where our formal library helps.
Starting from the algorithm obtained with kickmix, we instructed agents to start using our quantum algorithm specification and verification framework to further optimize the algorithm. For a while, agents were able to continue optimizing the circuit up to a cost of $415{,}200$ ($100$ qubits, $4{,}152$ $T$ gates, Fig.~\ref{fig:app_arcsin_circuit}a and Table~\ref{tab:app_arcsin_summary}), which continued using a restricted set of gates, and \textit{could} still be translated to kickmix. Subsequently, the agents truly took advantage of our library and produced circuits that \textit{could not} be expressed and efficiently simulated with kickmix (Fig.~\ref{fig:app_arcsin_circuit}b,c). 
This drove the cost down by a further $11\%$ to $369{,}800$ ($100$ qubits, $3{,}698$ $T$ gates), $290\times$ below the baseline~\cite{haner2018optimizing} (Fig.~\ref{fig:results}a,c).

In other words, for this problem, our agents using kickmix can do an efficient search in the restricted subspace without the overhead of a formal proof. Afterwards, our agents using formal verification lift the gate-set restriction and allow further optimization that would not be easily testable classically. Every circuit that can be simulated with kickmix can also be proven in Lean, and seeding a Lean discovery agent from a kickmix solution further improves the solution.

\begin{table}[t]
\addtokomafont{caption}{\footnotesize}
\caption{\textbf{Summary for fixed-point $\arcsin(y)$ on $[-1, 1]$ at $k = 25$.} 
The baseline evaluates the $(n, m, d) = (35, 3, 3)$ configuration~\cite{haner2018optimizing}, assuming $4\,T$ per Toffoli~\cite{gidney2018halving,jones2013toffoli}. 
We derive a lower bound on its spacetime cost by summing the two dominant reversible submodules on the $245$-qubit square-root workspace while neglecting persistent registers and the steps that combine the two parts. In the non-kickmix Lean~4 circuit, the $3{,}698$ $T$ gates comprise $776$ four-$T$ temporary-$\mathrm{AND}$ gates ($3{,}104\,T$) plus $594$ single-qubit $T$/$T^{\dagger}$ gates across $41$ inner-product gadgets.}
\label{tab:app_arcsin_summary}
\vspace{2pt}
{\fontsize{7.5}{9.2}\selectfont
\setlength{\tabcolsep}{3.2pt}%
\renewcommand{\arraystretch}{1.18}%
\begin{tabular*}{\textwidth}{@{\extracolsep{\fill}}llrrrr@{}}
\toprule
\textbf{Method} & $\boldsymbol{\varepsilon}$ & \textbf{Toffoli} & \textbf{$\boldsymbol{\mathcal{T}}$} & \textbf{$Q$} & \textbf{Spacetime cost ($Q\cdot \mathcal{T}$)} \\
\midrule
Baseline~\cite{haner2018optimizing} (lower bound) & $\sim 10^{-6}$ & $110{,}144$ & $440{,}576$ & $245$ & $1.08 \times 10^{8}$ \\
--  $\arcsin$ on $[-0.5, 0.5]$ & & $34{,}608$ & $138{,}432$ & ---& --- \\
--  $1/\sqrt{z}$ & & $75{,}536$ & $302{,}144$ & --- & --- \\
\midrule
Kickmix & $1.25 \times 10^{-7}$ & $1{,}038$ & $4{,}152$ & $100$ & $415{,}200$ \\
\textbf{Non-kickmix (Lean~4)} & $1.25 \times 10^{-7}$ & $776$ & $3{,}698$ & $100$ & $\mathbf{369{,}800}$ \\
\bottomrule
\end{tabular*}}
\end{table}

\subsection{Interpretation}
Fig.~\ref{fig:app_arcsin_circuit} provides an interpretation of the algorithms that we obtained, and shows where the optimization using the Lean verifier obtained an improvement.

Since $\arcsin$ is an odd function, i.e. $z(-y)=-z(y)$, we can perform the calculation only for positive inputs, and flip the output register according to the sign. 
Subsequently, we also split off the linear part, $z(|y|) = |y| + r(|y|)$ and approximate the remainder $r(|y|) \approx \arcsin(|y|) - |y|$  using a separate quadratic polynomial on $440$ intervals of $[0, 1]$.
Each interval is chosen to have a width roughly proportional to its distance to $1$, so on each interval the square root is smooth, a second order polynomial fits well, and the fit error shrinks when getting closer to $1$.
The intervals get narrower toward $|y| = 1$, where $\arcsin(|y|) \approx \frac{\pi}{2} - \sqrt{2(1 - |y|)}$ grows steeply.  
The closest input to $1$ on the fixed-point grid is $2^{-23}$ away, so finitely many intervals cover all inputs. The point $|y| = 1$ is handled separately.
Subsequently, the circuit looks for the interval that contains $|y|$ and stores its index in a $9$-bit register $s$. 
It does so by using a barrel shift, i.e. it counts the leading zeros of $1 - |y|$, and shifts the magnitude bits left using controlled swaps to form the local offset $t$. 
A lookup block then computes the interval's coefficients $A$, $B$, $C$ for $s$, and squarer and adder circuits add $A + Bt + Ct^2$ to $b$.
As an optimization, the $19$ high bits of $A$ are added to $b$ and erased so that the remaining coefficient bits can be computed on the same qubits. Similarly, $C$ and $t^2$ are erased midway so their qubits can be reused for the rest of $+Bt$. 
Measurements with $Z$ and $CZ$ corrections erase the coefficients and all other scratch qubits at zero $T$ cost. 
Finally, the circuit is made reversible by undoing the shift and the sign conversion, restoring $y$ and returning all ancillas to $|0\rangle$.

As shown in Fig.~\ref{fig:app_arcsin_circuit}, a large amount of the total kickmix $T$ gates are spent on the coefficient lookup phase. 
Specifically, each coefficient bit is a fixed function of the interval number $s$, given by a table with one entry per interval. 
Any such table can be written as an XOR of products of bits of $s$, e.g. $s_1 s_4 \oplus s_2 s_3 s_7 \oplus s_5$. 
The kickmix circuit first builds a few products that many coefficient bits share, and then builds each coefficient bit by XOR-ing in the products it needs using a $4\,T$ AND circuit for each product.

The non-kickmix circuit replaces most of these bitwise product blocks with $41$ inner-product circuits that need roughly half as many $T$ gates. 
Indeed, each block puts its coefficient qubit in an equal superposition of $0$ and $1$ states, writes the products as a relative phase between the two branches using $T$ and $T^\dagger$ gates, and then makes the branches interfere to get the appropriate bit.
However, each inner-product circuit leaves a sign $\pm 1$ that depends on the specific $s$, which is cancelled across each lookup pass with $Z$ and $CZ$ gates at zero $T$ cost.  
Because each inner product writes directly into its target coefficient qubit and needs no extra scratch qubits, the circuit can postpone more coefficient bits of $B$ and $C$ until after $A$ has been added to $b$ and cleared, leaving more spare qubits for that adder and reducing the lookup-and-add stage from $1{,}348$ to $894$ $T$ gates. 
Together, these changes lower the full circuit from $4{,}152$ to $3{,}698$ $T$ gates on the same $100$ qubits.

\section{Stopping Power in Inertial-confinement Fusion}
\label{app:physics}
\label{app:fusion}
\subsection{Hamiltonian and specification}
\label{app:fusion_spec}

Following Eqs.~(1)--(7) and Tables~1--2 of ref.~\citenum{rubin2024stopping}, the system consists of $\eta = 218$ electrons ($\zeta_e = -1$, mass $1$) and one $\alpha$-particle projectile ($\zeta_{\mathrm{proj}} = +2$, mass $M_{\mathrm{proj}} = 7{,}294.29954171\,m_e$, matching $3.9726$ proton masses in Table~2 of ref.~\citenum{rubin2024stopping}) in a periodic cubic cell of volume $\Omega = 2{,}419.68282\;a_0^3$ containing $216$ fixed protons at arbitrary positions $R_\ell \in \mathbb{R}^3$. In plane-wave first quantization, each electron momentum coordinate is stored in $n_p = 6$ signed-magnitude qubits and each projectile coordinate in $n_n = 8$ signed-magnitude qubits~\cite{su2021firstquant,rubin2024stopping}, giving $Q_{\mathrm{sys}} = 3(\eta n_p + n_n) = 3{,}948$ system qubits. The valid grid subspace excludes the signed-magnitude negative zero $\ket{1,0\cdots 0}$ from every coordinate, leaving integer momenta in $G = [-31, 31]^3$ for each electron ($63^3$ plane waves, covering the $53^3$ classical time-dependent density-functional grid of Table~1 in ref.~\citenum{rubin2024stopping}) and $\tilde{G} = [-127, 127]^3$ for the projectile, at wavevectors $k_\nu = 2\pi \nu / \Omega^{1/3}$. The projectile's momenta are taken relative to its initial mean momentum $k_{\mathrm{proj}}$, with $k_{\mathrm{proj}}\Omega^{1/3}/(2\pi) \in \mathbb{Z}^3$ for compatibility with the cell's periodicity and $\norm{k_{\mathrm{proj}}}_2 \le 29{,}376$~a.u.\ ($v_\alpha \approx 4$~a.u.).

On valid grid states, the discretized Hamiltonian in atomic units is
\begin{equation}
\begin{aligned}
  H &= \sum_{i=1}^{\eta}\sum_{p\in G} \frac{\norm{k_p}_2^2}{2}\ket{p}\!\bra{p}_i
  + \sum_{p\in\tilde{G}} \frac{\norm{k_p - k_{\mathrm{proj}}}_2^2}{2M_{\mathrm{proj}}}\ket{p}\!\bra{p}_{\mathrm{proj}}
  + \sum_a \sum_{\substack{p,q\in G_a \\ p\neq q}} \zeta_a S(p-q) K(p-q)\ket{p}\!\bra{q}_a \\
  &\quad + \frac{1}{2}\sum_{a\neq b}\sum_{\nu\neq 0}\sum_{\substack{p\in G_a,\, q\in G_b \\ p+\nu\in G_a,\, q-\nu\in G_b}} \zeta_a\zeta_b K(\nu)\ket{p+\nu}\!\bra{p}_a\ket{q-\nu}\!\bra{q}_b,
\end{aligned}
\label{eq:sp_ham}
\end{equation}
and acts as zero outside the valid grid subspace. Here $a$ and $b$ range over the $\eta$ electrons ($G_a = G$, $\zeta_a = -1$) and the projectile ($G_{\mathrm{proj}} = \tilde{G}$, $\zeta_{\mathrm{proj}} = +2$), $K(\nu) = 4\pi / (\Omega\norm{k_\nu}_2^2)$, and $S(\nu) = \sum_{\ell=1}^{216} e^{-i k_\nu \cdot R_\ell}$.

For every proton configuration $\{R_\ell\}$, initial projectile momentum $k_{\mathrm{proj}}$, evolution time $t \in \{1,\dots,10\}$~a.u., and partition of the $218$ electron registers into spin-up and spin-down subsets, a candidate circuit over Clifford+$T$ and mid-circuit measurement using $m$ ancilla qubits must realize a single unitary $W$ that matches $e^{-iHt}$ up to a global phase $\varphi$, satisfying
\begin{equation}
  \norm{W\bigl(\ket{\psi}\otimes\ket{0}^{\otimes m}\bigr) - e^{i\varphi}\bigl(e^{-iHt}\ket{\psi}\bigr)\otimes\ket{0}^{\otimes m}}_2 \;\le\; \varepsilon
  \label{eq:ham_sim}
\end{equation}
at $\varepsilon = 10^{-2}$ for every normalized grid state $\ket{\psi}$ antisymmetric under exchange of same-spin electrons. All ancillas begin in $\ket{0}^{\otimes m}$ in each run, rather than carrying pre-prepared resource states (such as phase-gradient states) across runs. Every circuit is compiled into Clifford+$T$ gates, mid-circuit measurements and classical feed-forward, and checked at that gate level by the correctness proof. Where later gates are conditioned on measurement outcomes, we count the $T$ gates of the most expensive outcome branch. 

To extract the stopping power from the projectile's rate of energy loss, the protocol repeats time evolution $50$ times at each of $t = 1,\dots,10$~a.u.\ ($500$ runs) to estimate the mean projectile kinetic energy, omitting state preparation and readout as their cost is negligible. In this work, our Lean specification asks for the algorithm to prove correctness and cost for all $500$ circuits. However, in Sec.~\ref{sec:fusion} and Fig.~\ref{fig:applications} we report the average $T$ count over the circuits, since this more accurately reflects the size of circuit a fault-tolerant quantum computer would need to coherently execute. We apply this to both the baseline and our solution, so the factor of improvement reported is equal regardless. However, the baseline costing convention~\cite{rubin2024stopping} (total over all circuits) can be obtained by simply multiplying our arithmetic mean by 500.

\subsection{Baseline}
\label{app:fusion_baseline}

Table~2 of ref.~\citenum{rubin2024stopping} gives a momentum-space LCU 1-norm of $\lambda = 1{,}744{,}784.42$~Ha, a block-encoding cost of $C_{\mathrm{BE}} = 2.498\times 10^{4}$ Toffolis per walk step, and $Q_{\mathrm{base}} = 5{,}650$ logical qubits ($3{,}948$ system and $1{,}702$ ancilla qubits). Each electron--proton or electron--electron pair contributes $\sum_{\nu \in [-62,62]^3\setminus\{0\}} (\pi\Omega^{1/3}\norm{\nu}_2^2)^{-1} = 22.53$~Ha to $\lambda$, giving $1.06\times 10^{6}$~Ha from electron--proton attraction and $5.33\times 10^{5}$~Ha from electron--electron repulsion ($91.4\%$ of $\lambda$).

Generalized QSP~\cite{motlagh2023gqsp} implements $e^{-iHt}$ as a polynomial in a qubitized walk operator~\cite{low2019qubitization} that block-encodes $H/\lambda$, with one walk step per query. Evaluating the query count~\cite{berry2024doubling} in Eq.~(12) of ref.~\citenum{rubin2024stopping},
\begin{equation}
  N_{\mathrm{steps}}(t) = \lambda t + \frac{3^{2/3}}{2}(\lambda t)^{1/3}\log_2^{2/3}(1/\varepsilon),
  \label{eq:sp_steps}
\end{equation}
at $\varepsilon = 10^{-2}$ and summing over the $50$ runs at each of $t = 1,\dots,10$~a.u.\ gives $4.80\times 10^{9}$ walk steps, or $1.20\times 10^{14}$ Toffolis over all $500$ runs.
This is an average of $9.60\times 10^{6}$ walk steps and $N_{\mathrm{Toff}}^{\mathrm{base}} = 2.40\times 10^{11}$ Toffolis per run, so $\mathcal{T}_{\mathrm{base}} = 4 N_{\mathrm{Toff}}^{\mathrm{base}} = 9.59\times 10^{11}$ $T$ gates per run and $Q_{\mathrm{base}}\cdot \mathcal{T}_{\mathrm{base}} = 5.42\times 10^{15}$ per run ($2.71\times 10^{18}$ over $500$ runs), the number presented in Fig.~\ref{fig:results}a. The same study~\cite{rubin2024stopping} also quotes an eighth-order product-formula estimate ($1.40\times 10^{15}$ Toffolis on $6{,}170$ qubits over $500$ runs), which we do not use as a baseline since it needs $11.7\times$ more Toffolis than the QSP estimate above.

\subsection{Main improvements}
\label{app:fusion_improvements}

Section~\ref{sec:fusion} summarizes the two changes that account for most of the reduction (Fig.~\ref{fig:applications}b,c). Here we describe each one alongside the baseline construction it replaces, and Appendix~\ref{app:fusion_resources} gives the resulting resource counts.

The baseline applies the electron--proton and electron--electron terms in momentum space, where each pair contributes $22.53$~Ha to $\lambda$ (Appendix~\ref{app:fusion_baseline}). Our solution instead Fourier transforms all electrons to a real-space grid each step to apply the electron--proton, electron--projectile and electron--electron interactions, while the kinetic and projectile--proton terms stay in momentum space. This basis change alone halves the contribution of each pair to $\lambda$, to $11.3$~Ha, but that reduction is canceled out by the extra cost of applying the walk step in real space. However, a further reduction follows by being in the position basis, where the solution can exploit Pauli exclusion. A subroutine checks every electron against each grid point near a given electron or proton, and each electron found there kicks back a phase of $i$ in one LCU branch and $-i$ in the other. Averaging the two branches gives $\cos(\pi n/2)$ for $n$ electrons on the point, which equals $1-n$ for $n \le 2$ and has magnitude at most one for any $n$. The Lean solution also proves that states antisymmetric under exchange of same-spin electrons have at most two electrons on any grid point, allowing it to satisfy the specification while not faithfully applying the $n > 2$ interaction terms. Therefore, the circuit reproduces the interaction up to a constant, while each such term adds its potential to $\lambda$ once rather than once for every electron that could occupy the point. At separations beyond $\approx 0.6\;a_0$, where the potential is below $1.4$~Ha, the interaction is too weak for the savings to outweigh the cost of the subroutine. The pair potential is therefore range-separated at $1.4$~Ha, with the subroutine applied only to the short-range part and the long-range part summed over all pairs, each contributing $0.74$~Ha (Fig.~\ref{fig:applications}b).

The second most significant improvement found, in terms of cost reduction, relates to the electron--proton interaction. Protons do not sit on grid points, so to apply this term the circuit shifts all electrons by $-R_\ell$ to put each proton on a grid point, and then shifts them back. In the plane-wave basis this shift is a phase set by the total electron momentum $P$. Earlier real-space solutions computed and uncomputed $P$ around each shift, summing all $218$ electron momenta four times per walk step, which made up about a fifth of the $T$ gates per step. Only the protons and the projectile change $P$, and both act in the real-space step, so the final circuit keeps $P$ in a $42$-qubit register between walk steps, clearing it before the real-space step and rebuilding it after. This halves the sums and cuts the $T$ gates per step by $10\%$ (Fig.~\ref{fig:applications}c).

\subsection{Resource counts}
\label{app:fusion_resources}

Our solution uses $Q = 4{,}338$ logical qubits: $3{,}948$ system qubits and $390$ clean ancillas, against the baseline's $1{,}702$. The ancillas are a $42$-qubit register that keeps the total electron momentum between walk steps, a $48$-qubit phase-gradient register prepared in each run, $87$ control and data qubits for the block encoding, and $213$ workspace qubits that every subroutine borrows and returns to $\ket{0}$. The improvements above lower $\lambda$ to $1.356\times 10^{5}$~Ha. Each run of length $t$ is split into $t$ segments of $1$~a.u., each implemented by generalized QSP~\cite{motlagh2023gqsp,berry2024doubling} of degree $135{,}867$, giving an average of $7.47\times 10^{5}$ walk steps per run.

Each walk step uses roughly $141{,}500$ $T$ gates. One-time initialization adds $59{,}600$ $T$ gates per run. Averaged over $t = 1,\dots,10$~a.u., this gives $1.06\times 10^{11}$ $T$ gates and $Q\cdot \mathcal{T} = 4.59\times 10^{14}$ per run, or $229{,}357{,}900{,}876{,}653{,}900$ over all $500$ runs. Compared with the baseline, this is $12.8\times$ fewer walk steps and $1.30\times$ fewer qubits at $1.42\times$ the $T$ gates per step, an $11.8\times$ lower cost (Fig.~\ref{fig:applications}a).

\section{\texorpdfstring{Non-Born--Oppenheimer Dynamics of $\mathrm{C_{23}H_{20}N_3O}$}{Non-Born-Oppenheimer Dynamics of C23H20N3O}}
\label{app:nonbo}
\subsection{Hamiltonian and specification}
\label{app:nonbo_spec}

Following Reaction~5 in Eqs.~(2)--(10) and~(23)--(25) and Tables~I and~IV of ref.~\citenum{pocrnic2026prebo}, the system consists of $\eta = 234$ particles of composition $\mathrm{C_{23}H_{20}N_3O}$: $47$ nuclei ($23\;{}^{12}\mathrm{C}$, $20\;{}^{1}\mathrm{H}$, $3\;{}^{14}\mathrm{N}$, and $1\;{}^{16}\mathrm{O}$, with charges $\zeta_i = Z_i \in \{6, 1, 7, 8\}$) followed by $187$ electrons ($\zeta_i = -1$). Each Cartesian coordinate is stored as a signed $n_g = 8$-bit integer $q_{i,w} \in [-128, 127]$ on $Q_{\mathrm{sys}} = 3\eta n_g = 5{,}616$ system qubits, representing position $r_{i,w} = q_{i,w}\Delta$ in a cubic box of width $L = 44$~\AA\ ($83.15\;a_0$) with grid spacing $\Delta = L / (2^{n_g} - 1) \approx 0.326\;a_0$. Electron masses are $m_i = 1$, and nuclear masses $m_i$ (not stated in the baseline~\cite{pocrnic2026prebo}) are set in electron-mass units from the $^{1}\mathrm{H}$, $^{12}\mathrm{C}$, $^{14}\mathrm{N}$, and $^{16}\mathrm{O}$ isotopic atomic masses minus $Z_i m_e$.

The discretized non-Born--Oppenheimer Hamiltonian is $H = T + V$, with
\begin{equation}
  T = \frac{2\pi^2}{L^2}\sum_{i=1}^{\eta}\frac{1}{m_i}\,\mathrm{QFT}_i^\dagger\Biggl(\sum_{k\in G}\norm{k}_2^2\,\ket{k}\!\bra{k}_i\Biggr)\mathrm{QFT}_i, \qquad
  V = \frac{1}{2}\sum_{i\neq j}\frac{\zeta_i\zeta_j}{\Delta\max\!\bigl(\norm{q_i - q_j}_2, \tilde{\Gamma}_{ij}\bigr)},
  \label{eq:nonbo_ham}
\end{equation}
where $G = [-128,127]^3$, $\mathrm{QFT}_i$ is the quantum Fourier transform on the three coordinates of particle $i$, and the squared Coulomb saturation distance is $\tilde{\Gamma}_{ij}^2 = 8$ when both $i$ and $j$ are nuclei ($n_\Gamma = 3$ in Table~IV of ref.~\citenum{pocrnic2026prebo}) and $\tilde{\Gamma}_{ij}^2 = 1$ otherwise, capping electron interactions at $1/\Delta$ and nuclear repulsion at $1/(\sqrt{8}\Delta)$.

A candidate circuit over Clifford+$T$ and mid-circuit measurement must realize a single clean-ancilla unitary $W$ satisfying Eq.~\eqref{eq:ham_sim} at $\varepsilon = 10^{-2}$ for $t = 1$~fs on every state of the system register. Matching the baseline~\cite{pocrnic2026prebo}, initial-state preparation and observable readout are not costed, since in end-to-end fault-tolerant simulations of photochemical dynamics, real-time Hamiltonian evolution dominates initial-state preparation by orders of magnitude~\cite{eklund2026endtoend}. All ancillas begin in $\ket{0}^{\otimes m}$, so resource states such as phase-gradient states must be prepared within the circuit. Every Toffoli is compiled into Clifford+$T$ gates and checked at that gate level by the correctness proof, and where later gates are conditioned on measurement outcomes we count the $T$ gates of the most expensive branch.

The baseline instead bounds only the block-encoding error $\norm{A - e^{i\varphi}e^{-iHt}}_{\mathrm{op}} \le \epsilon_{\mathrm{BE}}$ (Appendix~C of ref.~\citenum{pocrnic2026prebo}), where $A = (I \otimes \bra{0^m})\,W\,(I \otimes \ket{0^m})$ is the part of the circuit that starts and ends with the ancillas in $\ket{0^m}$. This leaves out the ancilla error, the part $\ket{\psi_\perp}$ of the output in which the ancillas are not returned to $\ket{0^m}$. The two parts are orthogonal, so the error in Eq.~\eqref{eq:ham_sim} splits as
\begin{equation}
  \norm{W(\ket{\psi}\otimes\ket{0^m}) - e^{i\varphi}(e^{-iHt}\ket{\psi})\otimes\ket{0^m}}_2^2 = \norm{(A - e^{i\varphi}e^{-iHt})\ket{\psi}}_2^2 + \norm{\ket{\psi_\perp}}_2^2,
  \label{eq:block_to_clean}
\end{equation}
and any circuit meeting Eq.~\eqref{eq:ham_sim} at $\varepsilon$ also meets the block-encoding criterion at $\epsilon_{\mathrm{BE}} = \varepsilon$. For the baseline QSP circuit, the polynomial approximation error bounds the ancilla error~\cite{low2019qubitization,motlagh2023gqsp,berry2024doubling} by $\norm{\ket{\psi_\perp}}_2^2 \le 2\epsilon_{\mathrm{exp}}$, so with the error split in Table~IV of ref.~\citenum{pocrnic2026prebo} ($\epsilon_{\mathrm{exp}} = \epsilon_{\mathrm{BE}}/12.95$) it is only guaranteed to meet Eq.~\eqref{eq:ham_sim} at $\sqrt{\epsilon_{\mathrm{BE}}^2 + 2\epsilon_{\mathrm{exp}}} \approx 0.041$, four times our target.

\subsection{Baseline}
\label{app:nonbo_baseline}

Table~I of ref.~\citenum{pocrnic2026prebo} reports $N_{\mathrm{Toff}}^{\mathrm{pub}} = 2.73 \times 10^{11}$ Toffolis per femtosecond on $Q_{\mathrm{pub}} = 6{,}198$ qubits at $\epsilon_{\mathrm{BE}} = 10^{-2}$, a spacetime cost of $6.77 \times 10^{15}$. Each walk step uses about $9.0 \times 10^{4}$ $T$ gates, of which $52\%$ go to the swap network, $25\%$ to preparing the LCU weights and $22\%$ to evaluating $1/r$ (Fig.~5a of ref.~\citenum{pocrnic2026prebo}). These give the baseline values of $46.8\text{k}$, $22.5\text{k}$ and $19.8\text{k}$ $T$ gates per step in Sec.~\ref{sec:photochem} and Fig.~\ref{fig:applications}e--g.

The baseline costs the time evolution as the number of walk steps times the cost of one step, $N_{\mathrm{Toff}} = D_{\mathrm{old}}(\tau, \epsilon_{\mathrm{exp}}) \cdot \mathcal{T}_H$ (Eqs.~(C1) and~(C4)--(C5) of ref.~\citenum{pocrnic2026prebo}), where the number of steps is the degree of the generalized QSP polynomial~\cite{motlagh2023gqsp,berry2024doubling}. Two of the same authors tighten this degree bound, removing the factor $e/2 \approx 1.36$ in front of $\tau$ (Theorem~1 of ref.~\citenum{pocrnic2026constants}, via Kapteyn's inequality),
\begin{align}
  D_{\mathrm{old}}(\tau, \epsilon_{\mathrm{exp}}) &= \left\lceil \frac{e}{2}\,\tau + \ln\frac{2c}{\epsilon_{\mathrm{exp}}} \right\rceil, \nonumber\\
  d_{\mathrm{Thm\,1}}(\tau, \epsilon_{\mathrm{exp}}) &= \left\lceil \tau \cosh\!\left(\left[\frac{3}{\tau}\ln\frac{2}{\epsilon_{\mathrm{exp}}(1 - e^{-\theta_0})}\right]^{1/3}\right) \right\rceil,
  \label{eq:nonbo_degrees}
\end{align}
where $\tau := \alpha_H t$, $c = 4(\sqrt{2\pi}\,e^{1/13})^{-1} \approx 1.4776$ and $\theta_0 = \bigl[\frac{3}{\tau}\ln(2/\epsilon_{\mathrm{exp}})\bigr]^{1/3}$. Theorem~2 there further tightens the subleading term via Watson's inequality. That work~\cite{pocrnic2026constants} does not apply this bound to the system studied here, but it improves the same QSP construction, so we include it so that our solutions are compared against the strongest available baseline.

The tighter bound only reduces the number of walk steps. The error split $t\epsilon_H + \epsilon_{\mathrm{exp}} \le \epsilon_{\mathrm{BE}}$ fixes $\epsilon_H$ independently of the polynomial degree, so the $T$-gate count $\mathcal{T}_H$ per step and the qubit count $Q_{\mathrm{pub}} = 6{,}198$ are unchanged. At the reported $\alpha_H = 2.16 \times 10^{5}$~Ha ($\tau \approx 8.93 \times 10^{6}$ at $t = 41.34$~a.u.) and $\epsilon_{\mathrm{exp}} = \epsilon_{\mathrm{BE}} / 12.95$, Theorem~2 lowers the number of walk steps from $D_{\mathrm{old}} = 12{,}136{,}779$ to $d_{\mathrm{Thm\,2}} = 8{,}930{,}416$ ($8{,}930{,}877$ with Theorem~1). Rescaling the published Toffoli count by this ratio gives our baseline,
\begin{align}
  N_{\mathrm{Toff}}^{\mathrm{base}} &= N_{\mathrm{Toff}}^{\mathrm{pub}}\,\frac{d_{\mathrm{Thm\,2}}}{D_{\mathrm{old}}} \approx 2.01 \times 10^{11}, \nonumber\\
  \mathcal{T}_{\mathrm{base}} &= 4N_{\mathrm{Toff}}^{\mathrm{base}} \approx 8.04 \times 10^{11}, \nonumber\\
  Q_{\mathrm{pub}}\cdot\mathcal{T}_{\mathrm{base}} &\approx 4.98 \times 10^{15},
  \label{eq:nonbo_base_cost}
\end{align}
the number presented in Fig.~\ref{fig:results}a. At this large $\tau$ the subleading terms barely matter: using Theorem~1 instead, or simply dividing the published Toffoli count by $e/2$, gives the same cost to three significant figures.

\subsection{Main improvements}
\label{app:nonbo_improvements}

Section~\ref{sec:photochem} summarizes three changes to the walk step that account for almost all of the reduction in $T$ gates per step relative to the baseline (Fig.~\ref{fig:applications}e--g). Here we describe each one alongside the baseline construction it replaces, and Appendix~\ref{app:nonbo_resources} gives the resulting $T$ counts.

First, to apply a two-body interaction, the baseline swaps the coordinates of the selected pair into the two particle registers on which its interaction circuit acts, applies the interaction there, and swaps them back afterward~\cite{pocrnic2026prebo}. Instead, our solution copies the selected coordinates into an ancilla register, leaving the particle register untouched. Because the ancilla holds only a copy, it can be cleared using mid-circuit Pauli-$X$ measurements and classically controlled $CZ$ corrections~\cite{gidney2018halving,berry2019qubitization}, as done for momentum bits copied in plane-wave kinetic terms~\cite{su2021firstquant} (Fig.~\ref{fig:applications}e).

Second, the potential $V$ in Eq.~\eqref{eq:nonbo_ham} requires computing $1/r = (r^2)^{-1/2}$ for $r = \norm{q_i - q_j}_2$. Because the coordinates are stored as integers, $r^2 = \sum_w (q_{i,w} - q_{j,w})^2$ can be obtained cheaply using integer subtraction and squaring. To get to $1/r$ without a square root, the baseline prepares an ancilla register in uniform superposition over integers $m < M$, assigning $+1$ to branches with $m^2 r^2 < M^2$ and an alternating sign $(-1)^m$ to the rest. The alternating terms cancel, so the average over branches approximates $1/r$, but the test requires multiplying $m^2$ by $r^2$, which dominates its cost~\cite{pocrnic2026prebo}. In this work, the range of $r^2$ is split into coarse intervals with rescaled position $\tilde{r}^2 \in [0,1)$, and $1/r$ on each interval is approximated by a cubic Bernstein polynomial $\sum_{k=0}^{3} \binom{3}{k} (\tilde{r}^2)^k (1-\tilde{r}^2)^{3-k} w_k$ ($512$ precomputed coefficients $w_k$ in total). Rather than multiplying powers of $\tilde{r}^2$, the circuit samples this polynomial directly~\cite{nacu2005coins,goyal2012bernstein}. It compares $\tilde{r}^2$ against three ancilla registers in uniform superposition and counts how many are smaller, which gives $k \in \{0,1,2,3\}$ with exactly the binomial weights above. Looking up $w_k$ then gives the polynomial in expectation (Fig.~\ref{fig:applications}f).

Finally, in the electron--nucleus terms of $V$, each nucleus $i$ is weighted by its integer charge $Z_i$. Rather than preparing these weights via coherent alias sampling~\cite{babbush2018linear} with amplitude amplification~\cite{pocrnic2026prebo}, a uniform superposition of basis states is prepared and a reversible classical circuit assigns $Z_i$ states to each nucleus $i$, as proposed for fixed nuclei~\cite{su2021firstquant} and for quantum nuclei~\cite{eklund2026endtoend}. Our solution applies this to every weight of the linear combination (Fig.~\ref{fig:applications}g).

\subsection{Resource counts}
\label{app:nonbo_resources}

The verified circuit found in this work uses $Q = 6{,}033$ logical qubits, consisting of $5{,}616$ system qubits and $417$ clean ancillas. Evaluating the $1$-norm at $L = 44$~\AA\ together with potential shifting gives block-encoding scale $\alpha = 8.045\times 10^{4}$~Ha and generalized QSP degree $d = 3{,}326{,}847$.

Each uncontrolled walk step uses $30{,}116$ $T$ gates ($30{,}132\;T$ on the two boundary controlled steps): $1{,}256\;T$ for reversible nuclear-charge and pair-weight preparation, $24{,}194\;T$ to copy selected particle coordinates into ancillas and $908\;T$ for $X$-basis measurement uncomputation ($25{,}102\;T$ total routing, against $46.8\text{k}\;T$ for the baseline swap network), $3{,}060\;T$ for the $512$-entry Bernstein $1/r$ evaluation, $446\;T$ for the QFT kinetic term, and $252\;T$ for the LCU reflection, plus $76\;T$ per step for QSP rotations and $245\;T$ of one-time overhead. The Lean solution fixes both the $3{,}326{,}848$ generalized QSP rotations~\cite{motlagh2023gqsp,berry2024doubling} and the Clifford+$T$ words that approximate them by existence proofs, the words through a derandomized form of unitary mixing over a finite table~\cite{campbell2017shorter,hastings2017turning}, so it certifies their $T$ count and error without listing them. The rotations are computable in time polynomial in $d$~\cite{haah2019product} but have not been computed at this degree. The same holds for the baseline~\cite{pocrnic2026prebo,pocrnic2026constants}, which uses the same doubled construction~\cite{berry2024doubling} and a larger degree polynomial which would require greater classical computation.
This totals $\mathcal{T} = 3{,}326{,}847 \times (30{,}116 + 76) + 2 \times 30{,}132 + 245 = 100{,}444{,}225{,}133 \approx 1.00\times 10^{11}$ $T$ gates and $Q\cdot \mathcal{T} = 605{,}980{,}010{,}227{,}389 \approx 6.06\times 10^{14}$ ($8.22\times$ below baseline).

\section{Cytochrome P450 Compound~I}
\label{app:p450}
\subsection{Hamiltonian and specification}
\label{app:p450_window}

The system is active space~X of Compound~I, with $\eta = 63$ electrons in $N = 58$ spatial orbitals~\cite{goings2022p450}. Under the Jordan--Wigner transformation each spin orbital $(p, \sigma)$ is stored on qubit $2p + \sigma$, giving $Q_{\mathrm{sys}} = 2N = 116$ system qubits.

GPU-accelerated density matrix renormalization group (DMRG) calculations now converge this $58$-orbital active space classically~\cite{menczer2024p450}, yet leave the physical spin-state ordering unresolved because dynamic correlation from omitted orbitals exceeds the sub-millihartree doublet--quartet gap~\cite{goings2022p450,menczer2024p450}. Active space~X nonetheless serves as the standard benchmark across fault-tolerant resource estimates~\cite{goings2022p450,caesura2025blissthc,low2025spectrum}, and circuit improvements to its DFTHC walk structure~\cite{low2025spectrum} carry over to the larger active spaces needed to capture dynamic correlation and resolve the spin ordering~\cite{menczer2024p450}.

The Hamiltonian is the $(R, B, C) = (9, 29, 14)$ double-factorized tensor-hypercontraction (DFTHC) factorization of Table~V in ref.~\citenum{low2025spectrum}, with the released tensors~\cite{rubin2025dfthcdata} rounded to $(b_{\mathrm{coeff}}, b_{\mathrm{rot}}) = (10, 15)$ bits:
\begin{equation}
  H_{\mathrm{DFTHC}} = \sum_{\sigma,p,q} h'_{pq}\, a_{p\sigma}^\dagger a_{q\sigma}
  + \frac{1}{2} \sum_{r=0}^{R-1}\sum_{c=0}^{C-1} \Biggl(W^{(rc)} I + \sum_{b=0}^{B-1} w_b^{(rc)} n_{u_b^{(r)}}\Biggr)^2,
  \label{eq:p450_dfthc}
\end{equation}
where $n_u = \sum_{\sigma} (\sum_p u_p a_{p\sigma})^\dagger (\sum_q u_q a_{q\sigma})$ for unit orbital vectors $u_b^{(r)} \in \mathbb{R}^{58}$ ($M = RB = 261$) and $W^{(rc)} = w_B^{(rc)} - \sum_{b<B} w_b^{(rc)}$, as follows from expanding $O_{\mathrm{SF},rc}$ in the DFTHC construction~\cite{low2025spectrum}. As in the released dataset~\cite{rubin2025dfthcdata}, Eq.~\eqref{eq:p450_dfthc} omits the scalar $-\frac{1}{2}\sum_{r,c} (W^{(rc)})^2$, and adding the classical Hartree--Fock shift $E_{\mathrm{HF}} - E_{\mathrm{HF}}^{\mathrm{DFTHC}}$ recovers the active-space total energy (Eqs.~42--43 of ref.~\citenum{low2025spectrum}). We define the problem to not allow for improvement in the DFTHC factors~\cite{rubin2025dfthcdata}, which reproduce the active-space coupled-cluster correlation energy to $0.28$~mHa, to separate quantum circuit optimization from classical tensor fitting.

We cost one run of phase estimation on the $116$ system qubits, started in a $63$-electron eigenstate $\ket{\psi}$ of $H_{\mathrm{DFTHC}}$ with energy $E$. Assuming an eigenstate as input loses no generality for resource estimation: phase estimation on any state $\ket{\phi}$ gives, with probability $|\langle\psi_j|\phi\rangle|^2$, the same outcomes~\cite{abrams1999eigenvalues} as on the eigenstate $\ket{\psi_j}$. In practice $\ket{\phi}$ only approximates the ground state $\ket{\psi_0}$, so finding $E_0$ takes on average $1/|\langle\psi_0|\phi\rangle|^2$ runs, each also paying for trial-state preparation~\cite{lee2023evaluating}. Like prior resource estimates~\cite{goings2022p450,caesura2025blissthc,low2025spectrum}, we do not count this overhead. The circuit uses Clifford+$T$ gates and mid-circuit measurements on the system and $m$ ancillas that start in $\ket{0}$. The specification applies to this gate-level circuit, so its error bound includes rotation synthesis and approximate state preparation as well as phase estimation. At the end, a $b$-qubit phase register among the ancillas is measured, and its outcome $y \in \{0, \dots, 2^b - 1\}$ is converted into the energy estimate
\begin{equation}
  \hat{E}(y) = E_{\mathrm{off}} + s\,\lvert\cos(\pi y / 2^{b})\rvert,
  \label{eq:p450_decoder}
\end{equation}
where $b$, $E_{\mathrm{off}}$ and $s$ are fixed numbers chosen with the circuit. This form covers both standard qubitization and sum-of-squares spectral amplification. For a walk eigenphase $\varphi$, the energy is $E = \lambda\cos\varphi$ in the first~\cite{babbush2018linear,low2019qubitization,lee2021thc,caesura2025blissthc} and $E = E_{\mathrm{SOS}} + \Lambda(1 + \cos\varphi)$ in the second~\cite{low2025spectrum,king2026sossa}. Since $\cos\varphi \le 0$ at the energies measured here, both read $E = E_{\mathrm{off}} + s\,\lvert\cos\varphi\rvert$, with $(E_{\mathrm{off}}, s) = (0, -\lambda)$ and $(E_{\mathrm{SOS}} + \Lambda, -\Lambda)$ respectively. When phase estimation applies the walk forward or backward depending on the phase register, $2\pi y/2^{b}$ estimates $\pm 2\varphi$, so Eq.~\eqref{eq:p450_decoder} estimates $E$.

If the specification only covered the ground state, a circuit with no gates and $s = 0$ would pass by setting $E_{\mathrm{off}}$ within $\varepsilon$ of the ground-state energy $E_0$. The specification therefore requires the same circuit and decoder to work for every eigenstate up to $E_{\mathrm{top}} = E_0 + \Delta$, with $\Delta = 0.1$~Ha and $\varepsilon = 10^{-3}$~Ha. For each $63$-electron eigenstate $\ket{\psi}$ with energy $E$, the outcome distribution $p(y \mid \psi)$ must satisfy
\begin{equation}
  \sum_{y} p(y \mid \psi)\,\bigl(\min(\hat{E}(y), E_{\mathrm{top}}) - \min(E, E_{\mathrm{top}})\bigr)^2 \;\le\; \varepsilon^2.
  \label{eq:measures_low_energies}
\end{equation}
This is the root-mean-square error behind the phase-estimation step counts of Eqs.~(21)--(23) in ref.~\citenum{babbush2018linear}. For an eigenstate inside the window, it bounds the root-mean-square error of the estimate by $\varepsilon$, with estimates above $E_{\mathrm{top}}$ counted as $E_{\mathrm{top}}$. For an eigenstate above the window, it only requires that estimates do not fall more than $\varepsilon$ below $E_{\mathrm{top}}$. A circuit that ignores $\ket{\psi}$ gives the same estimates for every eigenstate, and in root mean square these would have to lie within $\varepsilon$ of $E_0$ and at most $\varepsilon$ below $E_0 + \Delta$ at once. This is possible only if $\Delta \le 2\varepsilon = 2$~mHa, far below $\Delta = 0.1$~Ha. The window also spans the gap of about $75$~mHa between the quartet and sextet states found by DMRG (Fig.~4 of ref.~\citenum{menczer2024p450}). The tolerance $\varepsilon = 1$~mHa matches prior resource estimates on this active space, although separating the doublet and quartet states would require tighter tolerances.

All ancillas begin in $\ket{0}^{\otimes m}$, so resource states such as phase-gradient states must be prepared within the circuit. Every Toffoli is compiled into Clifford+$T$ gates and checked at that gate level by the correctness proof, and where later gates are conditioned on measurement outcomes we count the $T$ gates of the most expensive branch.

\subsection{Baseline}
\label{app:p450_baseline}

Our baseline starts from the sum-of-squares spectral amplification estimate in Table~V of ref.~\citenum{low2025spectrum} ($51{,}510$ walk steps at $9{,}535$ Toffolis/step on $1{,}150$ qubits, or a spacetime cost of $2.26\times 10^{12}$), where $H_{\mathrm{DFTHC}} - E_{\mathrm{SOS}} = H_{\mathrm{sqrt}}^\dagger H_{\mathrm{sqrt}}$ is block-encoded with norm $\Lambda$ and energy-window gap $E_{\mathrm{gap}} = E_{\mathrm{ub}} + \Delta - E_{\mathrm{SOS}}$ using the phase-estimation step count~\cite{babbush2018linear,low2025spectrum}
\begin{equation}
  \lambda_{\mathrm{eff}} = \sqrt{E_{\mathrm{gap}}(2\Lambda - E_{\mathrm{gap}})}, \qquad
  N_{\mathrm{steps}} = \left\lceil \frac{\pi \lambda_{\mathrm{eff}}}{2\varepsilon} \right\rceil.
  \label{eq:lambda_eff}
\end{equation}
Adding $\Delta$ to $E_{\mathrm{gap}}$ in that estimate~\cite{low2025spectrum} raises its step count from $51{,}510$ to $51{,}948$ and its cost to $2.28\times 10^{12}$ ($+0.85\%$).
The published construction~\cite{low2025spectrum} forms the one-body term $h'$ from the exact two-electron tensor $g$, while the squares in Eq.~\eqref{eq:p450_dfthc} use the factorized tensor $\tilde{g}$. The Hamiltonian it block-encodes therefore differs from the $(h, \tilde{g})$ Hamiltonian by a one-body term that lowers the low-lying energies by about $0.6$~Ha. Its $E_{\mathrm{gap}}$ subtracts the sum-of-squares bound of the block-encoded Hamiltonian from the sextet Hartree--Fock energy of the $(h, \tilde{g})$ Hamiltonian~\cite{rubin2025dfthcdata}, so it is about $0.6$~Ha larger than when both energies come from one Hamiltonian, which overestimates the cost. We form $h'$ from $\tilde{g}$, so that Eq.~\eqref{eq:p450_dfthc} is the $(h, \tilde{g})$ Hamiltonian up to a constant, whose CCSD(T) correlation energy is within $0.28$~mHa of that of $(h, g)$. This lowers $\Lambda$ from $97.4395$ to $97.3864$~Ha ($E_{\mathrm{SOS}} = -134.8874$~Ha), and with both energies evaluated on this Hamiltonian $E_{\mathrm{gap}} = 5.189$~Ha including $\Delta$ ($49{,}269$ steps). Bounding $E_0 \le E_{\mathrm{ub}} = -129.8443$~Ha with the quartet ($S = 3/2$) ROHF determinant instead lowers $E_{\mathrm{gap}}$ to $5.143$~Ha, giving $\lambda_{\mathrm{eff}} = 31.23$~Ha and $N_{\mathrm{steps}} = 49{,}056$.

Adding a coherent $16$-qubit phase-estimation register and storing each of the $57$ Givens angles in $13$ rather than $15$ bits gives $1{,}052$ qubits and a cost of $1.97\times 10^{12}$ before space--time tradeoffs. Thirteen bits suffice because the angles $\theta_p = 2\pi k_p / 2^{15}$ enter the orbital vectors~\cite{lee2021thc} as $\cos 2\theta_p$ and $\sin 2\theta_p$, and the overall sign of each vector does not affect $n_u$, so fixing that sign puts every angle index of $H_{\mathrm{DFTHC}}$ in $0 \le k_p < 2^{13}$. As suggested in prior work~\cite{caesura2025blissthc}, we use both these observations and integrate them into our baseline.
We then apply recent compilation options~\cite{low2026denser}, recomputing the $(G, r, c, \sigma, b = B)$ index registers from an $8$-bit outer index $x_o$ with minimum-Toffoli comparators~\cite{khattar2025rise} and loading the $57$ Givens angles in three batches of $19$ ($5$ QROM lookups per \textsc{select}) while reusing idle angle wires as clean-QROAM~\cite{berry2019qubitization} workspace. These options come from later work than the spectral-amplification study~\cite{low2025spectrum}, but we include them so that our solutions are compared against the strongest available baseline. On our $13$-bit instance we also pad $C$ rather than $R$ in $x_o$, skip the two always-zero addend bits in the $15$-bit phase-gradient adders~\cite{gidney2018halving}, and include $32$ Toffolis/step to control the walk reflections from the phase-estimation register, omitting approximate state preparation and rotation synthesis so $H_{\mathrm{DFTHC}}$ remains exact.
Minimizing $Q \cdot \mathcal{T}$ over QROAM block sizes and angle batch sizes gives $Q_{\mathrm{base}} = 481$ logical qubits and $C_{\mathrm{BE}} = 11{,}517$ Toffolis/step ($5.65\times 10^{8}$ Toffolis, $\mathcal{T}_{\mathrm{base}} = 4 C_{\mathrm{BE}} N_{\mathrm{steps}} = 2.26\times 10^{9}$ $T$ gates), for a baseline cost of
\begin{equation}
  Q_{\mathrm{base}}\cdot\mathcal{T}_{\mathrm{base}} = 1.087 \times 10^{12} \approx 1.09 \times 10^{12},
  \label{eq:p450_st_base}
\end{equation}
the number presented in Fig.~\ref{fig:results}a. At a fixed number of walk steps, the $13$-bit angles and these compilation options~\cite{low2026denser} together roughly halve the published cost~\cite{low2025spectrum}, from $1{,}150$ qubits and $9{,}535$ Toffolis per step to $481$ qubits and $11{,}517$, including the phase-estimation register. Fig.~\ref{fig:results}b shows the published estimate~\cite{low2025spectrum}, extended to the energy window, in gray and this baseline in red.

\subsection{Main improvements}
\label{app:p450_improvements}

Section~\ref{sec:p450} summarizes the main change in our solution, which replaces the angle lookups of the baseline with a one-hot register for the chosen orbital vector, and uses the same trick in the inner state preparation. Here we describe both alongside the baseline construction they replace, together with a smaller change to the spin swaps that the main text leaves out. Appendix~\ref{app:p450_resources} gives the resulting $T$ counts.

Each walk step block-encodes Eq.~\eqref{eq:p450_dfthc} with a state preparation and two \textsc{select} calls~\cite{low2025spectrum}. The outer stage of the state preparation picks a square $(r, c)$ or a one-body term, and the inner stage picks $b$, which fixes the orbital vector $u_b^{(r)}$. \textsc{select} rotates this vector into a single orbital with $57$ Givens rotations, acts on that orbital and rotates back~\cite{lee2021thc}.

In the baseline, each \textsc{select} call loads the $13$-bit Givens angles of the chosen vector into a $247$-qubit register, $19$ at a time with $5$ QROM lookups per call, and each rotation adds its angle into the phase-gradient register (Appendix~\ref{app:p450_baseline}). Our solution instead builds a $145$-qubit one-hot register in each \textsc{select} call, for $296$~$T$. Each of its qubits stands for two orbital vectors, so the register covers $290$ of them: the $261$ vectors $u_b^{(r)}$ and $29$ of the $58$ eigenvectors of $h'$. The other $29$ eigenvectors are handled by the second call of the step. A further qubit $s$ records which vector of the pair was chosen. Each angle bit then enters the phase-gradient adder of its rotation as
\begin{equation}
  (\alpha \oplus M_1)(\beta \oplus M_0) \oplus M_{01}, \qquad \beta = q_c s, \quad \alpha = q_c \oplus \beta,
  \label{eq:p450_angle_bit}
\end{equation}
where $q_c$ is the control qubit of the rotation and $M_0$, $M_1$ and $M_{01}$ are CNOT parities of the one-hot qubits whose first vector, second vector or both vectors have that bit set. At most one one-hot qubit is set, so $M_0 M_1 = M_{01}$, and Eq.~\eqref{eq:p450_angle_bit} is $0$ for $q_c = 0$ and the bit of the chosen vector for $q_c = 1$. The baseline already uses one AND gate per angle bit to control the rotation. Eq.~\eqref{eq:p450_angle_bit} also needs one AND gate per bit, plus one per rotation for $\beta$. With the one-hot register, the rotations and their angles cost $11{,}818$~$T$ per call rather than $18{,}844$~$T$, and the angle register is no longer needed.

The inner stage uses the same idea. It keeps a $126$-qubit one-hot register for $(r, c)$ and a $5$-bit binary register for $b$, so each alias-sampling lookup~\cite{babbush2018linear} costs $390$~$T$, against $1{,}968$~$T$ for the QROAM of the baseline. Together with the one-hot \textsc{select}, this accounts for $96\%$ of the $T$ gates saved per walk step (Table~\ref{tab:p450_step}).

Our solution also includes a second, smaller improvement, which halves the cost of the spin swaps. Rather than performing the basis rotations once for each spin, the baseline applies them to the spin-up qubits only and, controlled by a qubit in superposition, swaps the spin-down qubits into their place before and back after, with four blocks of $58$ controlled swaps per step~\cite{low2025spectrum}. Since swapping two qubits only flips the sign of their antisymmetric state $(\ket{01}-\ket{10})/\sqrt{2}$, each block can instead be implemented as a single controlled sign flip, applied when an odd number of the $58$ orbitals are in that state. After a Clifford change of basis to Bell pairs, the circuit computes this parity with $58$ Toffolis, or $232$~$T$, as many as one block of swaps, and erases it with $X$-basis measurements for free. Since the system register is untouched between consecutive \textsc{select} calls, one parity serves two swap blocks.

\subsection{Resource counts}
\label{app:p450_resources}

The best verified circuit uses $Q = 333$ logical qubits and $\mathcal{T} = 1{,}310{,}839{,}337 \approx 1.31\times 10^{9}$ $T$ gates, so $Q\cdot \mathcal{T} = 436{,}509{,}499{,}221 \approx 4.37\times 10^{11}$, $2.49\times$ below the baseline (Fig.~\ref{fig:results}a,b). This comes from $1.67\times$ fewer $T$ gates per walk step (Appendix~\ref{app:p450_improvements}), $1.44\times$ fewer qubits and $3\%$ fewer walk steps. The qubits are the $116$ system qubits, a $16$-qubit phase-estimation register, an $11$-qubit phase-gradient register, $17$ and $24$ qubits for the outer and inner state-preparation registers, and $149$ qubits of workspace shared by the subroutines. Most of the saving against the $481$ qubits of the baseline comes from dropping its $247$-qubit angle register.

The circuit runs $M = 47{,}487$ walk steps. We bound $E_0 \le E_{\mathrm{ub}} = -130.1526$~Ha with a $3{,}000$-determinant quartet wavefunction, whose energy is checked in Lean. This is $0.31$~Ha below the quartet ROHF bound of the baseline and gives $E_{\mathrm{gap}} = 4.835$~Ha and $\lambda_{\mathrm{eff}} = 30.30$~Ha, $3\%$ below the $31.23$~Ha of the baseline. A slightly tighter error bound for phase estimation than Eq.~\eqref{eq:lambda_eff} saves another $0.2\%$. The state preparation is approximate, so the block-encoded Hamiltonian differs from $H_{\mathrm{DFTHC}}$ by at most $4.0\times 10^{-7}$~Ha on the $63$-electron sector. This comes from the synthesized rotations in the state preparation, and the correctness proof adds it to the phase-estimation error to show that the total stays within $\varepsilon$.

\begin{table}[t]
\centering
\begin{tabular}{lrr}
\toprule
Part of a walk step & Baseline & This work \\
\midrule
Rotations and their angles & $37{,}688$ & $23{,}636$ \\
Index decoding & $264$ & $440$ \\
Spin swaps & $928$ & $464$ \\
Outer state preparation & $1{,}032$ & $598$ \\
Inner state preparation & $5{,}888$ & $2{,}264$ \\
\midrule
Total & $45{,}800$ & $27{,}402$ \\
\bottomrule
\end{tabular}
\caption{$T$ gates per walk step for the baseline and for a step inside the first eight blocks ($45{,}440$ steps) of the best verified circuit. Reflections and phase-estimation control are counted separately in the text.}
\label{tab:p450_step}
\end{table}

The walk steps run in $16$ blocks totaling $47{,}360$ steps, followed by $127$ single steps. As qubits of the phase-estimation register are prepared stage by stage, fewer remain free as extra workspace for \textsc{select}: $8$ blocks ($45{,}440$ steps in total) borrow $11$ unprepared qubits for $160$ workspace qubits at the $27{,}402$~$T$ per step of Table~\ref{tab:p450_step}, $1$ block of $1{,}024$ steps borrows $10$ qubits ($28{,}298$~$T$ per step), and $7$ blocks of $128$ steps borrow $7$ qubits ($31{,}010$~$T$ per step). On top of Table~\ref{tab:p450_step}, the baseline spends $268$~$T$ per step on reflections and phase-estimation control, for $46{,}068$~$T$ in total. Each of our blocks uses $84$~$T$ for each reflection between its steps and $232$~$T$ for an ordinary first swap block. The $127$ single steps cost $37{,}642$~$T$ each (including $418$~$T$ per inner lookup), because all phase-estimation qubits are then in use and \textsc{select} has to fit in $149$ qubits of workspace.
The reflections between blocks and between single steps, together with phase-estimation control, take $13{,}316$~$T$, with $31$ reflections controlled by the phase-estimation register. Preparing the initial state of the phase-estimation register takes $142{,}248$~$T$, the inverse QFT $12{,}603$~$T$ and the phase-gradient state $1{,}036$~$T$. Adding these up gives the $\mathcal{T} = 1{,}310{,}839{,}337$ quoted above.

\printbibliography[segment=\therefsegment, title={Appendix References}]
\end{refsegment}

\end{document}